\documentclass[aps,prd,twocolumn,groupedaddress,nofootinbib]{revtex4-1}

\usepackage{graphicx}
\usepackage[caption=false]{subfig}
\usepackage{dcolumn}
\usepackage{multirow}
\usepackage{tabularx}
\usepackage{bm}
\usepackage{hyperref}
\usepackage{amsmath}
\usepackage{amssymb}
\usepackage{comment}
\usepackage{color}
\usepackage{enumerate}
\usepackage{booktabs}
\usepackage{siunitx}
 \graphicspath{{figures/}}

\newcommand{\MeV}{\text{\,MeV}}

\newcommand{\eV}{\text{~eV}}

\newcommand{\cm}{\, \text{cm}}
\newcommand{\second}{\text{~s}}
\newcommand{\yr}{\text{~yr}}
\newcommand{\msun}{M_{\odot}}
\newcommand{\Hz}{\text{~Hz}}
\newcommand{\kHz}{\text{~kHz}}

\newcommand{\OO}{\mathcal{O}}
\newcommand{\PP}{\mathbb{P}}

\usepackage[normalem]{ulem}

\def\apj{\ref@jnl{ApJ}}                 

\newcolumntype{L}[1]{>{\hsize=#1\hsize\raggedright\arraybackslash}X}%
\newcolumntype{R}[1]{>{\hsize=#1\hsize\raggedleft\arraybackslash}X}%
\newcolumntype{C}[1]{>{\hsize=#1\hsize\centering\arraybackslash}X}%

\makeatletter
\newcommand{\thickhline}{%
    \noalign {\ifnum 0=`}\fi \hrule height 2pt
    \futurelet \reserved@a \@xhline
}
\newcolumntype{"}{@{\hskip\tabcolsep\vrule width 2pt\hskip\tabcolsep}}
\makeatother

\begin{document}

\title{Black Hole Superradiance Signatures of Ultralight Vectors}

\author{Masha Baryakhtar}
\email{mbaryakhtar@perimeterinstitute.ca}
\affiliation{Perimeter Institute for Theoretical Physics, Waterloo, Ontario N2L 2Y5, Canada}
\author{Mae Teo} 
\email{maehwee@stanford.edu}
\affiliation{Stanford Institute for Theoretical Physics, Stanford University, Stanford, California 94305, USA}
\author{Robert Lasenby} 
\email{rlasenby@perimeterinstitute.ca}
\affiliation{Perimeter Institute for Theoretical Physics, Waterloo, Ontario N2L 2Y5, Canada}

\date{\today}

\begin{abstract}
  The process of superradiance can extract angular momentum and energy
  from astrophysical black holes (BHs) to populate
  gravitationally-bound states with an exponentially large number of
  light bosons. We analytically calculate superradiant growth rates
  for vectors around rotating BHs in the regime where the vector
  Compton wavelength is much larger than the BH size. Spin-1 bound
  states have superradiance times as short as a second around stellar
  BHs, growing up to a thousand times faster than their spin-0
  counterparts. The fast rates allow us to use
  measurements of rapidly spinning BHs in X-ray binaries to exclude a
  wide range of masses for weakly-coupled spin-1 particles,
  $5\times 10^{-14} - 2\times 10^{-11}$~eV; lighter masses in the
  range $6\times 10^{-20} - 2\times 10^{-17}$~eV start to be
  constrained by supermassive BH spin measurements at a lower level of
  confidence. We also explore routes to detection of new vector
  particles possible with the advent of gravitational wave (GW)
  astronomy. The LIGO-Virgo collaboration could discover hints of a
  new light vector particle in statistical analyses of masses and
  spins of merging BHs. Vector annihilations source continuous
  monochromatic gravitational radiation which could be observed by
  current GW observatories. At design sensitivity, Advanced LIGO may
  measure up to thousands of annihilation signals from within the
  Milky Way, while hundreds of BHs born in binary mergers across the
  observable universe may superradiate vector bound states and become
  new beacons of monochromatic gravitational waves.

\end{abstract}

\maketitle

\section{Introduction}
\label{sec:intro}
Light, weakly-coupled new particles are a feature of many beyond
Standard Model (SM) physics scenarios \cite{Essig:2013lka,Arvanitaki:2009fg,Arvanitaki:2009hb}, and depending on how they
couple to the SM, there are a plethora of ways in which they can be
searched for experimentally. At an absolute minimum, any new particles
must interact gravitationally. However, the very high value of the
Planck scale results in very small rates for the purely-gravitational
production of new particles.

One way to get around very small production rates is to take advantage
of coherent enhancement. As in a laser, it can lead to exponential
amplification of a very small initial population (including vacuum
fluctuations). For light bosonic particles, such a setup can be
realized around a spinning black hole (BH), where gravitational bound
states of the bosonic field can extract energy and angular momentum
from the BH through `rotational superradiance', the wave analogue of
the Penrose process \cite{Zeldovich,Misner:1972kx,Starobinskii}. This
process results in the exponential growth of some bound states,
eventually spinning down the BH and creating a very high occupation
number bosonic `cloud' around it.

Using BH superradiance as a tool to search for new light bosons was
proposed in \cite{Arvanitaki:2009fg}, and spin-0 superradiance
dynamics and phenomenology have been studied in many
papers~\cite{Ternov:1978gq,Zouros:1979iw,Detweiler:1980uk,Dolan:2007mj,Arvanitaki:2010sy,
  Yoshino:2013ofa,
  Arvanitaki:2014wva,Brito:2014wla,Brito:2015oca,Arvanitaki:2016qwi}.
While there are superradiant (growing) bound states for any scalar
mass, the growth rates are very suppressed for Compton wavelengths
significantly larger or smaller than the BH size. Observationally, the
two main signatures are the lack of rapidly-spinning BHs, and the
monochromatic gravitational waves (GWs) sourced by the bosonic cloud
itself~\cite{Arvanitaki:2010sy, Arvanitaki:2014wva}. Precise
measurements of BH spins through X-ray observations constrain
sufficiently weakly coupled spin-0 particles over a wide mass range,
while near-term gravitational wave observations have the potential to
discover such particles at other masses~\cite{Arvanitaki:2010sy,
  Arvanitaki:2014wva,Arvanitaki:2016qwi}.

In this paper, we study the theory of spin-1 superradiance around BHs,
and use the associated astrophysical phenomenology to search for and
constrain gravitationally-coupled massive vector particles below
$10^{-11}$~eV. It is technically natural for vectors to have small
masses, and there are a number of string theory constructions which
result in very light and weakly coupled spin-1
particles~\cite{Goodsell:2009xc,Camara:2011jg}. Computationally, the
extra degrees of freedom in vector fields, as compared to scalar
fields, make solving for the behavior of bound states more
challenging. However, for vector Compton wavelengths larger than the
BH size, the bound states are `non-relativistic' and hydrogen-like,
and we analytically compute the growth and decay rates of such
states. Our calculation applies around Kerr BHs of arbitrary spin,
which is significant since astrophysical signatures from superradiance are
dominated by fast-spinning BHs.  The dynamics of light vector fields
around slowly rotating and/or Schwarzschild BHs have been studied in a
number of previous works, including a combination of analytic and
numerical techniques~\cite{Rosa:2011my,Pani:2012bp,Pani:2012vp} and
EFT operator matching~\cite{Endlich:2016jgc}.  Our results agree with
numerical computations in the close-to-Schwarzschild limit, and
explain the scaling of superradiance rates with vector mass, as found in the numerical
computations of~\cite{Rosa:2011my}.  Extrapolated to more relativistic
bound states, they are also in approximate agreement with
time-domain numerical results~\cite{East:2017mrj,Witek:2012tr}.  These results
are summarized in Section~\ref{sec:srrates} and discussed in detail in
Appendix~\ref{ap:rates}.

The fact that a vector bound state can carry spin angular momentum
means that bound states which carry no `orbital' angular momentum can
still grow by superradiant extraction of BH spin.  The hydrogenic
wavefunctions of such bound states have larger density near the BH
than superradiant light scalar bound states, since the latter must
carry orbital angular momentum. As a result, vector bound state growth
rates can be parametrically larger than those for scalars.
Consequently, vector superradiance can be fast enough to be
astrophysically relevant over a wider range of vector and BH
masses. In Section~\ref{sec:bounds}, we derive constraints on a range
of vector masses from existing measurements of rapidly rotating BHs,
and describe future signals that may be observed in statistical
analyses of BH spin-mass distributions in LIGO data.

In Section~\ref{sec:directsignatures}, we compute approximate GW
emission rates from the non-relativistic vector bound states, with
further discussion in Appendix~\ref{ap:gwrates}. We estimate the
expected number of sources that GW observatories such as Advanced LIGO
could observe in continuous wave searches for coherent, monochromatic
GWs. The higher density of the zero orbital angular momentum vector
bound state near the BH, where the potential is fast-varying, results
in a larger high-momentum component able to emit GWs, and gives a
parametrically larger emission rate than superradiant scalar bound
states. As a result, GW signals from vector bound states could be
visible over cosmological distances, from across a large fraction of
the observable universe.

In this paper, we consider light vectors with purely gravitational
couplings. If a vector has other couplings to itself or to SM matter,
our results apply if they are small enough such that gravitational
interactions with the BH remain dominant. We leave the question of how
large such couplings would have to be to change this story, and
whether such large couplings could have other observational
consequences, to future work.

\section{Vector bound states and superradiance}
\label{sec:boundstates}

\subsection{Spin-1 superradiance}

Superradiant scattering of massless spin-1 particles (e.g.\ the
photon) in a Kerr background was calculated by~\cite{Teukolsky:1974yv}. As for spin-0
particles, the condition for a mode to be amplified is for the
angular velocity of the BH horizon to be larger than the angular phase
velocity of the wave mode,
\begin{equation}
	\frac{\omega}{m} < \Omega_H \,,
	\label{eq:srcond}
\end{equation}
where $\omega$ is the frequency of the mode, $m$ its angular momentum
about the BH spin axis, and $\Omega_H$ the angular velocity of the
BH horizon. This condition on the energy to angular momentum ratio of the
wave can be derived from the black hole area theorem; superradiant
modes are those whose emission increases the BH area, so their
emission is allowed classically \cite{PhysRevD.7.949}.

Rotational superradiance of spin-1 particles can occur
around systems other than black holes. Historically, the first
example worked out was Zeldovich's calculation~\cite{Zeldovich} 
for superradiant scattering of EM waves from a conducting
cylinder. In general, whenever an object is able to absorb waves, and
the superradiance condition in Eq.~\ref{eq:srcond} holds (with
$\Omega_H$ taken to be the angular velocity of the object),
it will be thermodynamically favored for the object to spin down
by emitting superradiant wave modes.
For the gravitationally-coupled vectors we are considering, this means
that spinning objects other than BHs, such as neutron
stars, planets, etc., will also superradiate.
The reason why we focus on BHs is that the amplification rates
for superradiant vector modes around other objects will generally
be many, many orders of magnitude lower. This is because
other objects absorb gravitationally coupled vectors much less
efficiently than black holes, and spin much slower than a near-extremal
BH of comparable size.

Since there are no gravitational bound states of a massless vector
around a BH, superradiant amplification only occurs during a single
`pass', and amplifies the incoming energy by at most
$4.4\%$~\cite{Teukolsky:1974yv}. The `black hole bomb' thought
experiment of~\cite{Press:1972zz} evaded this limitation by
placing a mirror around the BH. A superradiant mode confined by the
mirror can then be amplified exponentially, eventually extracting an
$\OO(1)$ fraction of the BH's spin. The addition of a vector mass
effectively allows the BH's gravitational potential to act as a
`mirror', confining superradiant modes around the black hole and
enabling their exponential growth. The rest of this section will
investigate the dynamics of these gravitationally-bound states.

\subsection{Bound states}

We consider a free massive spin-1 field $A_\mu$ with
Lagrangian\footnote{We use the $-+++$ signature.}
\begin{equation}
\mathcal{L_{A}} = -\frac{1}{4} F_{\mu\nu}F^{\mu\nu} - \frac{1}{2} \mu^{2} A_{\mu}A^{\mu}\,.
\end{equation}
In this paper, we assume a Stuckelberg-type mass for the vector, i.e.\ there are no additional light degrees of freedom;
we comment briefly on the problems associated with a low-scale Higgs mechanism
in Section~\ref{sec:interactions}.
$A^\mu$ obeys the Proca equation of motion,
\begin{equation}
	D_\mu F^{\mu\nu} = \mu^2 A^\nu \,,
\end{equation}
where $F_{\mu\nu} \equiv \partial_\mu A_\nu - \partial_\nu A_\mu$
is the usual field strength.
This implies the `Lorentz condition' $D_\mu A^\mu = 0$.
To find the bound states of a massive vector around a BH,
we need to solve the Proca equation in the Kerr metric
(as we will see, the BH horizon means that there are generally
only metastable `quasi-bound' states \cite{Page:1976df}).
In general, this has to be done numerically. However,
for vectors with Compton wavelength large compared to the size
of the black hole, we expect the $r_g/r$ part of the metric
to be most important, where $r_g \equiv G M_{\rm BH}$.
In these circumstances, the bound states are
`non-relativistic' and hydrogen-like.

Non-relativistic bound states will oscillate with frequency $\omega
\simeq \mu$, where we write
\begin{equation}
	A^\mu(t,x) = \frac{1}{\sqrt{2\mu}} \left(\Psi^\mu(x) e^{- i \omega t} + 
	{\rm c.c.}\right) \,.
\end{equation}
Quasi-bound states will have complex $\omega$ (when we compare
$\omega$ to real quantities, e.g.\ inequalities such
as $\omega < m \Omega_H$, we implicitly mean the real part of $\omega$).
For $r \gg r_g$, we will 
assume that $\Psi^\mu$ varies slowly on scales $\mu^{-1}$,
i.e. that its momentum components are non-relativistic,
and also that the metric is close to flat.
Then, keeping only the terms 
that are not suppressed by small momenta,
the Proca equation becomes
\begin{equation}
	(\omega^2 - \mu^2) \Psi^\nu \simeq - \nabla^2 \Psi^\nu + \omega^2 (1 + g^{00}) \Psi^\nu.
\end{equation}
In the Kerr metric at $r \gg r_g$, $g^{00} \simeq -(1 - 2 r_g/r)$ so
we have a Schr\"odinger-type equation describing motion in a $1/r$
potential,
\begin{equation}
	(\omega-\mu) \Psi^\nu \simeq -\frac{\nabla^2}{2 \mu} \Psi^\nu + \frac{\alpha}{r} \Psi^\nu\,,
\end{equation}
where $\alpha\equiv r_g\mu$.

Using the Lorentz condition, $
\partial_t A_0 \simeq \partial_i A_i$,
we can solve for $\Psi_0$ in terms of $\Psi_i$, and find all of the
bound states by solving the $\Psi_i$ Schr\"odinger equation.  Since
the $1/r$ part of the potential is spherically symmetric, we can (at
leading order) separate $\Psi_i$ into radial and angular functions,
\begin{equation}
  \Psi_i = R^{n\ell}(r)Y_i^{\ell,jm}(\theta,\phi),
  \end{equation}
where the angular functions are `pure-orbital vector spherical harmonics'~\cite{thorne},
which
are eigenfunctions of the orbital angular momentum operator:
$-r^2 \nabla^2 Y_i^{\ell,jm} = \ell(\ell+1) Y_i^{\ell,jm}$.
The $j$ label corresponds to the total angular momentum of the state,
while $m$ is the total angular momentum along the $z$ axis, and $\ell \in \{j-1,j,j+1\}$
is the `orbital angular momentum' (see
App.~\ref{ap:rates}).\footnote{Our notation differs from some existing literature:
  \cite{Rosa:2011my} mis-identifies $\ell$ as the total angular momentum
  of the bound states,
  and~\cite{Rosa:2011my,Pani:2012bp,Pani:2012vp} have $\ell$ and $j$
  swapped compared to our usage. Our notation is the same as \cite{Endlich:2016jgc}.
  The basis used in \cite{Rosa:2011my} is the `pure-spin vector spherical harmonics' of~\cite{thorne};
  the linear combinations of these found in \cite{Rosa:2011my} for non-relativistic bound
  states at long distances are precisely the pure-orbital harmonics.
}

The equation for the radial wavefunction $R^{n\ell}(r)$ is the same as for the
scalar Coulomb problem, so the bound state radial wavefunctions
are hydrogenic, labelled by orbital angular momentum
$\ell$ and overtone number $n$.
The leading-order energy levels are hydrogen-like,
\begin{equation}
	\omega \simeq \mu \left(1 - \frac{\alpha^2}{2 (n + \ell + 1)^2}\right) \,.
\end{equation}

\begin{figure}[t]
\includegraphics[width = .62 \columnwidth]{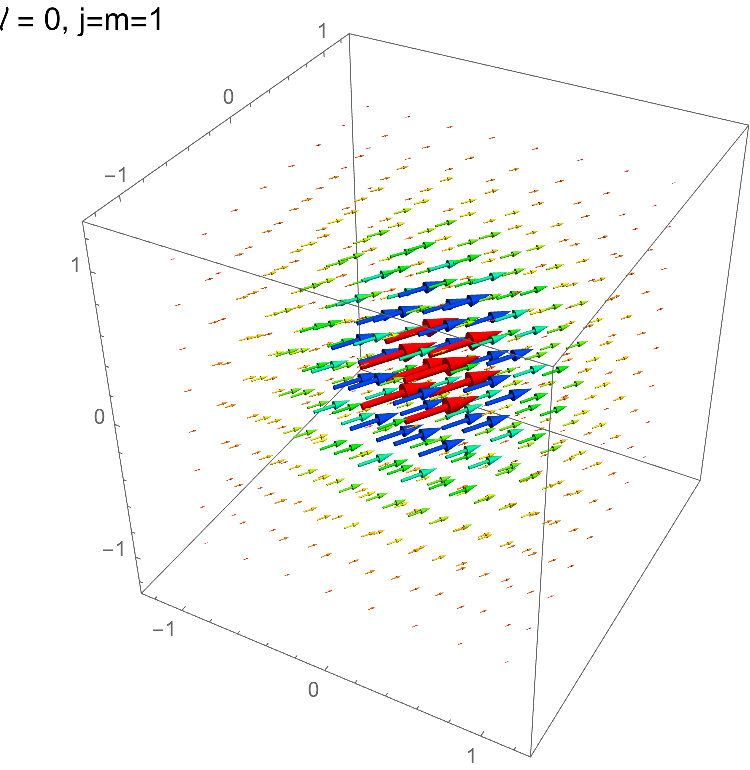}
\includegraphics[width = .62 \columnwidth]{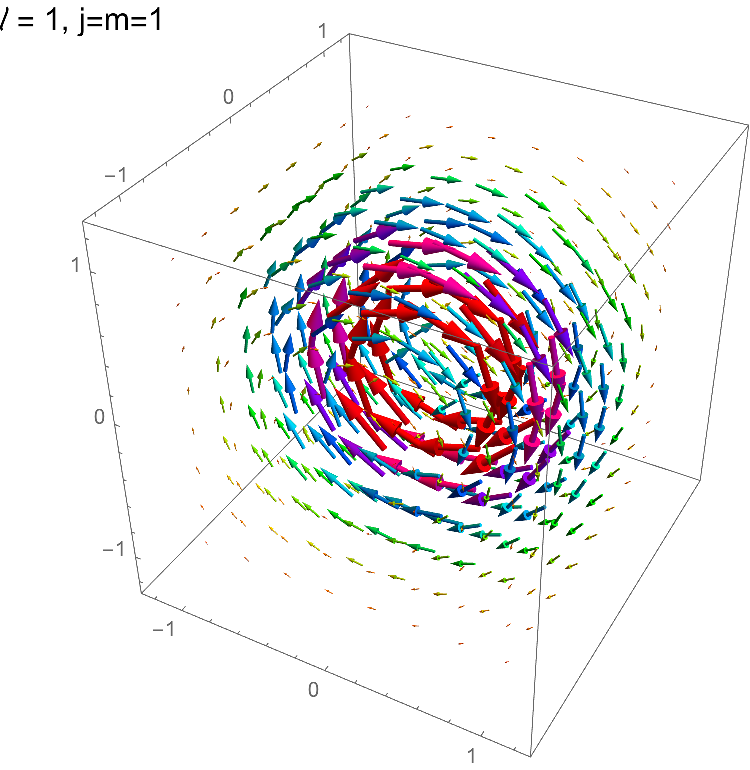}
\includegraphics[width = .62 \columnwidth]{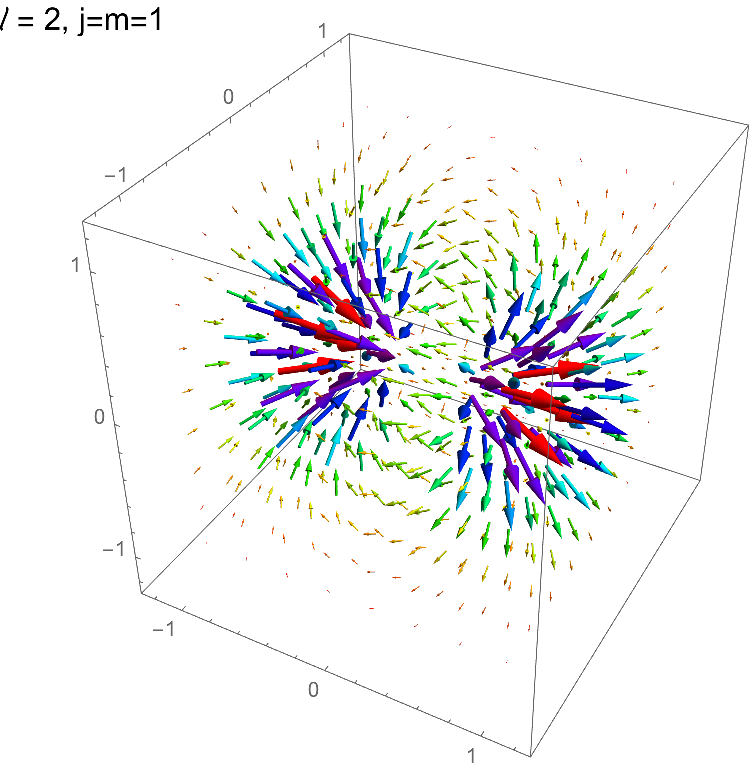}
\caption{The $A_i$ field at a moment in time for the
  lowest-radial-overtone $j=1$, $m=1$ hydrogenic bound states, with $\ell=0,1,2$
  from top to bottom.  The size and color of an arrow correspond to
  the magnitude of $A_i$ at that point.  The change in the field over
  time corresponds to rotating each figure about its vertical axis,
  with angular frequency $\omega \simeq \mu$.  }\label{fig:Afield}
\end{figure}

Figure~\ref{fig:Afield} shows the $A_i$ fields at a moment in time,
for a set of $j=1$ bound states. These illustrate how the
amplitude varies with radius in a hydrogen-like way (controlled by $\ell$),
and how the total angular momentum is a combination
of spin and orbital components.
As we will see in the next section, the fastest-growing bound
state around a fast-spinning BH will be the $\ell=0$, $j=m=1$ state
with $n=0$.
The real time-dependent fields for this state
have leading-$\alpha$ form
\begin{equation}
	A_i = \frac{1}{\sqrt\pi \mu^{1/2} a^{3/2}} e^{-r/a} 
	\begin{pmatrix}
		-\cos\omega t \\ -\sin \omega t \\ 0
	\end{pmatrix}\,,
\end{equation}
\begin{equation}
	A_0 = \frac{1}{\sqrt\pi \mu^{3/2} a^{5/2}} e^{-r/a} \sin\theta \cos (\omega t - \phi)\,,
\end{equation}
where $a \equiv 1/(\mu \alpha)$ is the characteristic size of the bound state.

\subsection{Superradiance rates}
\label{sec:srrates}

If a bound state satisfies the $\Omega_H > \omega/m$ superradiance
condition, then --- assuming that gravitational perturbations to the Kerr
background are small --- it will grow with time, exponentially increasing in amplitude
as it extracts energy and angular momentum from the BH.
In particular, this process is possible for any mass $\mu$ (though for large
$\mu$, only large-$m$ modes will satisfy the superradiance condition, Eq.~\ref{eq:srcond}).
However, for superradiance to be important for astrophysical BHs,
the growth rate of bound states needs to be fast, at the very
least compared to the age of the BH. As in the case of scalar bound 
states~\cite{Dolan:2007mj,Arvanitaki:2010sy},
it will turn out that the growth rate is a rapidly-varying function of $\alpha$,
falling off as a high power of $\alpha$ for $\alpha \ll 1$,
and exponentially suppressed for superradiant states with $\alpha \gg 1$.

There are no full analytic solutions for scalar or vector bound states in a general Kerr background, and in most cases, no approximate
analytic solution that is close to the true solution at all radii.
So, to obtain bound state growth and decay rates, 
one either needs to solve the wave equation numerically, or
match together approximate solutions with different regimes of validity.
In Appendix~\ref{ap:rates}, we review these different approaches,
and perform an analytic matching calculation which gives
the leading form  of the growth and decay rates in the small-$\alpha$ approximation.
Far away from the BH, the bound state wavefunctions are hydrogen-like,
as described above. Close to the BH, the mass term in the Proca wave equation
becomes a sub-leading correction, allowing us to use the behaviour
of massless waves in a Kerr background to determine the energy flux
across the horizon.
Matching these two regimes together, the leading-$\alpha$ growth rate for a vector
bound state scales as
\begin{equation} \label{eq:scalingvector}
	\Gamma \sim \alpha^{2j + 2\ell + 5} (m \Omega_H - \omega)\, ,
\end{equation}
(where we have also taken $m \Omega_H - \omega$ to be a small parameter,
to show the behavior near the $\omega = m \Omega_H$ crossover point)
compared to the scalar form 
\begin{equation} \label{eq:scalingscalar}
	\Gamma_{\rm scalar} \sim \alpha^{4\ell + 5} (m \Omega_H - \omega) \,.
\end{equation}
In both cases the rate scales as $\alpha^{2\ell + 2j + 5}$ --- the
difference in the vector case is that we do not always have $j=\ell$.
Roughly speaking, the more localized the bound state 
wavefunction is close to the BH, and the smaller the total
angular momentum barrier it faces to reach the horizon,
the larger the growth or decay rate.

\begin{figure}[t]
	\includegraphics[width=.99\columnwidth]{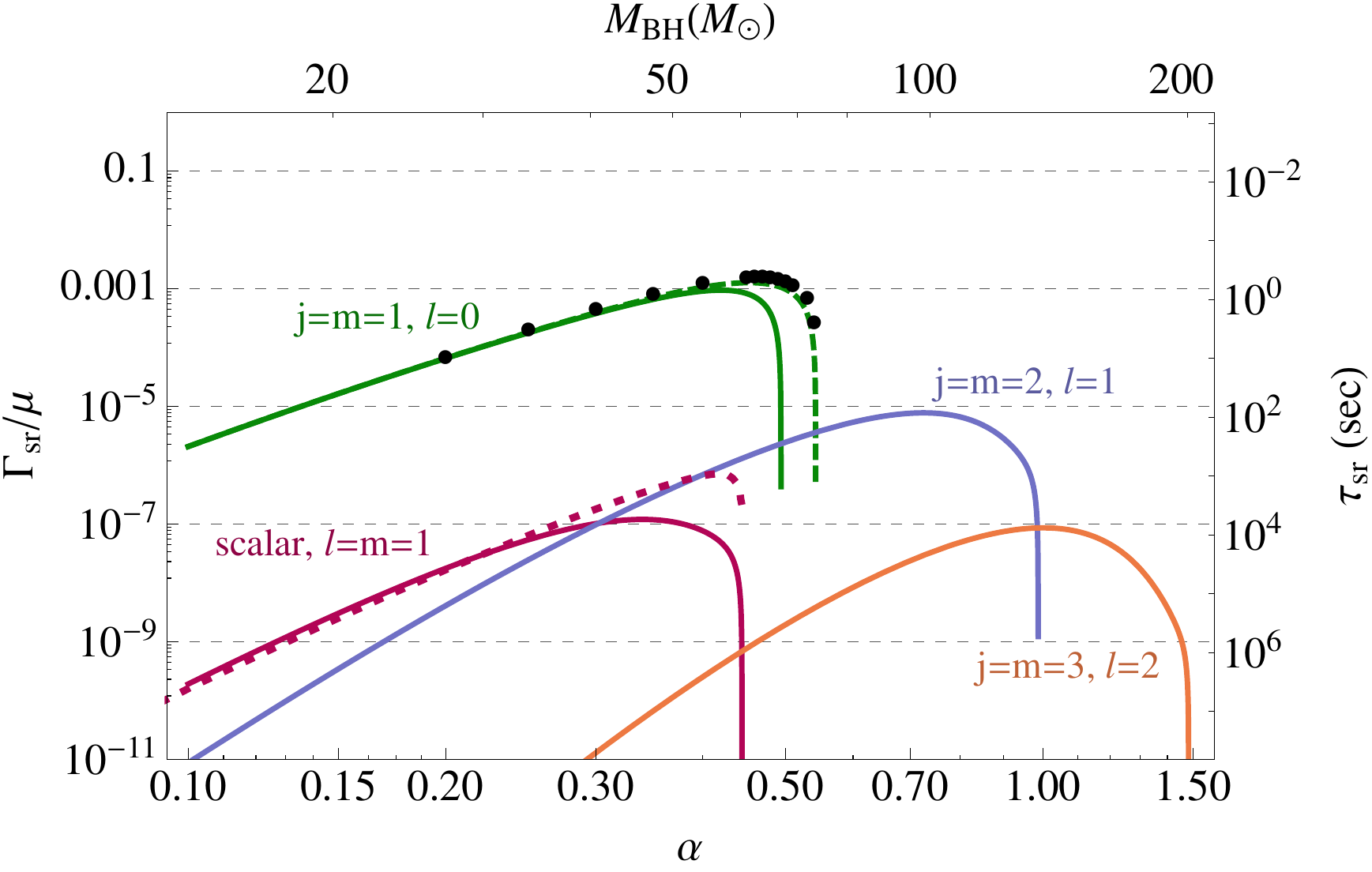}
	\caption{Analytic approximations to superradiance rates for
          the fastest-growing vector bound states, at each value $m$ of
          angular momentum about the BH spin axis, for a Kerr BH with
          $a_* = 0.99$ (upper solid curves). The black data points are
          full numerical results from \cite{East:2017mrj}. For the
          highest superradiance rate, including the subleading
          correction to bound state energy found in
          \cite{East:2017mrj} improves the fit to numerical results at
          high $\alpha$ (dashed green curve). The lower red curves
          show the numerical results~\cite{Dolan:2007mj} (dotted) and
          analytic approximation (dashed) for the superradiance rate
          of the fastest-growing scalar bound state.  The top axis
          shows the BH mass corresponding to a given $\alpha$ for
          vector mass $\mu = 10^{-12} \eV$.  The right-hand axis shows
          the growth e-folding time in seconds, again for
          $\mu = 10^{-12} \eV$.  The curve for the $\ell = 0$,
          $j = m = 1$ mode is the full leading-$\alpha$ rate, along
          with higher-order corrections from $m \Omega_H - \omega$
          terms (see Appendix~\ref{ap:rates}).  The curves for the
          $\ell,j,m = 1,2,2$ and $\ell,j,m = 2,3,3$ modes show the
          sub-component of the superradiance rates due to Poynting
          flux through the BH horizon. These are underestimates of the
          full leading-$\alpha$ rates (Appendix~\ref{ap:rates}), but
          have the correct scaling in the $\alpha \ll 1$ limit.}
	\label{fig:vrates}
\end{figure}

The difference between scalar and vector rates has the important
consequence that the fastest-superradiating mode for a vector, the
$\ell = 0, j = m = 1$ mode, has leading-$\alpha$ growth rate
\footnote{These are the growth rates for the energy density of the
  bound states (equivalently, for their `occupation number'); the
  growth rate for the field amplitude is a factor $1/2$ smaller,
  with ${\rm Im} (\omega) = \Gamma/2$.
  }
\begin{equation}
	\Gamma \simeq 4 a_* \alpha^6 \mu \,,
\end{equation}
where $a_* \equiv J_{BH} / (G M_{BH}^2) \in [0,1)$ is the BH spin, which is related to the horizon angular velocity by
$\Omega_H = \frac{1}{2} \left( \frac{a_*}{1+\sqrt{1-a_*^2}} \right) \mu \alpha^{-1}$.
In contrast, the fastest-growing level for a light scalar, the $\ell=m=1$ level,
has $\Gamma \simeq \frac{1}{24} a_* \alpha^8 \mu$.	Vector bound states
can therefore, in the non-relativistic limit, grow significantly faster
than scalar ones, as illustrated in Figure~\ref{fig:vrates}.

The strong $\alpha^6$ dependence of the superradiance
rate implies that, for a given BH, vectors with $\mu \ll r_g^{-1}$
will have very small superradiance rates.
For $\mu \gg r_g^{-1}$, states must have $m \gg 1$ to be superradiant.
The ingoing component must therefore tunnel through a large
angular momentum barrier, which results in an exponential suppression
of the tunnelling rate~\cite{Zouros:1979iw,Arvanitaki:2010sy}.
Thus, vector masses much larger than $r_g^{-1}$ also
have very small superradiance rates. Combining these results,
we expect astrophysically significant bound state growth rates
to occur for vectors whose Compton wavelength is of approximately
the same order as the size of the BH.

Figure~\ref{fig:vrates} shows the superradiance rates for the
fastest-growing bound states at each $m$ value, around a
very-fast-spinning ($a_* = 0.99$) Kerr BH, illustrating the
properties discussed above. For comparison, we also
show the superradiance rate for the fastest-growing scalar
mode, including both the small-$\alpha$ analytic approximation 
and numerical results from~\cite{Dolan:2007mj}.
\cite{East:2017mrj} performs a full numerical computation
of the (time-domain) evolution of vector bound states in a Kerr background.
Comparing their results for the fastest-growing vector
mode to our analytic calculation, we find good agreement
for the superradiant growth rate of the fastest-growing level
(to within $\OO(1)$ across the entire range).

\subsection{Black hole evolution by superradiance}
\label{sec:history}

The growth of superradiant bound states can begin spontaneously,
without the need for any pre-existing density 
of the vector field. Though the calculations in this paper
are all on the level of classical fields, at the quantum
level one can view superradiance as amplifying vacuum fluctuations,
with decoherence turning the resulting state into a mixture
of coherent, classical-like bound state oscillations.
Consequently, while the light, weakly-coupled vector fields we consider
could be dark matter candidates, superradiance will occur
whether or not there is a pre-existing astrophysical abundance.

If the superradiance time of the fastest-growing bound state --- with
$j=m=\ell+1$--- is significantly smaller than the age of the BH, then
its exponential growth will extract enough energy and angular momentum
from the BH to bring $\Omega_H$ down to the superradiance boundary for
that level.  For a BH starting with spin $a_{*,0}$, if the cloud
extracts a fraction $\Delta a_* / a_{*,0}$ of its initial angular
momentum, the final occupation number\footnote{The bound states grow
  into coherent classical oscillations of the vector fields, so are
  not Fock states with a definite occupation number. However, the
  occupation numbers involved are so large that the fractional
  variance in occupation number for the coherent states is tiny, so we
  can sensibly discuss bound state occupation numbers.  } of the cloud
will be
\begin{equation}
	N_m \simeq \frac{G M_{BH}^2 \Delta a_*}{m}
	\simeq \frac{10^{77} }{m} \left(\frac{M_{BH}}{10 M_\odot}\right)^2 \frac{\Delta a_*}{0.1}\,;
	\label{eq:nmax}
\end{equation}
this behavior is borne out by full numerical time-domain simulations
\cite{East:2017ovw}. Since the angular momentum of fast-spinning
stellar-mass BHs is so high, it takes the cloud
$\sim \log (N_m) \sim 180$ e-folds of growth, starting from vacuum
fluctuations, to extract an order-1 fraction of the BH's angular
momentum. For small $\alpha$, this corresponds to the extraction of a
small proportion of the black hole's mass, since
$\Delta M_{BH} \simeq \mu N_m \simeq \alpha M_{BH} \Delta a_* / m$,
and $\alpha/m < \frac{1}{2} \frac{a_*}{1 + \sqrt{1 - a_*^2}}$ for a
superradiant level. If annihilation rates of vectors to gravitational
waves (Section~\ref{sec:directsignatures}) are sufficiently rapid, the
cloud may not reach maximum size before annihilations begin to deplete
it; however, we do not expect this to be the case in the
non-relativistic limit.

\begin{figure}[t]
\includegraphics[width = .99 \columnwidth]{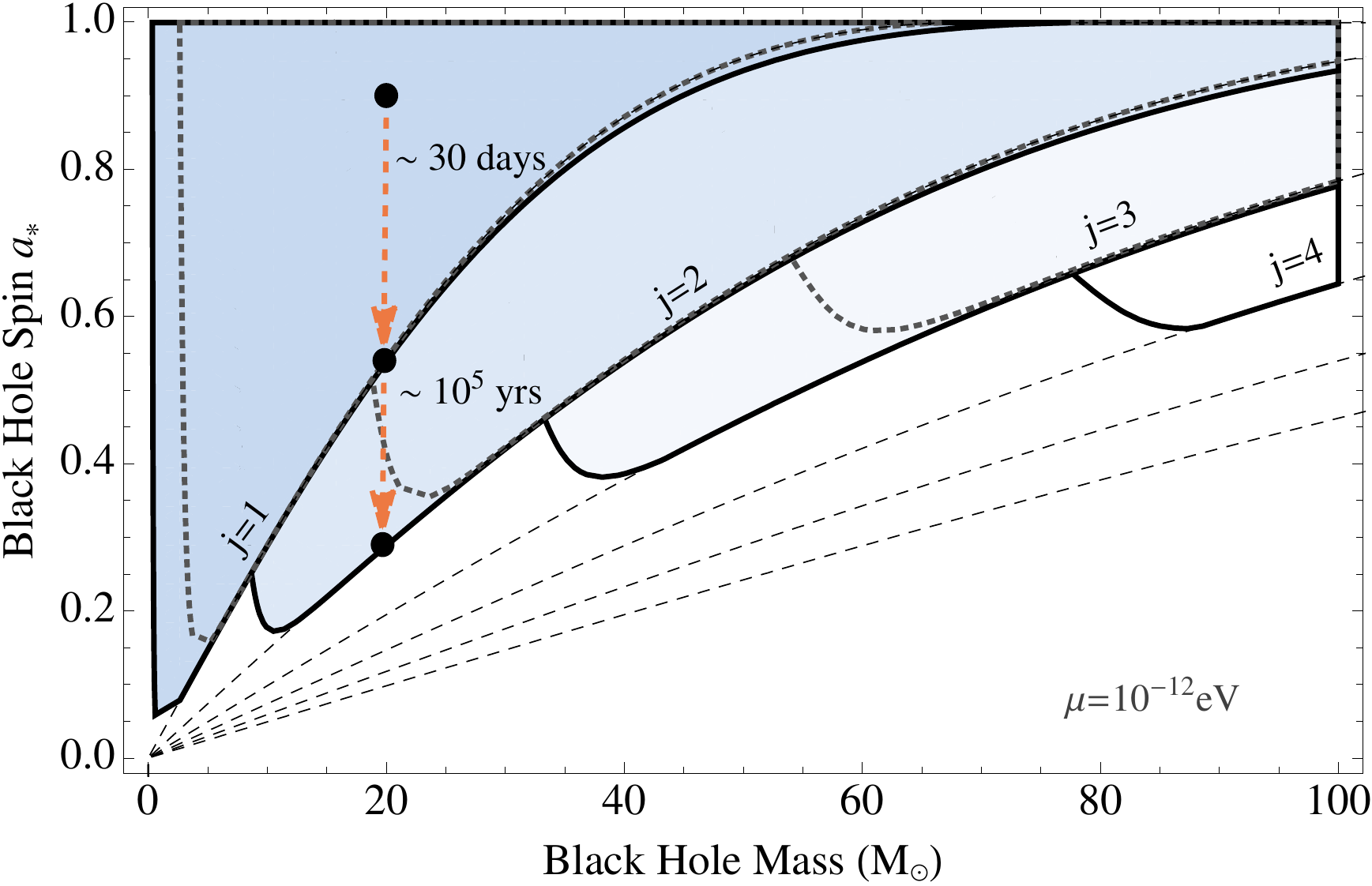}
\caption{Effect of superradiance on a BH due to a vector (bold) or
  scalar (dotted) with mass $\mu=10^{-12}$ eV. Shaded regions above
  the lines correspond to BH parameters that result in spin-down
  within a typical binary lifetime
  $\Gamma_m > \frac{1}{\tau_{bh}}\log N_m$, shown for
  $\tau_{\text{bh}} = 5 \times 10^{6}$ years, for levels with total
  angular momentum $j=1$ to $j=4$. We also show an example evolution
  in the spin-mass plane of a $20M_{\odot}$ BH with initial spin
  $a_{*}=0.9$. The dashed
	lines correspond to the superradiance boundaries for
	each $m$ value.}\label{fig:regge}
\end{figure}

After the growth of a given level stops, higher-$m$ modes will still be
superradiant. If the fastest-growing of these has a sufficiently small
superradiance time, the process will repeat, with the BH being spun
down to the next superradiance boundary.  Since it takes $\sim 180/m$
e-folds of growth for a bound state level to extract a significant
fraction of a roughly-stellar-mass BH's spin, a BH
spends most of its lifetime very near each superradiance boundary,
with the transits between them occurring mostly in the last $e$-fold of
each level's growth.

Figure~\ref{fig:regge} illustrates this evolution for a BH
with initial mass $20 M_\odot$ and initial spin $a_* = 0.9$, assuming
the existence of a weakly-coupled vector of mass $\mu = 10^{-12} \eV$
(giving $\alpha = 0.15$). The strong dependence of the superradiance
rate on $m$ means that, though the $m=1$ and $m=2$ levels grow quickly
compared to astrophysical timescales, the $m=3$ level does not have
time to grow in the age of the universe, so the BH spends
almost its entire lifetime on the $m=2$ superradiance boundary, at
$a_* \simeq 0.3$ and $M \simeq 18.5 M_\odot$. The solid curves plotted
in Figure~\ref{fig:bhpoints} are the isocontours of
$\Gamma_m = \frac{1}{\tau_{bh}}\log N_m$, where $\tau_{bh}$ is the
lifetime of the BH (here taken to be $5 \times 10^6 \yr$). If a black
hole is born above these curves, then it will be spun down by the
growth of successive superradiant levels until it reaches a
superradiance boundary on or below the curves. If it is born below the
curves, it will be unaffected by
superradiance. Figure~\ref{fig:bhpoints} illustrates the results of
these processes on a selection of initial BH masses and spins, showing
how superradiance depopulates the region above the superradiance rate
contours, and leaves a population of BHs scattered along the
superradiance boundary curves (with the slight complication that the
figure considers BHs in binary systems, where a sufficiently close
companion can disrupt superradiance, as discussed in
Section~\ref{sec:perturb}).

As we will see in Section~\ref{sec:directsignatures}, bound states are
depleted by annihilations to gravitational radiation once they have
grown to large occupation number. In most circumstances, the
difference between superradiant growth rates for successive $m$ is
large enough that, on a timescale short compared to the growth
timescale of the next level, the previous level's occupation number
has mostly annihilated away. Once the
next level has extracted enough spin from the BH, the previous level
goes from being almost on the superradiance boundary, to decaying into
the BH. The result is that the small occupation number that did not
annihilate away falls back into the BH, delaying the spin-down process
very slightly (though not making a significant impact).\footnote{This
  pattern of levels sequentially growing from, and then falling back
  into, a BH, is explored by~\cite{Bosch:2016vcp,Sanchis-Gual:2016tcm}   for the case of charged field superradiance of a Reissner-Nordstrom BH.}

\subsection{Perturbations}
\label{sec:perturb}

The presence of a perturbation to the BH gravitational potential can
affect superradiance by mixing a growing level with a decaying one. In
particular, a binary companion sources a dipole-type gravitational
potential (to leading order in the ratio of BH size to companion
distance), which can mix levels which have a small energy splitting
and different angular momentum numbers.

For vectors, the $j=1, \ell=0$ level --- which dominates the
phenomenology --- is especially robust against perturbations because
there are no bound states that decay with a rate faster than its
superradiance rate. Furthermore, it is the state with the lowest
energy, with a large $\OO(\alpha^2)$ energy difference from other
levels, which further suppresses the mixing. For example, for binary
systems the largest mixing term is the dipole term with
$\Delta j = \Delta \ell =1$. So, at leading order, the fastest-growing
$j=1$ level can only mix with the $N=2$, $j=0$, $\ell=1$, $m=0$ state,
where $N = n+\ell + 1$ is the principal quantum number.  Because these
states have different energies ($\Delta N \neq 0$), the mixing is
further suppressed, and an external perturbation has to be close to
order one to significantly affect whether the state is
superradiant. For typical binary systems, this translates into the
condition that the cloud is sufficiently tightly bound,
$\alpha \gtrsim 10^{-3}$, and does not affect the signals or
constraints. More details for this estimate, as well as perturbations
of $j>1$ levels, are covered in Appendix~\ref{ap:mixing}.

\subsection{Effects of interactions}
\label{sec:interactions}

The BH evolution described above only takes into account
the gravitational interactions of the light vector.
The vector may also have non-gravitational interactions with SM matter.
As long as such couplings always have sub-dominant effects to the
gravitational potential of the BH, superradiance proceeds
broadly as we have described.

There is also the possibility of the vector having non-gravitational
interactions with other non-SM states --- for example, if it
gets its mass via a Higgs mechanism. Such states will induce
self-interactions of the vector, and if the superradiant cloud
reaches a high enough density, may be produced on-shell
(in analogy to Schwinger pair production).
The energy density of a cloud that has extracted $\OO(1)$ 
of a BH's spin is
\begin{equation}
 \rho \! \sim \! \frac{\alpha M}{a^3}\! \sim
 \!\frac{\alpha^7}{G^3 M^2}\! \sim \! \left(20 \MeV\right)^4
 \left(\frac{10 M_\odot}{M}\right)^2 \left(\frac{\alpha}{0.2}\right)^7
 \! ;
\end{equation}
so, for the purely gravitational interactions to dominate, the new
states must be sufficiently heavy and/or weakly coupled.  If there is
some astrophysical abundance of states which interact with the vector,
their interactions with the cloud could also change the evolution
around BHs.

\section{Black hole spin measurements}
\label{sec:bounds}
\subsection{X-ray Binaries and Supermassive Black Holes}
\begin{figure}[t]
  \centering
    \includegraphics[width=.99\linewidth]{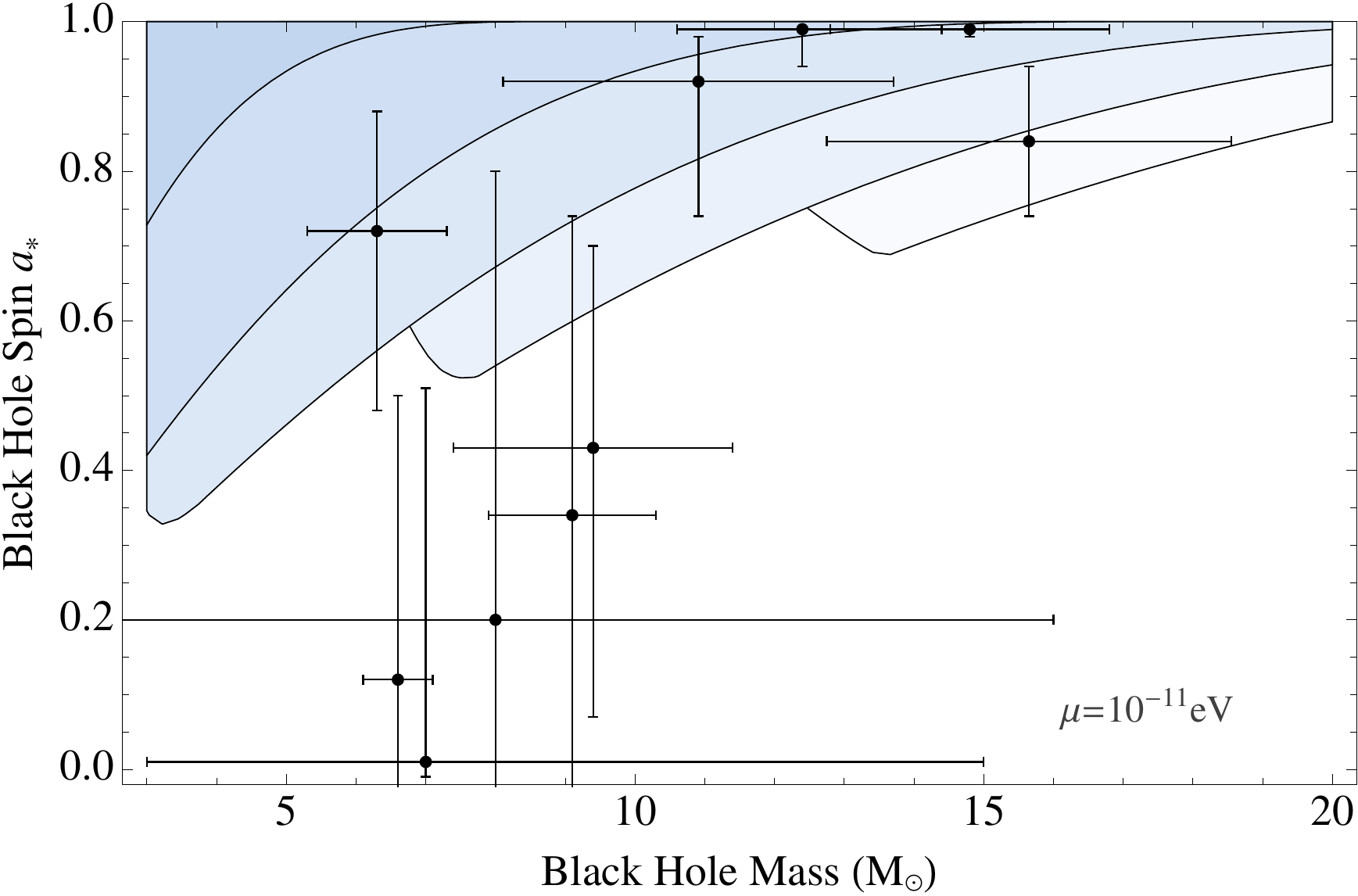}\\
   \caption{The areas above the curves correspond to initial BH
	parameters affected by superradiance within $5\times 10^{6}$ years, in the presence of a vector with mass $\mu=10^{-11}\eV$. The data points are stellar BH measurements with $2\sigma$ errors. We use the five with the highest spin measurements (listed in Table~\ref{tab:stellar}) to constrain vector masses --- BH parameters
	above the curves, such as the two fastest-spinning BHs in this Figure, rule
	out a given vector mass.}
  \label{fig:superimposed}
\end{figure}
In recent years, an increasing number of stellar
BH~\cite{Miller:2014aaa} and supermassive BH (SMBH) spins \cite{Reynolds:2013qqa} have
been measured. The stellar BHs are in binary systems with a companion
star. For both stellar BHs and SMBHs, their accretion disks are used to
extract information about their spin. The methods used fall into two
categories: continuum fitting \cite{McClintock:2013vwa} and X-ray relativistic reflection
\cite{Reynolds:2013qqa}, which measure properties of the accretion disk to find the
radius of the innermost stable orbit and extract the BH spin.
Because a BH that satisfies the superradiance condition can lose
its spin quickly on astrophysical time scales, these spin measurements
place new limits on vector masses.

Similar to Fig.~\ref{fig:regge}, Fig.~\ref{fig:superimposed} shows
regions where superradiance of the $j=1,...,5$ levels for a $\mu =
10^{-11}$~eV vector will spin down a BH. Each region satisfies the superradiance
condition, and also has a superradiance rate fast enough to grow a
maximally-filled cloud within the relevant BH time scale, $\tau_{bh}$;
$\tau_{bh}$ varies between systems so the regions shown in Figure
\ref{fig:superimposed} are approximate.
$\tau_{bh}$ is the shortest time scale on which superradiance can be disrupted:
for stellar BHs, we use the shorter of the age and the shortest
timescale on which the spin can change by accretion, while for SMBHs,
the accretion timescale is the relevant one. We conservatively use
$\tau_{bh} <\tau_{\text{Salpeter}} /10 $, to account for possible
periods of super-Eddington accretion, where $\tau_{\text{Salpeter}} \sim 4.5\times 10^{7}$ years \cite{Shankar:2007zg,Shapiro:1983du}.

Figure \ref{fig:superimposed} also shows BH spin and mass data with
$2\sigma$ errors.\footnote{For the more slowly spinning BHs,
we do not have a reliable 2$\sigma$ estimate so we assume a Gaussian
error on the mass and spin for visualization.} A BH excludes the
vector mass if it lies in the shaded region, within experimental
error. The faster superradiance rates of vectors compared to scalars
allows more BHs to be useful in excluding vector masses. The
two most rapidly spinning BHs clearly exclude this mass for the
vector particle.

\begin{figure*}[t]
\centering
\includegraphics[width = .991 \columnwidth]{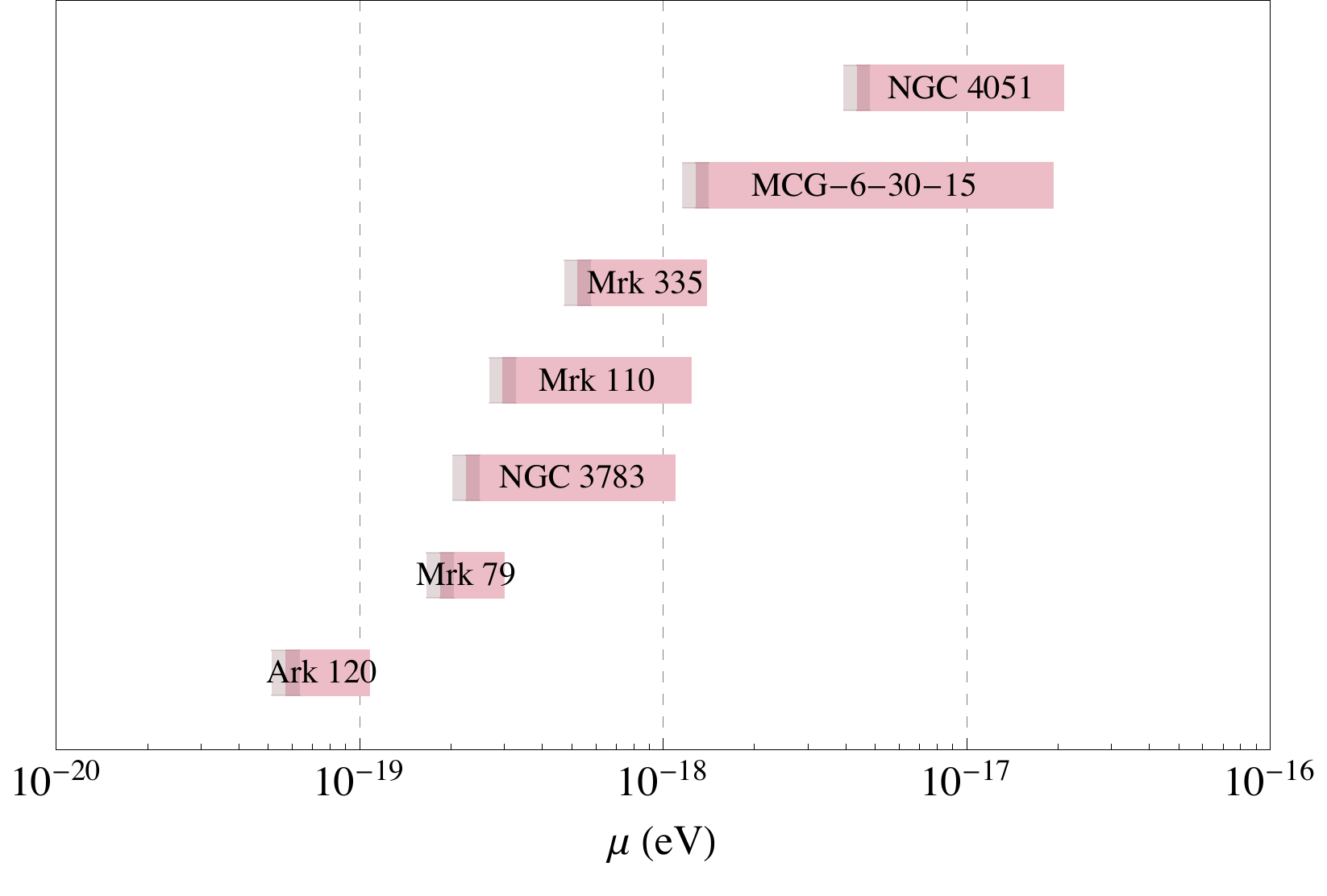}
\centering
\includegraphics[width = .991 \columnwidth]{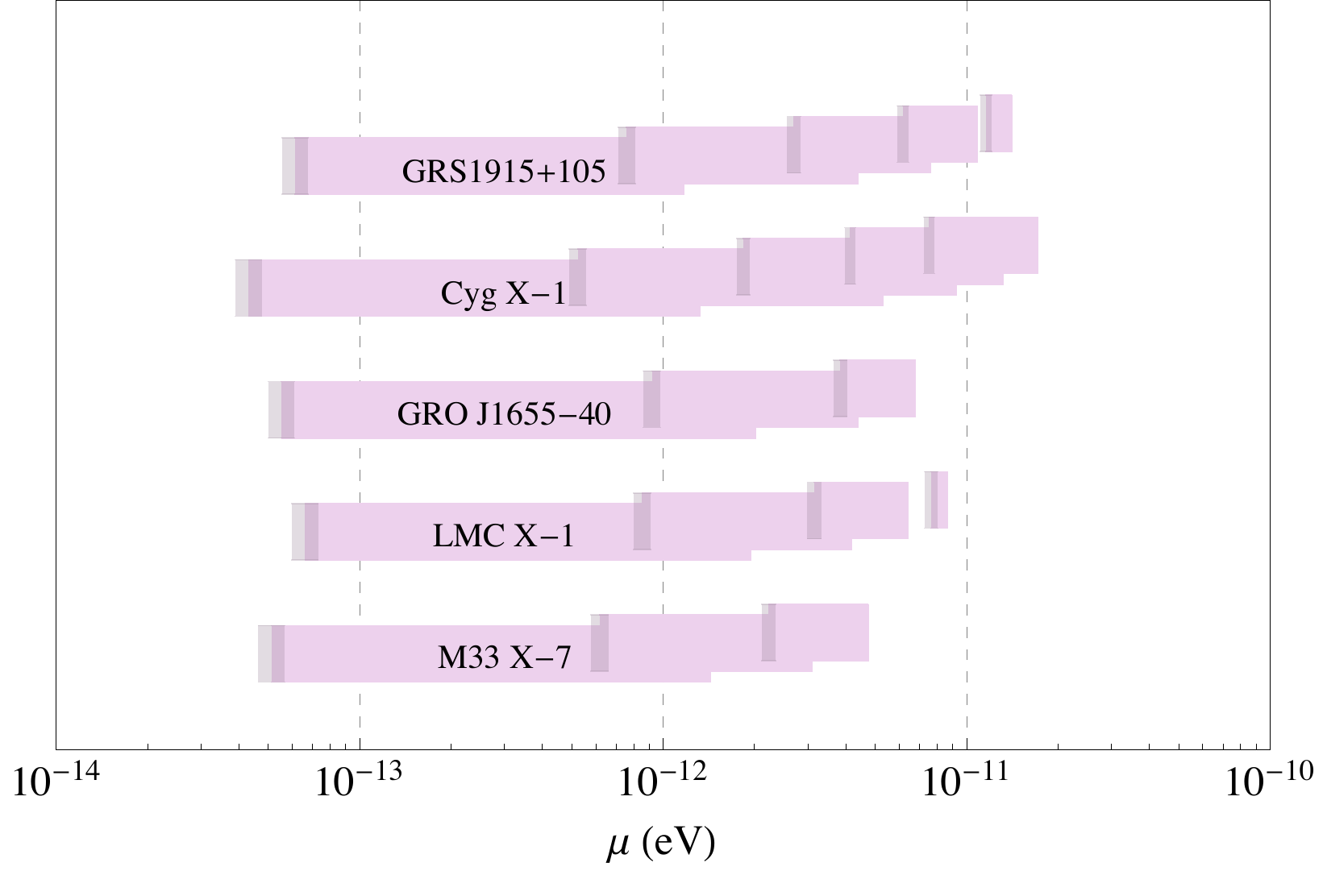}
\caption{ \textit{Left:} Constraints on mass of vectors derived from
  quickly-rotating supermassive BHs (at $90\%$ confidence). Only the
  $j=1$ level is used to set a constraint. \textit{Right:} Constraints
  on mass of vectors derived from quickly-rotating stellar-mass BHs
  (at $2\sigma$). Each rectangle corresponds to a $j$-level which sets
  a constraint, starting with $j=1$ on the left.  The gray bands
  account for theoretical uncertainty in the superradiance rates; the
  right and left edges of these bands are set by superradiance
  rates $\frac{1}{2}\times$ and $2\times$ the analytic value,
  respectively.  }\label{fig:BHspinlim}
\end{figure*}

As the mass of the vector increases, the affected regions shift to
the left, towards lower BH masses. For a given $j$ level and a given BH data point,
the superradiance condition is no longer met at some maximum vector
mass. Decreasing the mass of the vector causes the affected regions to
shift to the right. The minimum BH mass for exclusion is determined
by requiring that 
the superradiance rate remains fast enough to grow a maximally-filled
cloud within the relevant BH time scale, 
\begin{equation}
  \Gamma_{\text{sr}}
  \tau_{\text{bh}} \geq \log N_{\text{m}}\,.
\end{equation}

In the presence of non azimuthally-symmetric perturbations to the cloud, e.g.\
from the accretion disk or the companion star, a given level can mix with decaying bound states, which can relax the constraint at low vector mass~\cite{Arvanitaki:2014wva}. However, for the stellar mass BH systems we consider, the perturbations
are a sub-leading effect, as discussed in Appendix~\ref{ap:mixing}.

The resulting constraints are shown in Figure \ref{fig:BHspinlim}.
In Figure \ref{fig:BHspinlim} \textit{(right)},
each stellar BH places a limit on the vector mass. The overlapping
rectangles correspond to different $j$ levels which set a constraint.
For example, the BH Cyg X-1 places limits on vector masses with
the $j=1,...,5$ levels. 

For SMBHs shown in Figure \ref{fig:BHspinlim} \textit{(left)}, we conservatively
use the $j=1$ level to set constraints. The $j>1$ levels may place an
additional constraint at higher vector masses. However, the density of the accretion disk grows quickly with BH mass, so the effect of accretion disks around SMBHs on the cloud is relatively larger than those around stellar mass BHs. Although the mass in the accretion disk is small compared to the BH mass, estimating the size and form of the
perturbation relies on details of the disk \cite{Arvanitaki:2014wva}.
The $j=1$ level is robust against small perturbations, since the
decay rate of the fastest decaying bound state also scales as $\sim
\alpha^{7}$, similar to the $j=1$ superradiance rate; we save the
extension of limits using higher levels to future work.

The constraints set using spin measurements depend on our computation
of the superradiance rates (see Appendix~\ref{ap:rates}) and thus have
some theoretical uncertainty. In Figure \ref{fig:BHspinlim}, we indicate
the uncertainty with gray bands. The right and left edges of these gray
bands are set by superradiance rates $\frac{1}{2}\times$ and $2\times$
the calculated value. The right edge of each $j$ level is cut off by the
superradiance condition; one source of uncertainty is the energy of the
bound state that goes into the superradiance condition. We expect the
uncertainty in the energy to be less than the lowest order correction in
$\alpha$, $\delta \omega \sim \frac{\alpha^{2}}{2N^{2}} \mu$.

In Table \ref{tab:stellar}, we list details about the stellar BHs used
to set limits in Figure \ref{fig:BHspinlim}. In addition to their
spins, which are determined by both methods, these BHs have precise
measurements for their masses, as well as estimated ages for the
binary systems. They set limits on the vector mass in the range
$5\times 10^{-14}$~eV~$< \mu <2\times 10^{-11}$~eV. The lower limit is
about an order of magnitude better than that for scalars
($6\times 10^{-13}$~eV -- $2\times 10^{-11}$~eV)
\cite{Arvanitaki:2014wva}, by virtue of the faster superradiance rates
for vectors and weaker level-mixing of the leading $j=1,\ell=0$ bound
state.

Details about the SMBH masses and spins that we used to set limits in
Figure~\ref{fig:BHspinlim} are listed in Table \ref{tab:SMBH}. We
assume approximately Gaussian errors to estimate the $90\%$ interval
on the masses. We use one-tenth of the Salpeter time to set the
constraints, ensuring that superradiance is the fastest process
affecting the BH spin evolution. The SMBHs exclude the mass range
$6\times 10^{-20}$~eV~$< \mu < 2\times 10^{-17}$~eV (with the
exception of a small gap around $1\times 10^{-19}$~eV).\footnote{This mass range is slightly less
  constraining than found in \cite{Pani:2012vp}; we use superradiance
  rates for the $j=1,\ell=0$ mode (while \cite{Pani:2012vp} use the
  slower $j=1,\ell=1$ mode for the conservative
  bound), take into
  account the $\sim 180$ e-folding times needed to grow the cloud, and
  use an updated set of BH measurements (which includes a new BH
  \cite{Reynolds:2013qqa}, but has a smaller spin for Fairall 9 than
  used in \cite{Pani:2012vp}). In addition, we assume purely gravitational
  interactions, while \cite{Pani:2012vp} consider the SM photon with a
  mass, the dynamics of which would be strongly affected by the plasma
  surrounding a SMBH.} We consider
these limits preliminary since the properties of SMBHs are less well-known, and the spin measurements only come from X-ray reflection
\cite{Reynolds:2013qqa}.

As more BHs are measured with better precision, we expect the limits to extend further. 
\begin{table*}[t!]
  \begin{center}
    \begin{tabular}{c|  c | c | c | c  |c |c}
      \hline
      \# &Object & Mass ($M_{\odot}$) & Spin & Age (yrs) & Period
      (days)
      &$M_{\rm{comp. \, star}}$ ($M_{\odot}$) \\
      \hline
       1 &\, \,GRS1915+105 \,& $12.4^{+2.0}_{-1.8}$ \cite{0004-637X-796-1-2} & $>0.95$ \cite{Reid:2014ywa,Steiner} & $3-5 \times 10^{9}$
      \cite{dhawan2007kinematics}& 33.85 \cite{Steeghs:2013ksa} &0.47
      $\pm$ 0.27  \cite{Steeghs:2013ksa} \\
       2 &Cyg X-1 & $14.8 \pm 1.0$ & $>0.99$ \cite{2014ApJ...790...29G}& $4.8-7.6\times
      10^{6}$ \cite{Wong:2011eg} &5.599829 \cite{Gou:2011nq}& 17.8  \cite{Gou:2011nq}\\
       3 &\,GRO J1655-40 \,& $6.3\pm 0.5$ & $0.72^{+.16}_{-.24}$ \cite{Steiner} & $3.4-10 \times 10^{8}$ \cite{Willems:2004kk}&2.622 \cite{Willems:2004kk}& 2.3 - 4  \cite{Willems:2004kk}\\
     4 &LMC X-1 & $10.91 \pm 1.4$ & $0.92^{+.06}_{-.18}$  \cite{Gou:2009ks} & $5-6 \times
      10^{6}$ \cite{Orosz:2008kk}& 3.9092 \cite{2011ApJ...742...75R}&
      31.79$\pm$3.48 \cite{2011ApJ...742...75R} \\
      5 &M33 X-7 & $15.65\pm 1.45$ & $0.84^{+.10}_{-.10}$\cite{Steiner} & $2-3 \times 10^{6}$ \cite{Orosz:2007ng} & 3.4530 \cite{0004-637X-646-1-420} & $\gtrsim  20 $ \cite{0004-637X-646-1-420} \\
           \hline
    \end{tabular}
  \end{center}
  \caption{Stellar-mass BHs that set limits on vectors as shown in 
    Figure~\ref{fig:BHspinlim} (data compiled in \cite{McClintock:2013vwa}
    unless otherwise specified). Mass and spin errors are quoted at
    $1\sigma$ 
    and $2\sigma$, respectively~\cite{Steiner}. We
    use the lower continuum-fitting spin values for GRO J1655-40
    \cite{Reynolds:2013qqa}, and
    $\tau_{bh}=\tau_{\mathrm{Salpeter}}/10$ for GRS1915+105
    to set a conservative
    limit.}
  \label{tab:stellar}
\end{table*}

\begin{table}[t]
  \begin{center}
    \begin{tabular}{ c|c | c | c }
      \hline
      \#&Object & Mass ($10^6 \, M_{\odot}$) & Spin \\ 
      \hline
           1& Ark 120 & $150\pm 19$ & $0.64^{+0.19}_{-0.11}$ \\
           2& Mrk 79 & $52.4\pm 14.4$ & $0.7\pm 0.1$ \\
             3 &NGC 3783 & $29.8\pm 5.4$ & $>0.88$ \\
             4 &Mrk 110 & $25.1\pm 6.1$ & $>0.89$ \\ 
                5&Mrk 335 & $14.2 \pm 3.7$ & $0.83^{+ 0.09}_{-0.13}$ \\
                 6&\,MCG-6-30-15 \,& $2.9^{+0.18}_{-0.16}$ & $>0.98$ \\ 
            7&NGC 4051 & $1.91\pm 0.78$ & $>0.99$ \\
      \hline
    \end{tabular}
  \end{center}
  \caption{
    Supermassive BHs that set limits on vectors
    as shown in Fig.~\ref{fig:BHspinlim}  (compiled in
    \cite{Reynolds:2013rva,Reynolds:2013qqa}; our analysis excludes
    BHs without a mass error estimate). The mass and spin errors are
    quoted at $1\sigma$ and $90\%$ confidence interval, respectively. }
  \label{tab:SMBH}
\end{table}

\subsection{Binary black hole mergers}
\label{sec:spinstats}

Black hole spins have only recently been measured in another way: via
the GW signals from binary BH (BBH) mergers, as observed at
aLIGO~\cite{TheLIGOScientific:2016pea}.  Compared to the spin
measurements in X-ray binary systems, the GW measurements of the two
pre-merger spins have very large errors, (roughly
$\sigma_{a_*} \gtrsim 0.3$~\cite{TheLIGOScientific:2016pea}).
However, as the sensitivity of gravitational wave observatories
improves, they will detect an increasing rate of such
mergers. Advanced LIGO may build up a catalogue of hundreds of spin
measurements --- at design sensitivity, aLIGO is expected to detect
80-1200 BBH merger events per year of data~\cite{Abbott:2016nhf,
  TheLIGOScientific:2016htt,TheLIGOScientific:2016pea}.  This raises
the possibility that, while individual measurements are not very
informative, the statistical properties of the whole set may provide
evidence for the altered BH mass vs. spin distribution predicted by
superradiance (as proposed for spin-0 superradiance signals
in~\cite{Arvanitaki:2016qwi}).

\begin{figure*}[t]
\centering
\includegraphics[width = .99 \columnwidth]{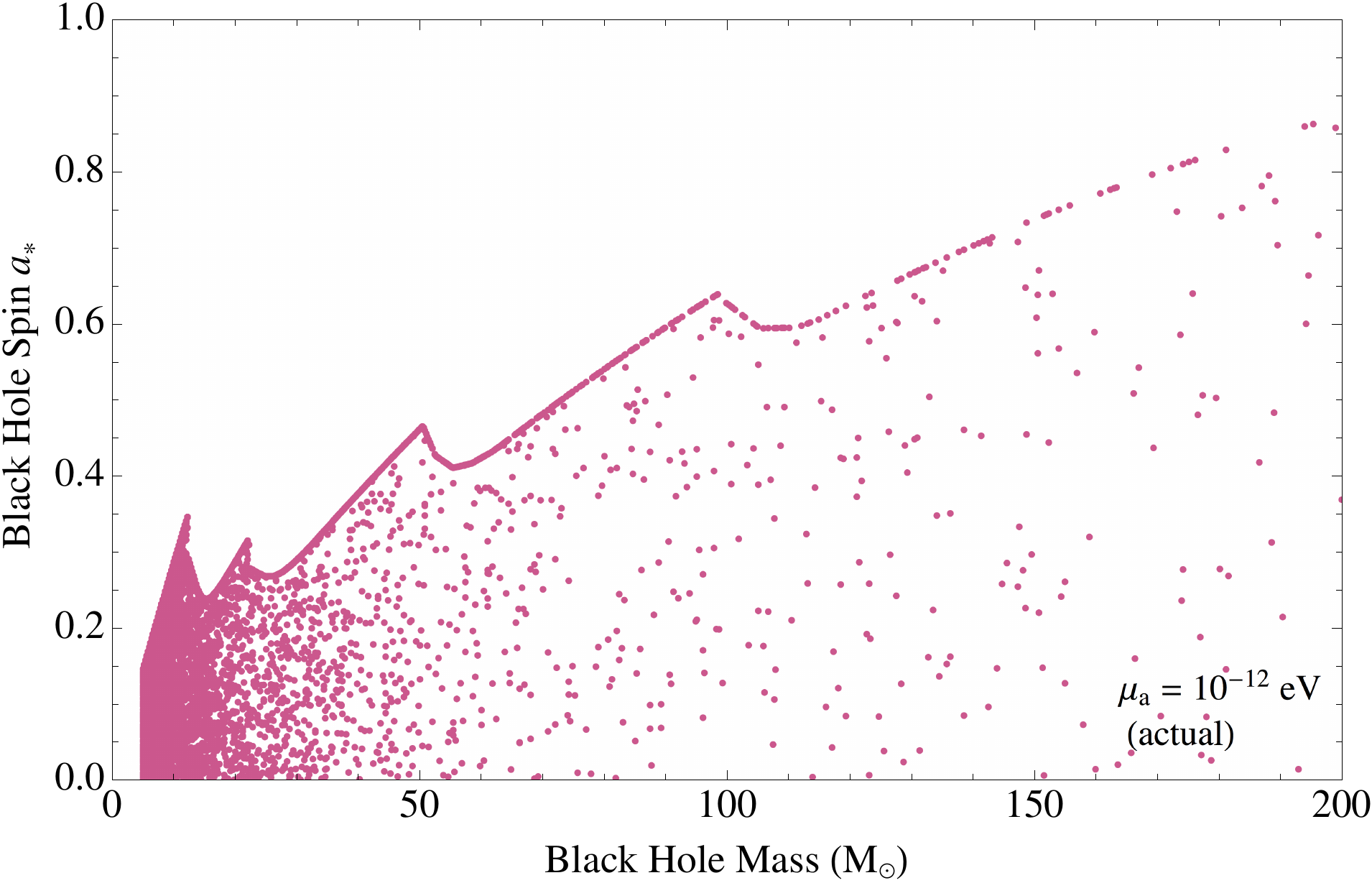}
\centering
\includegraphics[width = .99 \columnwidth]{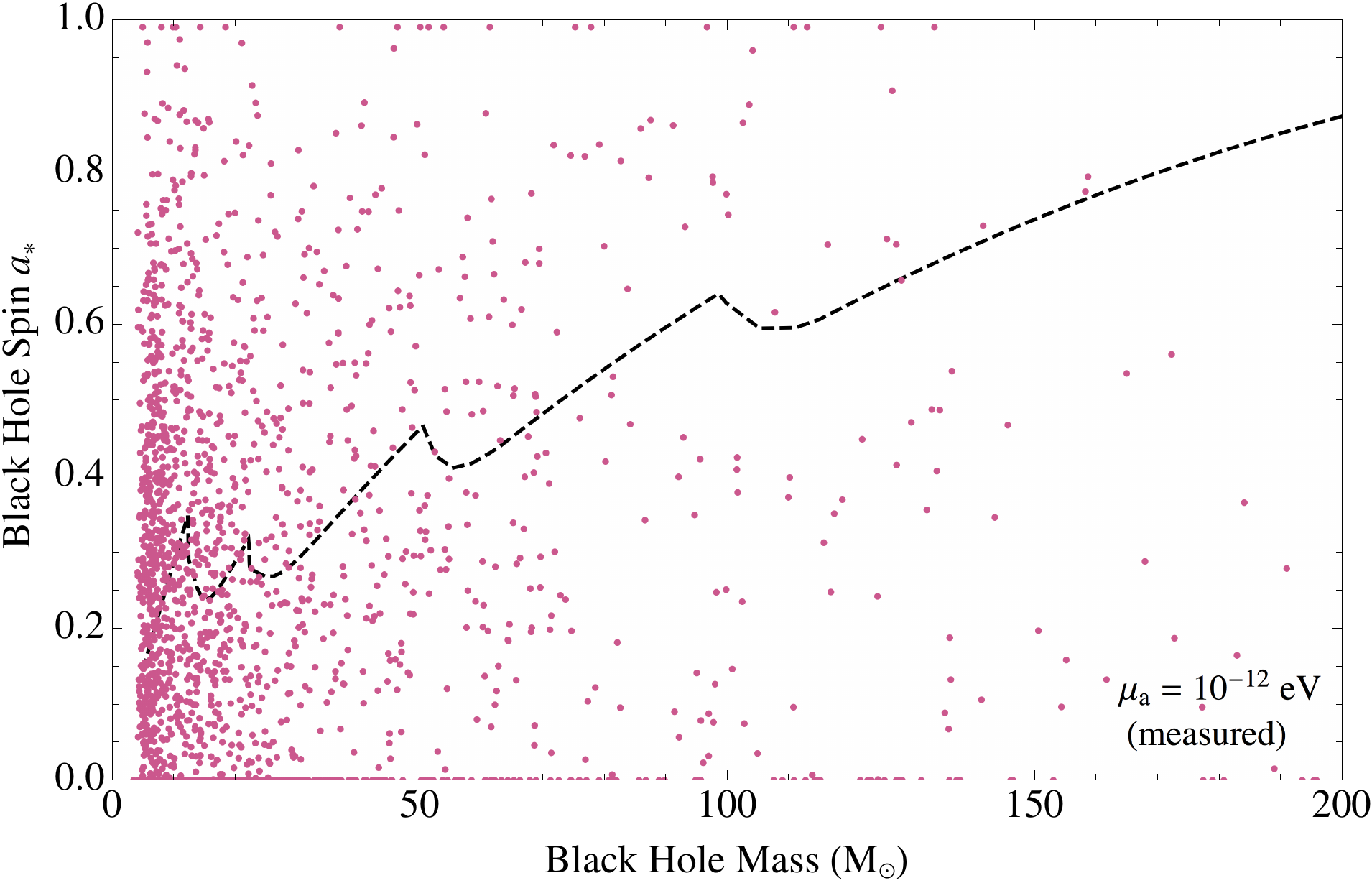}
\caption{Expected distribution of spins and masses of merging BHs in
  the presence of a gravitationally-coupled vector of mass
  $10^{-12}\eV$ (left), and as would be observed at Advanced LIGO
  (right).  We assume typical values of $\sigma_M/M\sim10\%$
  measurement error in the mass and $\sigma_{a_*}\sim 0.3$ error in
  the spin~\cite{Vitale:2014mka,*Vitale,Vitale:2016avz}.  The
  distributions assume that all BBHs form at a distance such that they
  take $\sim 10^{10}$ years to merge. The theoretical curves shown
  (black) are boundaries of the regions where superradiance
	spins down BHs within $10^{10}$~yrs, 
 and where the effect of the
  companion BH perturbation does not significantly affect the
  superradiance rate.}\label{fig:bhpoints}
\end{figure*}

To give an example, Fig.~\ref{fig:bhpoints} shows the effects of BH
spin-down through a weakly-coupled vector of mass $10^{-12}$~eV,
assuming (for illustrative purposes) that BHs are born with a uniform
spin distribution. As described in Section~\ref{sec:history}, black
holes that are born with high spin end up on one of the superradiance
boundary lines. For levels with $j>1$, the mixing due to the companion
BH may be significant, and can limit the growth of the cloud at
small $\alpha$ (see Appendix~\ref{ap:mixing}); this effect is included in the curves of
Fig.~\ref{fig:bhpoints}.\footnote{Since bound levels will, in general,
  annihilate away most of their energy to gravitational radiation
  (Section~\ref{sec:directsignatures}) before the BHs merge, we do not
  need to worry about the gravitational effects of the clouds once the
  BHs get close to each other.}  This neat picture is spoiled by the
large spin measurement errors, which blur the measured distribution 
(Fig.~\ref{fig:bhpoints}, right panel) --- however, there is still an overdensity
along the superradiance boundary lines, and an under-density above the
superradiance rate isocontours.

In reality, it is unlikely that BHs are born with a uniform spin
distribution. We also have no way of knowing the formation history for
a given BH binary; any given pair may not have been able to
superradiate, due to merging too quickly or forming too close
together. A proper estimate of the statistical detectability of
superradiance-induced spindown would, in addition to taking into
account detector-related measurement errors, need to scan over an
astrophysically-plausible ranges of BH formation histories and birth
distributions in the mass-spin plane. 

We will not attempt such an estimate here. Instead, we show
projections for the number of events required to detect that the BH
spin distribution varies with mass (as per Fig.~\ref{fig:bhpoints}),
without the statistical test assuming a particular form for the spin
distribution.  Such variation could come from astrophysical
mechanisms, as well as from superradiance. Our estimates simply
indicate the number of events required to discover mass-dependent
structure in the BH mass-spin distribution --- if this were to be
seen, a detailed study of possible astrophysical explanations would be
required.  In constrast, if we had strong priors on the astrophysical
processes involved, then significantly fewer events may be necessary
to point to superradiance; for example, simply seeing a deficit of
high-spin BHs may suffice. 

\begin{figure}[t]
	\includegraphics[width = 0.99 \columnwidth]{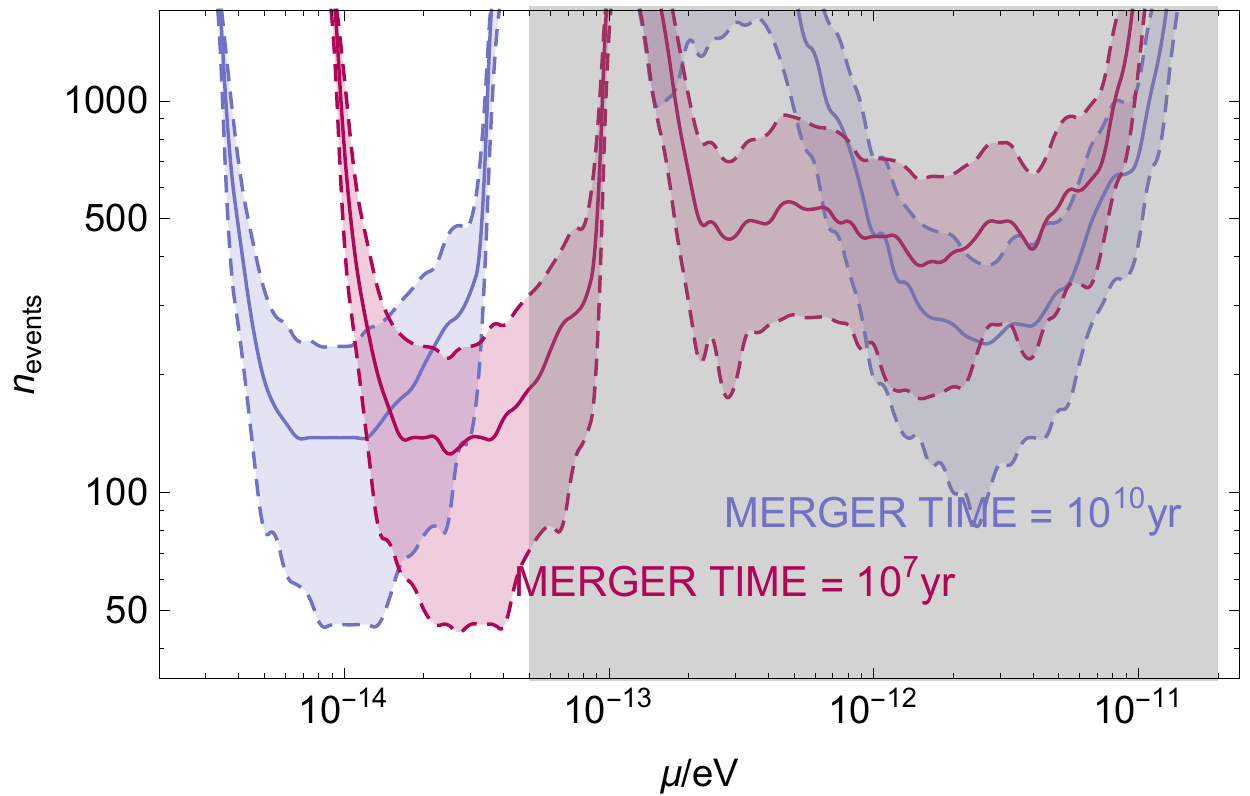}
	\caption{Number of observed events required to show variation
          of the BH spin distribution as a function of BH mass,
          assuming the presence of a gravitationally-coupled vector of
          mass $\mu$. Blue (red) curves correspond to BHs taking
          $10^{10}$~yrs ($10^7$~yrs) to merge. The solid curves shows
          the median number of events required to reject the
          separable-distribution hypothesis at $2 \sigma$, assuming
          the initial mass-spin distribution from
          Fig.~\ref{fig:regge}; the upper/lower dashed curves show
          the upper/lower quartiles, respectively. The test statistic
          used is the Kolmogorov-Smirnov distance between the spin
          distributions outside and inside a given BH mass range,
          maximized over choice of mass range.  The shaded region
          corresponds to the exclusion from BH spin measurements in
          X-ray binaries, as in Fig.~\ref{fig:BHspinlim}.  The raised
          segment at $\mu \sim 10^{-13} \eV$ occurs since the
          spin-down process is so efficient that BHs over the entire
          mass range considered are spun down to within measurement
          error of zero, i.e. there is no mass-dependent structure of
          the spin distribution.  }\label{fig:stats}
\end{figure}

Figure~\ref{fig:stats} shows our estimates for the number of events
required to detect mass-dependent structure in the BH spin
distribution.  We have assumed a uniform initial spin
distribution~\cite{Vitale:2014mka,*Vitale}, and a power-law BH mass
distribution $\rho(M)\propto~M^{-2.35}$~\cite{Abbott:2016nhf}, as were
assumed in LIGO analyses.  Our measurement error estimates are based
on studies of intermediate mass BBHs; at design-sensitivity LIGO/Virgo
detectors, one expects to obtain a 90\% confidence interval of width
$|\Delta a_*| < 0.8$ for total masses up to ~ $600 \msun$,\footnote{For the most pessimistic case of equal BH masses and
  misaligned spins; better measurements are possible for
  dissimilar masses or aligned spins.} and a $\sim10\%$ error in mass
determination for an order-one fraction of primary BH masses
\cite{Vitale:2016avz}.  These measurement errors are broadly
comparable to those in the spin and mass posterior distributions
published by the LIGO collaboration for the three BBH events observed
in the O1 aLIGO  run~\cite{TheLIGOScientific:2016pea}.

The `gap' at intermediate masses in Figure~\ref{fig:stats} occurs
because superradiance there is \emph{too efficient} --- if such a
vector existed, practically all BHs in the observable mass range would
be spun down to very small spins. As a result, there would be no
mass-dependent structure in the BH spin distribution, and it would not
be distinguishable from all of the BHs in such binaries having been
born with small spin.  Of course, this would be interesting in
itself.  Figure~\ref{fig:stats} illustrates that there is a strong
possibility of observing structure in the BH mass-spin distribution
for vector masses an order of magnitude below the exclusion from the
X-ray binary spin measurements (Section~\ref{sec:bounds}), after
$\sim$ hundreds of observed events. Such a statistical test would be
an interesting hint of a light vector's presence, and could
not on its own easily be used to exclude or confirm it.

As shown in Fig.~\ref{fig:stats}, reducing the assumed merger time
from $10^{10}\yr$ to $10^7\yr$ (the range suggested by BBH formation
models~\cite{TheLIGOScientific:2016htt}) reduces the sensitivity of
the statistical search at small vector masses, since superradiance
becomes too slow to spin down low-mass BHs, and the effect of
perturbations for $j>1$ levels is increased.  Conversely, third-generation
gravitational wave observatories may achieve much higher
signal-to-noise BBH merger detections, and consequently much higher
precision progenitor spin measurements (e.g. a $90\%$ interval of
$|\Delta a_*| < 0.1$ for a majority of high-SNR events
\cite{Vitale:2016icu}).  Along with the increased event rate, the
higher precision would improve the reach of the statistical search,
and enable detailed investigation of any features found by
Advanced LIGO.

\section{Gravitational wave signals}
\label{sec:directsignatures}

So far, we have considered the evolution of a massive vector field on
a fixed background. However, since superradiance can extract 
$\OO(\alpha)$ of the BH's energy into the bound states, the
gravitational effects of the bound states may not be negligible.  In
particular, since superradiance builds up a coherent classical wave in
the bound states, it results in an oscillatory stress energy tensor,
which sources gravitational radiation.  As we will see in this
Section, this gravitational radiation provides one of the most
important potential observational signatures of a superradiant cloud.
Such signatures were analyzed for the case of scalar bound states
in~\cite{Arvanitaki:2009fg,Arvanitaki:2010sy,
  Arvanitaki:2014wva,Arvanitaki:2016qwi}. Here, we will discuss the
corresponding signals for vectors.

Decomposing $A_\mu$ into a combination of oscillations in different
bound state levels,
\begin{equation}
	A_\mu = \frac{1}{\sqrt{2 \mu}} \sum_i \sqrt{N_i} \left(\Psi^i_\mu e^{- i \omega t}
	+ {\rm c.c.}\right) \,,
\end{equation}
the stress energy tensor is
\begin{widetext}
\begin{equation}
	T_{\mu\nu} \supset \mu^2 A_\mu A_\nu + \dots
	= \frac{1}{2 \mu} \sum_{i,j} \sqrt{N_i N_j} \left(\underbrace{\Psi^i_\mu \Psi^j_\nu
	e^{-i (\omega_i + \omega_j)t}}_{\text{`annihilations'}} + \underbrace{\Psi^i_\mu {\Psi^j_\nu}^* e^{- (\omega_i 
	- \omega_j)t}}_{\text{`transitions'}} + {\rm h.c.} \right) + \dots \,,
\end{equation}
\end{widetext}
where we have only displayed the $A_\mu A_\nu$ term to 
demonstrate the schematic form of $T_{\mu\nu}$. The components of
$T_{\mu\nu}$ with quadrupole or higher angular dependence, and
non-zero time dependence, source gravitational radiation in the usual
way. At the particle level, the high-frequency
$\omega_i + \omega_j \simeq 2 \mu$ terms correspond to annihilations
of two bound-state vectors to a single graviton, while the
$\omega_i - \omega_j \sim \alpha^2 \mu$ terms correspond to
transitions from a higher-energy level to a more deeply bound one.

Both of these processes result in monochromatic gravitational waves,
emitted in a characteristic angular pattern from the BH, at a
frequency related to the mass of the vector. Gravitational wave
observatories, present and future, can search for such signals.

\subsection{Transitions}

Transition signals will be sizable only in the case where two levels
are populated simultaneously at nearly maximum cloud sizes. Since the
occupation number attained by slower-growing levels is exponentially
suppressed by the difference between their superradiance rates and
that of the fastest-growing level, this only occurs for levels where
the superradiance rates of the `excited' and `ground' states are
similar.  That is, we need $\Gamma_g \sim \Gamma_e$, where the excited
state is the one with a larger real energy component,
$\omega_e > \omega_g$; furthermore, the peak signal strain scales and
the superradiance rate~\cite{Arvanitaki:2010sy, Arvanitaki:2014wva}.

The frequency of the emitted radiation is suppressed by the
$\alpha^2/N^2$ splitting between levels, so its wavelength is long
compared to the cloud size.  As reviewed in Appendix~\ref{ap:gwrates},
this means that it is simple to reliably estimate the rate of GW
emission due to transitions.  This is approximately given by the
quadrupole formula, so is dominantly affected by the size of the
cloud, controlled by $\ell$. For the vector case, the
fastest-superradiating level at any time is one with
$j = \ell + 1$, with growth rate
$\Gamma \sim \alpha^{4\ell + 6} \mu$. In comparison to the scalar
case, where the fastest-growing level has
$\Gamma \sim \alpha^{4\ell + 4}\mu$, a vector bound state with
equivalent transition rate (i.e.\ the same $\ell$) has a smaller
superradiance rate than its scalar equivalent.  For a given vector
mass, we therefore expect to observe fewer transition signals than we
would for a scalar of the same mass.  Since Advanced LIGO has marginal sensitivity to scalar transition signals, with 
optimistic assumptions giving $\sim 
0.3$ detectable signals~\cite{Arvanitaki:2014wva}, we do not expect it to be sensitive to
transition signals from vector superradiance.

Nevertheless, it is possible that at large $\alpha$ the superradiance
rates have a different dependence on $n$ than the form we can extract
in the $\alpha \ll 1$ limit.  For example, in the scalar case,
numerical results~\cite{Yoshino:2015nsa} seem to indicate that the
$n=0$ level for $\ell=m=3$ has a superradiance rate that crosses over
with those for other $n$ at large $\alpha$ and $a_*$. This may result
in other levels giving rise to transition signals, which could be more
promising for observational purposes. We leave the investigation of
such issues to future work.

\subsection{Annihilation rates}

Annihilation signals have the advantage that they only require a large
occupation number in a single level, so can occur without the
coincidence of superradiance rates required for transitions.  However,
as reviewed in Appendix~\ref{ap:gwrates}, the annihilation rates are
parametrically smaller than transition rates, and are rather more
subtle to evaluate.  The momentum of the emitted
gravitational wave is $\sim 2 \mu$, which is much larger than the
`typical momentum' in a bound state, $\sim \alpha \mu$; the
high-momentum behaviour of the bound state is dominated by the
discontinuity from the $e^{-r/a}$ wavefunction dependence near the
origin. This gives an emission rate that scales as a higher power of $\alpha$
compared to a quadrupole formula estimate, and also
means that the small-$r$ distortion of the gravitational wave in the
BH background can affect the rates at the same order as the flat-space
estimate.

In the flat-space approximation, we obtain an emitted GW
power of $ P \simeq \frac{32}{5} \alpha^{12} \frac{G N^2}{r_g^4}$ for
the fastest-growing $\ell=0$, $j=m=1$ level.  As discussed in
Appendix~\ref{ap:gwrates}, and similarly to the scalar case, it is
necessary to take into account the modified propagation of the
gravitational wave due to the $r_g/r$ part of the metric in order to
obtain the leading-$\alpha$ rate. We estimate that this can increase
the power to $\sim 10 \times$ larger than the flat-space result,
\begin{equation}
	P \sim 60\alpha^{12} \frac{G N^2}{r_g^4}.
\label{eq:annpower}
\end{equation}
This compares to a GW power of $\sim 0.02 \alpha^{16} G N^2 r_g^{-4}$
for the fastest $\ell=m=1$ scalar bound state~\cite{Brito:2014wla}.
For both scalars and vectors, the GW power emitted from a bound state
scales as $\alpha^{4\ell + 12} G N^2 r_g^{-4}$, with the faster rates
possible for vectors reflecting the existence of a $\ell=0$
superradiant level.

We use Eq.~\eqref{eq:annpower} as the central value for reach and
event rates and take the difference between the flat- and curved-space
rates as an estimate of the theoretical uncertainty in the event
rates plots.  Numerical simulations of the GW emission rate
from the fastest-growing level~\cite{East:2017mrj} approximately agree with
our estimate, though they do not extend to small enough $\alpha$ to
see the $\alpha^{12}$ scaling in the $\alpha \ll 1$ limit.

The angular distribution of GW emission is set, at leading order in
$\alpha$, by the form of the $T_{ij}$ source.  For a $j=m=1$ state,
this corresponds to an emission profile that is the same as for a
standard quadrupole GW source, such as a spinning neutron star with an
asymmetry.  The emitted power per solid angle is
\begin{equation}
	\frac{dP}{d\Omega} \propto (1 + 6 \cos^2\theta_k + \cos^4 \theta_k) \,,
\end{equation}
where $\theta_k$ is the angle of the line-of-sight from the BH's spin axis.
While the power varies by a factor of a few across different angles, 
this does not make a qualitative difference to the phenomenology,
and we ignore this variation in the following sections.

\subsection{Annihilation signals at LIGO}
\label{sec:anligo}

The monochromatic GW signals from bound state annihilations
around stellar-mass BHs are in the right frequency range to be detected
at Advanced LIGO. A vector of mass $\mu$ results
in a GW signal at frequency
\begin{align}
f_{\rm GW} &= \frac{2 \omega}{2\pi} \simeq  \frac{\mu}{\pi} \left(
	1 - \frac{\alpha^2}{2 (\ell + 1 + n)^2}\right) \\
	&\simeq 1.3 \kHz \,  \frac{\alpha}{0.2} \, \frac{10 M_\odot}{M} \,.
\end{align}
In particular, whatever the mass of the BH, the GW signals from a
given vector will be clustered within the angular frequency interval
$\sim (1.7\mu, 2\mu)$; numerical results indicate that
this lower bound is correct even for relativistic bound states
\cite{East:2017mrj}.  Writing the emitted GW power as
$P \equiv \Gamma_{\rm ann} N^2 \omega_{\rm GW}$, the occupation number
$N$ of the cloud evolves as
$\dot N = \Gamma_{\rm sr}N - \Gamma_{\rm ann} N^2$.  When the
annihilation term dominates, as it generally will after the cloud has
reached maximum size and spun down the BH, $N$ decreases as
\begin{equation}
N(t) = \frac{N_{0}}{1+\Gamma_{\text{ann}}N_{0}t}\,,
\end{equation}
where $N_0 \equiv N(t=0)$.
The observed strain therefore decreases like $1/t$ for $t \gg (N_{0} \Gamma_{\rm ann})^{-1}$, with the timescale of the signal set by
\begin{equation} \label{eqn:tsig}
t_{\rm sig}\simeq	(N_{m} \Gamma_{\rm ann})^{-1} \simeq 80 \text{ sec } \left(\frac{0.2}{\alpha}\right)^{11}
	\left(\frac{M}{10 M_\odot}\right) \,.
\end{equation}
Hence, signals can last anywhere between seconds and millions of years,
depending on the value of $\alpha$.

The mass of the cloud itself contributes to its own gravitational binding
energy. As the cloud annihilates away, this contribution to the binding
energy decreases,
leading to a positive frequency drift of the annihilation
signal,
\begin{equation}
\dot{f} \simeq 10^{-8}\mathrm{Hz/sec}\times
\left(\frac{\mu}{10^{13}\,\mathrm{eV}} \right)\left(\frac{\alpha}{0.2} \right)^3 \frac{250\,\mathrm{hr}}{t_{\rm sig}}\,.
\end{equation}
 Monochromatic searches may become difficult with large frequency
drifts. On the other hand, a positive frequency drift is quite unusual
for an astrophysical source and could be used to distinguish the signal
from e.g.\ monochromatic radiation from rotating neutron stars.

\subsection{Annihilation reach}
\label{sec:reach}

\begin{figure}[t]
\includegraphics[width = .99 \columnwidth]{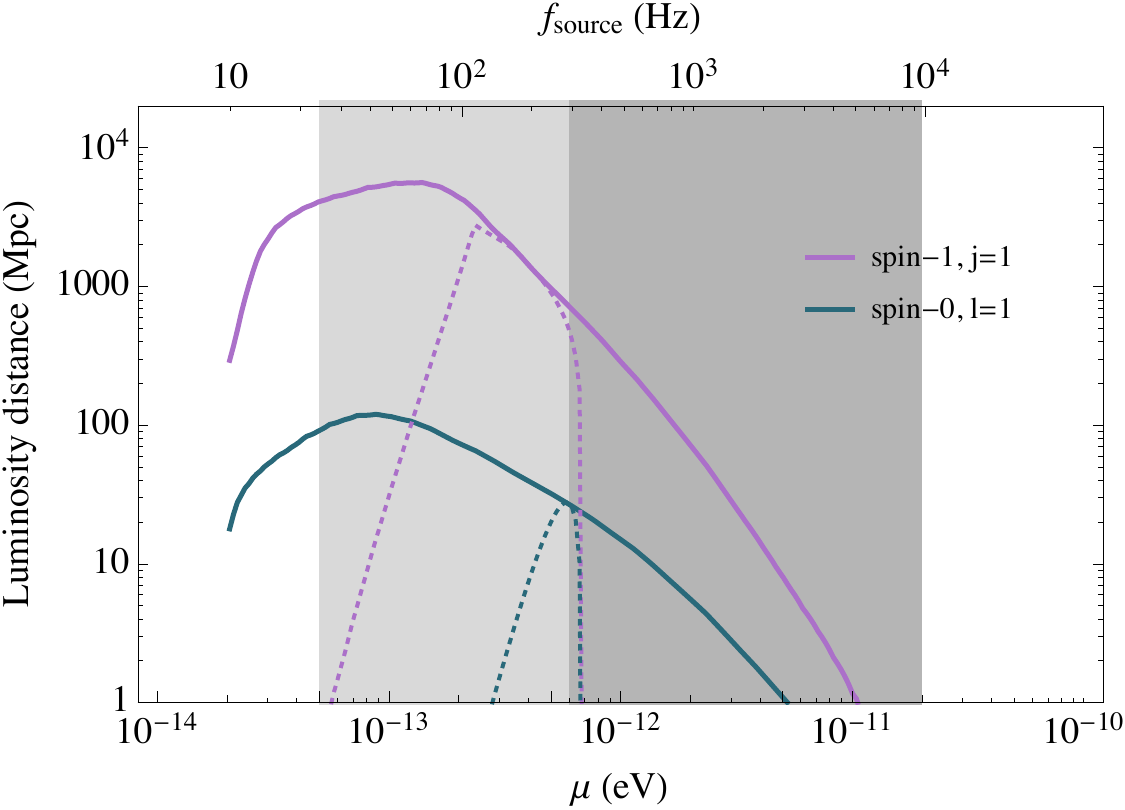}
\caption{Maximum distance at which design-sensitivity Advanced LIGO
  could detect monochromatic GWs from a superradiant cloud around a BH
  with $a_* = 0.9$. The solid curves correspond to the reach for annihilations
	of the
  $\ell = 0, j=m=1$ vector bound state (purple), and the
  $\ell = m = 1$ scalar bound state (green). For each particle mass,
  we choose a BH mass that gives the strongest GW signal, resulting in
  optimal $\alpha$ values 0.2 for vectors and 0.28 for scalars.  The
  dotted lines indicate the reach for an example BH mass of
  $M_{\rm BH} = 62 M_\odot$, corresponding to the final BH formed in
  the GW150914 BBH merger observed by Advanced
  LIGO~\cite{Abbott:2016blz} (it most likely has spin $\sim 0.7$,
  which reduces the reach for higher vector masses). Lighter (darker)
  shading indicates the mass range excluded by spin measurements for
  vectors (scalars). Source frequency is shown on the top axis; the
  frequency detected on Earth is reduced by redshift, which
  is significant for signals originating from a luminosity
  distance further than $\sim$~Gpc.}\label{fig:reach}
\end{figure}

The rotation and motion of the Earth complicates the search for
narrow-band GW signals at terrestrial GW observatories. A
monochromatic signal from a distant source is subjected to a
time-dependent redshift in the detector frame, with this time
dependence set by its sky location. As a result, it is not possible to
find such signals by simply Fourier-transforming the strain data;
instead, it is necessary to scan over each possible frequency and sky
position for the source~\cite{Wette:2009uea}.  This is a
computationally demanding process, and if the source is not
well-localized, it is generally too expensive to integrate coherently
over the entire observation time.  Instead, LIGO
searches~\cite{TheLIGOScientific:2016uns} break up the observation
time into shorter segments and integrate coherently within each
segment, combining the results from each segment incoherently. This
results in a strain sensitivity (for given desired signal-to-noise
ratio) of
\begin{align}
h_{\text{det}} = \mathrm{SNR}\, C_{\text{tf}} \frac{\sqrt{S_{h}}}{N_{\text{seg}}^{1/4} T_{\text{coh}}^{1/2}}\,,
\end{align}
where $N_{\rm seg}$ is the number of segments, $T_{\rm coh}$ is
the length of each segment, $S_h$ is the detector's noise
spectral sensitivity~\cite{AmaroSeoane:2012je}, and $C_{\rm tf}$ is
the `trials factor', which quantifies the look-elsewhere effect inherent
in searching over many bins in frequency and sky position.
This should be compared to a GW signal strain at the detector of 
\begin{align}
h=\left(\frac{4G P}{r^{2}\omega_{\rm GW}^{2}} \right)^{\frac{1}{2}}\,,
\end{align}
for a gravitational wave source with power $P$ and angular frequency
$\omega_{\rm GW}$ at a distance $r$ away from the Earth.

Figure~\ref{fig:reach} shows the maximum distance at which Advanced
LIGO (at design sensitivity) could detect annihilation signals from
bound states around a BH with $a_* = 0.9$.  This distance is set by
the strength of annihilations from the $\ell = 0, j = m = 1$ bound
state, which is the fastest-annihilating
superradiant state.  The reach is calculated assuming a large dataset
of 121 segments of 250 hour coherent integration times, and trials
factor of 20; the most recent search uses 90 segments of 60 hour
coherent times \cite{TheLIGOScientific:2016uns}.

As Figure~\ref{fig:reach} illustrates, the existence of the $\ell = 0$
superradiant state means that vector annihilation signals can be seen
from much further away than those from scalar states.  The converse of
this is that the signal lifetimes for vector annihilations are
correspondingly shorter (since they radiate away approximately the
same overall energy from the cloud).\footnote{The brightest signals
  are typically shorter than the coherent integration time
  $T_{\text{coh}}$. The reach is thus penalized by a shorter
  integration time. Consequently, increasing the annihilation power
  increases the signal strain but reduces the detector strain
  sensitivity by the same amount, so the reach at fixed $\alpha$ stays
  the same even if we account for uncertainty in the annihilation rate
  by varying its value from $\times 10^{-1}$ to $\times 10^{1}$.  }
Because the reach for $j=1$ annihilations extends to cosmological
distances, we account for the signals red-shifting due to Hubble
expansion. The detected frequencies are reduced compared with the
source frequency,
$f_{\text{observed}} = (1+z)^{-1} f_{\text{source}}$, and signal times
are increased by the same factor. Between $10^{3}$ - $10^{4}$ Mpc in
luminosity distance, the signal frequencies are $1.2$ - $2.5$ times
lower. The effect of redshift changes the peak of the signal reach
toward higher values of $\mu$.

\subsection{All-sky searches}

\begin{figure}[t]
\includegraphics[width = .99 \columnwidth]{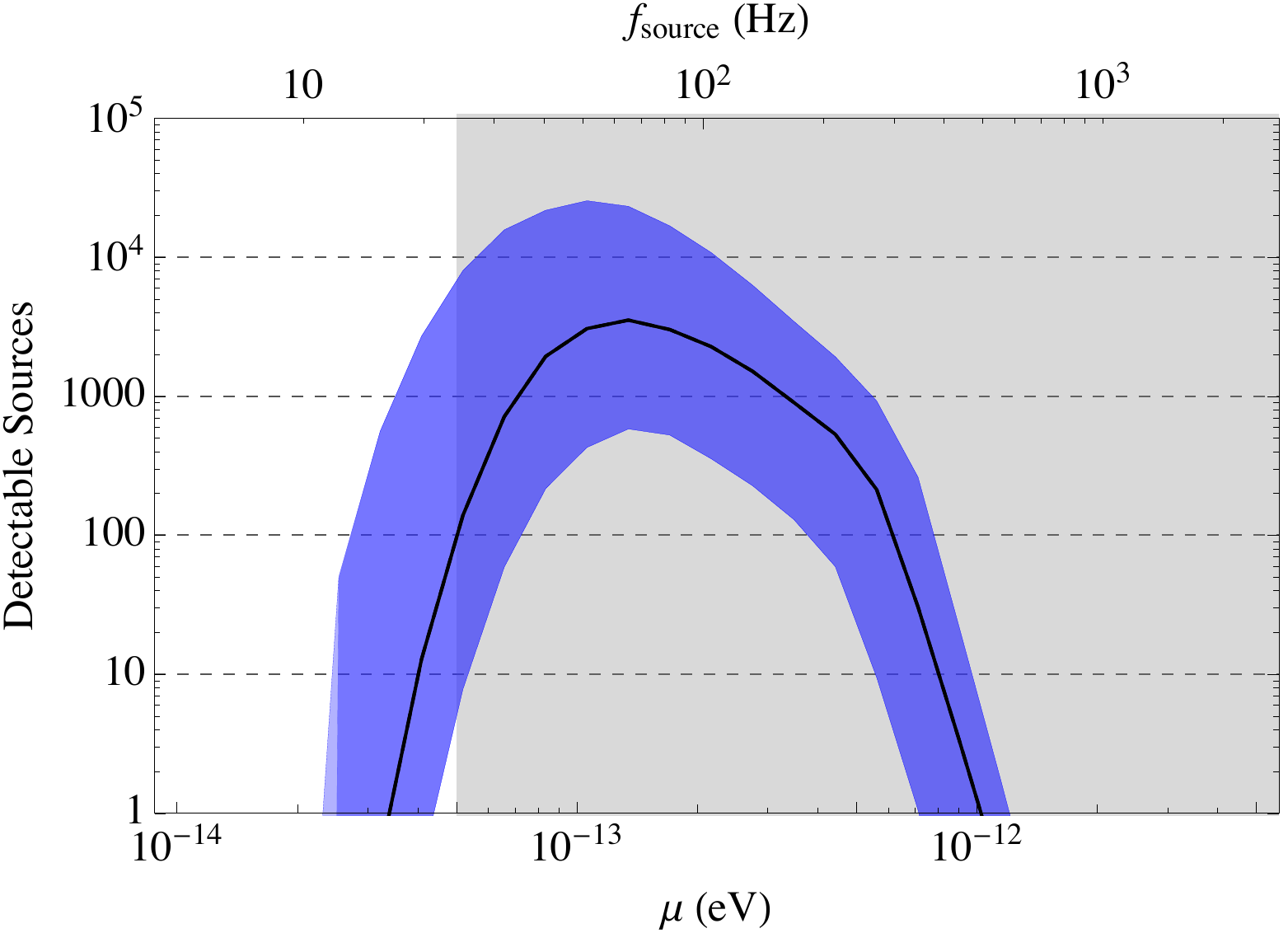}
\caption{ Expected number of detectable vector $j=1$ annihilation
  sources at Advanced LIGO
	(at design sensitivity)
  in a blind monochromatic GW search with $T_{\text{coh}} = 250$ hr,
	$N_{\text{seg}} = 121$, and $C_{\text{tf}} = 20$ (see text for definitions).  Virtually all
  events originate from within the Milky Way. The two bands (dark
  blue, light blue) assume a BH mass distribution with a maximum mass
  $M_{\text{m}} = (80,160) M_{\odot}$; in addition, each band is
  bounded by optimistic and pessimistic estimates from astrophysical
  and theoretical uncertainties. The central values (for both
  $M_{\text{m}} $) are given by the solid
  curve. The shaded region is excluded by X-ray binary spin
  measurements.
}\label{fig:blindsearch}
\end{figure}

Advanced LIGO performs all-sky searches for monochromatic
gravitational radiation from unknown sources
(e.g.~\cite{TheLIGOScientific:2016uns}), and vector annihilation
signals provide an excellent target for such searches.
Figure~\ref{fig:blindsearch} shows our estimates for the number of
events that might be observed by Advanced LIGO, using the assumptions
detailed in Appendix~\ref{ap:events}.  To account for the large
uncertainties regarding the population of astrophysical BHs, we show a
band bounded by optimistic and pessimistic estimates for BH mass and
spin distributions, as well as their formation rates.  We also account
for theoretical uncertainties in our estimate of the annihilation
rates by including a range of annihilation power a factor of 10 above
and below the estimate in Eq.~\eqref{eq:annpower}. The resulting
uncertainty is much smaller than the uncertainty in astrophysical
populations.  The non-detection of annihilation signals in an all-sky
search at LIGO can in principle constrain vector masses;
however, most of the parameter space has already been excluded by BH
spin measurements.

While, as illustrated in Figure~\ref{fig:reach}, sufficiently strong
annihilation signals could be visible from cosmological distances,
such signals last for only a short time. At smaller vector masses, for
which strong signals only occur around heavy BHs (i.e.\ at $\alpha$
not too small), such signals are probably rare enough that it is
unlikely a suitable BH was born at precisely the right time for its
annihilation signals to reach us now (though as we discuss below, BBH
mergers may give rise to a suitable population of heavy, fast-spinning
BHs). Consequently, the blind search event rates are dominated by
longer, quieter signals originating from inside our galaxy. A typical
such signal, for a vector mass of $\mu \sim 10^{-13} \eV$, comes from
a BH of mass $\sim 20 M_\odot$, corresponding to a small
$\alpha \sim 10^{-2}$, and consequently a very long signal time
$\sim 10^8 \yr$.

At larger vector masses ($\gtrsim$ few $\times 10^{-13} \eV$), lighter
and more abundant BHs are able to give strong annihilation signals, so
the expected rate for these is large enough that extra-galactic
signals dominate the blind search.  The typical frequency drift of the
short, bright signals outside the Milky Way (MW) is
$\sim 10^{-7}$~Hz/s, which falls well outside the current range
covered by the aLIGO search; in Fig.~\ref{fig:blindsearch}, we only
include signals with frequency drift smaller than $2\times 10^{-11}$
Hz/s (see App.~\ref{ap:freqdrift}). With this constraint, virtually
all events from beyond the MW are cut off for having a too-large
frequency drift, leaving only events from the MW.  Vector masses for
which such distant events could dominate the expected event rate are
disfavored by X-ray spin measurements. 

Finally, given a single vector particle, we expect the fractional
variation in signal frequencies to
be at most $\sim 17\%$ (ignoring redshift), set by the fractional binding energy,
$\sim\frac{\alpha^2}{2(n+\ell+1)^2}+\OO(\alpha^3)$~\cite{East:2017mrj}. This
clustering of multiple monochromatic signals would be another handle
pointing toward a particle source of the GW signal. For signals from
the MW, the typical signal has $\alpha\ll 0.1$, so the monochromatic
lines would appear very close together, within $\delta f/f <10^{-3}$.

In addition to monochromatic signals from individual BHs, there will
also be a `stochastic background' from signals too weak to be
individually resolved. As for the resolvable signals, at low vector
masses we expect this stochastic background to be dominated by signals
from BHs within the galaxy. Due to the very good frequency resolution
of GW detectors such as Advanced LIGO, individually-resolvable signals
will stand out well above such a background, but 
whether such a background may be observable warrants further
investigation. 

\subsection{Follow-up Searches}
\label{sec:followup}

The blind search described above fits well into the existing LIGO
search strategy. Another interesting category of search is to look for
monochromatic GWs from the product of a BBH or BH -- neutron star
(BH-NS) merger, as proposed for scalar superradiance
in~\cite{Arvanitaki:2016qwi}. At design sensitivity, Advanced LIGO may
observe a number of BH-NS mergers, and up to hundreds of BBH mergers a
year \cite{TheLIGOScientific:2016htt,Belczynski:2015tba}, each of
which could result in the perfect source for superradiance signals: a
new, isolated, rapidly spinning BH.  If its spin is high enough, and
there is a light vector in the right mass range, the new BH will
superradiate a cloud of vectors which will subsequently annihilate
into gravitational radiation. The cloud will begin to grow once the
final BH closely approximates the Kerr metric; this happens quickly
since the ringdown time is parametrically short compared to the
superradiance time. From the properties of the new BH, and the
following monochromatic GWs, it would be possible to check that such a
signal was compatible with superradiance.

\begin{figure}[t]
	\includegraphics[width = 0.99 \columnwidth]{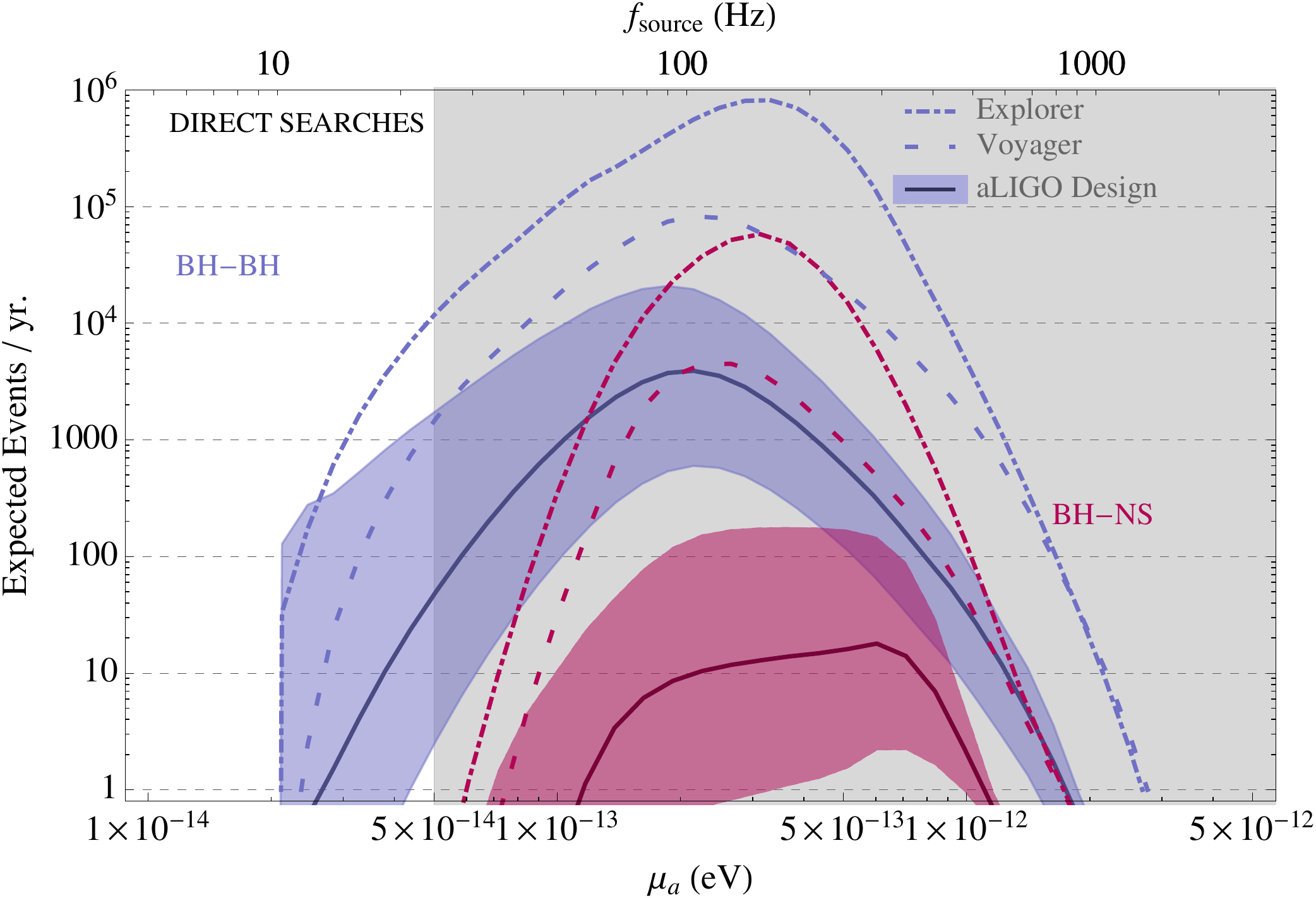}
	\caption{Annihilation events visible in follow-up searches at
          Advanced LIGO and future observatories from products of
          BH-NS mergers (red) or BBH mergers of equal mass (blue).  We
          take a power-law mass distribution with varying exponential
          suppression at $M>60 M_{\odot}$ and flat spin
          distribution of the merging BHs (and $a_*=0$ for the
          NS). The bands span the merger rate uncertainty given
          GW150914 for BBH \cite{Abbott:2016nhf} and a factor of $100$
          uncertainty in the annihilation rates. For BH-NS
          uncertainties in rates from simulations dominate (we take
          V4l and V2l in \cite{Belczynski:2015tba}). We assume a
          coherent integration time of the signal duration, or up to
          10 days for BBH and 1 year for BH-NS. Shaded region is as in
          Fig.~\ref{fig:blindsearch}.}\label{fig:followup}
\end{figure}

Figure~\ref{fig:followup} shows the expected number of `follow-up
search' events that could be seen at Advanced LIGO and future GW
observatories. The signals which dominate the event rates are very
bright, visible up to the edge of the reach sensitivity
(Fig.~\ref{fig:reach}), corresponding to redshifts of $z\sim 1-2$ with
aLIGO design sensitivity. On the other hand, the signals are quite
short (see Eq. \ref{eqn:tsig}), and fall outside of the current LIGO
search strategy~\cite{TheLIGOScientific:2016uns}. The signals also
have a large frequency drift of up to $\sim 10^{-5} \Hz/\sec$, which
may make a search more computationally
challenging~\cite{Wette:2009uea} (see App.~\ref{ap:freqdrift}).  The
events visible at vector masses below the region constrained by X-ray
binaries rely on the presence of a population of heavy BHs
($M>100\msun$). We include an estimate of the uncertainty by using a
power-law mass distribution with a varying exponential suppression of
BH populations at high mass
($M>60\msun$)~\cite{Belczynski:2015tba}. Recent aLIGO searches put the
first upper bound on merger rates of intermediate-mass BHs, which is
$\mathcal{O}(1)$ consistent with the upper end of the power-law mass distribution at
$M \sim 100\msun$~\cite{Abbott:2017iws}; the BBH mass function will be
better constrained with further measurements, narrowing the range of 
expected event rates. In the case of BH-BH mergers, Advanced LIGO
design reach for the merger signal and the follow-up signal are
comparable; for BH-NS mergers, the rates may be limited by the reach
to the merger, which we take to be $0.927$~Gpc \cite{Abadie:2010cf}.

 For both types of mergers, the event will be localized on the sky,
 and the follow-up search can be focused on a particular region. For
 BH-BH mergers there is likely no electromagnetic counterpart, so the
 localization is poor \cite{Abbott:2016iqz}. For BH-NS events, there
 may be an electromagnetic counterpart, which would localize the event
 well enough to eliminate the need to scan over sky location. We thus
 assume longer integration times for the BH-NS follow-up search,
 although the signal length limits the overall coherence
 time for the brightest signals~\cite{Wette:2009uea}.

GW150914, the first observed BBH merger, resulted in a final BH mass
of $62.2^{+3.7}_{-3.4}~M_{\odot}$ and spin of $0.68^{+.05}_{-.06}$, at
a luminosity distance of $440^{+160}_{-180}$~Mpc~\cite{Abbott:2016izl}. For an
optimal vector mass of $\sim 2\times 10^{-13}\eV$ the following
annihilation signal would have been visible from as far as $\sim 1$~Gpc;
however this mass is excluded by X-ray binary measurements.  A
lighter vector with mass outside the exclusion would only be visible
at distances of less than $\sim 1$~Mpc. As shown in
Fig.~\ref{fig:reach}, a rapidly spinning final BH would be
visible for a wider range of vector masses.

As discussed above, although there is theoretical uncertainty in the
GW annihilation rates, this does not translate into a large
uncertainty in the event rates: the larger strain is compensated by
shorter signal times, and therefore shorter coherent integration
times. If the annihilation rate is more than a factor of a few larger
than Eq.~\eqref{eq:annpower}, the cloud does not reach maximum size,
which leads to smaller than expected signal power.

In Fig.~\ref{fig:followup}, we assume the binary formation mechanism
does not allow for superradiance, and estimate the spin of the final
BH using~\cite{Buonanno:2007sv} starting with an initial flat
distribution of spins. For the BBH merger, if the two initial BHs have
spin close to zero, the final BH will have spin $a_*\sim 0.68$ for
equal mass BHs~\cite{Buonanno:2007sv}; an initial distribution with
zero spin reduces the number of BBH follow-up events by a factor of
$\sim 3$.

\section{Discussion}
\label{sec:conclusion}
Black hole superradiance provides a unique probe of new light,
weakly-coupled bosons.  In this paper, we have derived constraints and
estimated future signals for a new, light, gravitationally-coupled
vector particle. Existing X-ray measurements of spins of stellar
BHs set stringent constraints on the parameter space,
excluding nearly three orders of magnitude in mass below
$2\times 10^{-11}$~eV; if confirmed, supermassive BH spin measurements
would constrain another three orders of magnitude at smaller 
masses. Continuous GW signals from annihilating vector clouds could be
bright enough to be seen across cosmological distances; these signals
can be seen in Advanced LIGO all-sky searches, as well as in follow-up
searches after BBH and BH-NS merger events. The strong spin
constraints, combined with poor LIGO sensitivity at low frequencies,
leave a range of a factor of a few in vector mass that could be
observed through GWs.
 
Superradiance may spin down a significant fraction of the pre-merger
BHs, giving a statistical signal. The statistical search has the
additional benefit that it only depends on measuring the broadband
burst from the BBH merger, extending its sensitivity to masses an
order of magnitude below X-ray spin constraints. The lack of aLIGO
sensitivity to frequencies below $\sim 10 \Hz$ means that GW searches
cannot probe vector masses too far below the X-ray binary bound. As
discussed in Section~\ref{sec:spinstats}, as long as high-mass BBH
events are seen, detector simulations indicate that reasonably
accurate spin measurements should be possible --- the question then
becomes how many of the merging BHs have such high masses.

As we have explored in this paper, the observational signatures of
light vector superradiance around astrophysical BHs are analogous to
those of light scalars. The parametrically faster superradiance and GW
emission rates of vector bound states alter the relationships between
different signals and constraints.  For scalars, the all-sky
monochromatic GW search for superradiant clouds is the most promising
observational route~\cite{Arvanitaki:2016qwi}.  For vectors, the
stronger exclusions coming from spin measurements in X-ray binaries
mean that there is only a narrow mass window in which such searches
could discover a light vector (Figure~\ref{fig:blindsearch}). The
strong constraints derived in this paper give one handle to
distinguish vector and scalar signals: observing monochromatic
radiation pointing to a particle of mass in the range
$5\times 10^{-14}-6\times 10^{-13}$ eV would suggest scalar rather
than vector superradiance. In addition, the vector $j=1,\ell=0$ mode
is more tightly bound to the BH, so while for scalars we
expect a spread of $\lesssim 4\%$ in frequency below $2\mu_a$ due to
BH binding energy, vector monochromatic lines will cluster within a
broader range, of up to $\sim 17\%$ (not taking into account the
red-shifting of signals). This difference will only be apparent
for large-$\alpha$ signals, such as those giving rise to extra-galactic
sources. In all-sky searches where weak, low-$\alpha$ signals dominate,
the expected frequency differences between different sources are
much less than $1\%$ for both
scalars and vectors.

Follow-up searches for newly formed BHs are more promising for vectors
than scalars at current GW observatories, due to the larger reach.
Signals could be seen from distances beyond a Gpc --- however, their
short duration and relatively large positive frequency derivative
requires a dedicated LIGO analysis. Measuring the parameters
of the newly formed BH determines, for a given particle
mass and spin, the expected superradiance and annihilation rates.
It would therefore be possible to check whether any
follow-up GW observations were consistent with vector or
scalar superradiance.

Statistical searches for structure in the mass/spin distribution of
merging BBHs are similarly powerful for scalars and vectors, through
the more stringent X-ray spin limits on vectors mean that sensitivity
in the non-excluded region relies on the observations of heavier BBH
mergers.  The existence of the superradiant $\ell = 0$ mode for a
vector means that BH spin-down can be significantly more efficient.
Because of slower superradiance rates and a larger impact of companion
perturbations, we do not expect scalar superradiance to reduce BH
spins close to zero on astrophysical timescales, while this could be
possible for vectors.

Our discussion of GW signals was limited to those from stellar-mass
BHs. Lower-frequency observatories, both space- and ground-based
\cite{Dimopoulos:2008sv,Graham:2012sy,Audley:2017drz},
could extend the reach to higher black hole masses, and therefore
lower vector masses. Once lower-frequency GW detectors such as LISA
\cite{Audley:2017drz} are operational, mergers involving SMBHs will
become observable, as well as possible monochromatic signals from
low-mass vector clouds. Estimating the sensitivity of LISA to
monochromatic signals from vector clouds, we find that, if clouds
carrying a significant fraction of a SMBH's spin can be built up, then
their GW signals could be visible from most of the way across the
observable universe. The relatively more complicated environments of
SMBHs and the importance of processes of accretion and mergers on SMBH
spin evolution~\cite{Arvanitaki:2010sy} requires further dedicated
study for reliable event rate estimates.

A number of papers have considered whether exponential amplification
though superradiance may occur via the SM photon. It was claimed
in e.g.~\cite{Pani:2012vp} that, if the SM photon had a very small mass,
then the growth of gravitationally bound photon states could lead to
significant spin-down of supermassive BHs.  However, this analysis
neglects the effects of the surrounding plasma on the photon equation
of motion, which for the extremely low photon masses considered,
$\lesssim 10^{-18} \eV$, will be very important (for example, the
plasma frequency corresponding to a typical interstellar medium
electron density of $0.01 \cm^{-3}$ is
$\simeq 4 \times 10^{-12} \eV$).

Another possibility is that, even for a zero-mass SM photon, the effects
of an astrophysical plasma medium on the photon's dispersion relation
could result in superradiant amplification of bound states.
The original `black hole bomb' paper~\cite{Press:1972zz} noted that, if
there was an `evacuated cavity' in the plasma around a BH, this could
act as a `mirror' to confine superradiant waves. However,
we are not aware of plausible astrophysical mechanisms to generate
suitable `cavities'.
More recently, papers including~\cite{Pani:2013hpa}
and~\cite{Conlon:2017hhi} have claimed that for BH within
astrophysical plasmas (the hot early universe plasma
for~\cite{Pani:2013hpa}, and the diffuse interstellar medium
for~\cite{Conlon:2017hhi}), the `effective plasma mass' could
gravitationally confine EM oscillations around the BH, leading to
exponential superradiant growth. 

Both of these papers make the
assumption that EM oscillations within a uniform plasma are
governed by the Proca equation,
$D_\mu F^{\mu\nu} = \omega_p^2 A^\nu$.  However, only the
transverse photon modes have a dispersion relation approximately
corresponding to a free particle with mass $\omega_p$ --- the
longitudinal mode in a non-relativistic plasma propagates at a much
lower speed $\OO(T/m_e)$ \cite{Raffelt:1996wa}.  As reviewed in
Appendix~\ref{ap:rates}, only the `axial' ($j=\ell$) bound states for a
massive vector correspond to a superposition of purely transverse
waves, so we do not expect the bound state properties for polar states
of a Proca field to reflect the behaviour of the SM photon in a
plasma. 
Also, an increased plasma density close to the BH
(as expected from e.g.\ accretion onto the BH) will generally
result in a suppressed bound state density close to the BH,
leading to suppressed growth rates.
It should be kept in mind that there are other astrophysical
situations in which energy and angular momentum could be extracted
from Kerr BHs via electromagnetic effects, e.g.\ the Blandford-Znajek
process~\cite{Blandford:1977ds}. However, more careful calculations
would be required to determine whether setups closer to those considered
in this paper could be realised for the SM photon.

In this paper, we have only considered the gravitational interactions
of light vectors. Since the vector masses we are interested in are so
small, the couplings of the vector to SM matter are generally
constrained to be significantly sub-gravitational in
strength~\cite{Adelberger:2003zx,Wagner:2012ui}. The effects of such
couplings should therefore be a small perturbation to the
gravitational effects we have worked out.  An exception is if
long-distance fifth-force or equivalence principle tests do not
apply. A clear example is a very light kinetically-mixed dark photon,
for which the effective coupling to SM matter is suppressed by at
least $(m_{A'} / \omega_p)$, where $m_{A'}$ is the dark photon mass,
and $\omega_p$ is the `effective plasma frequency' of the SM medium.
The charge density around Earth-based experiments is high enough that
they impose weak constraints on the kinetic mixing at the very light
vector masses we consider~\cite{Essig:2013lka}.  In contrast, the BH
superradiance signals in this paper will apply for any kinetic mixing
below some upper threshold.  Calculating this threshold, and whether
any additional observational signals arise near to it, would be very
interesting, though we leave it to future work.

\section{acknowledgements}
We thank Joseph Bramante, Savas Dimopoulos, Sergei Dubovsky, Luis
Lehner, and Riccardo Penco for many informative and clarifying
discussions, Junwu Huang, Anthony Lasenby and Ken Van Tilburg for
comments on the manuscript, Sam Dolan for correspondence regarding
vector bound states around Schwarzschild BHs, Warren Morningstar for
BH spin measurements discussions, and especially Asimina Arvanitaki
for conversations throughout the completion of this work and comments
on the manuscript. We are grateful to William East and Solomon Endlich
for many extensive conversations about their respective spin-1
superradiance rate calculations. This research was supported in part
by Perimeter Institute for Theoretical Physics. Research at Perimeter
Institute is supported by the Government of Canada through the
Department of Innovation, Science and Economic Development and by the
Province of Ontario through the Ministry of Research, Innovation and
Science. M.T. is supported in part by the Stanford Graduate
Fellowship.

\vspace{1em}


\appendix

\section{Superradiance rates for non-relativistic bound states}
\label{ap:rates}

As mentioned in Section~\ref{sec:boundstates}, there are no
analytic approximations to scalar or vector bound states in a Kerr
background that are valid at all radii.
Finding bound state superradiance rates, which depend on the wavefunction
near the horizon, therefore requires either a numerical solution
of the wave equation, or some matching between approximate
solutions with different regimes of validity.

Both approaches have been used to find scalar superradiance rates. The
Klein-Gordon equation is separable in the Kerr background, so the
problem can be reduced to solving the one-dimensional radial equation. This
has been done numerically in a number of papers
(e.g.~\cite{Dolan:2007mj}).
At small $\alpha$, \cite{Detweiler:1980uk} finds approximate
solutions at large and small radii in terms of hypergeometric
functions, and matches at intermediate radius to obtain the
superradiance rates to leading order in $\alpha$ (as discussed below, there appears
to be a factor of 2 error in~\cite{Detweiler:1980uk}'s calculation).

We use a similar matching argument to find the growth/decay rates
for vector bound states at small $\alpha$. Compared to the scalar case,
the problem is complicated further by the fact that the Proca
equation is not separable in the Kerr background.
Instead, we find separable approximations which apply at large and small
radii, which can be matched in an intermediate regime.

In the limit of small mass, the Proca field can be decomposed into two
transverse modes, which obey the photon EoM, and a longitudinal mode,
which obeys the massless KG equation.  
The evolution of massless scalars, and the behavior of some of the
gauge-invariant functions of the photon field,
can be 
solved exactly on the Kerr background via the Teukolsky
equation~\cite{Teukolsky:1973ha}. This helps us because, for
$r \ll \mu^{-1}$, the mass term in the Proca wave equation is sub-dominant,
and solutions to the full wave equation are well approximated
by solutions to the massless equations.
Conversely, for $r \gg r_g$, the wavefunction is hydrogenic in the small-$\alpha$
limit, as reviewed in Section~\ref{sec:boundstates}.
Consequently, if the hydrogenic wavefunction in the regime
$r_g \ll r \ll \mu^{-1}$ can be approximated as a superposition
of massless
transverse and scalar modes, then we can use the behavior
of these massless modes to find the energy flux across the BH
horizon for the bound state.

The absorption probability for a long-wavelength massless (bosonic)
spin-$s$ wave carrying total angular momentum $j$
is~\cite{Page:1976df}
\begin{widetext}
\begin{equation}
\mathbb{P}_{\rm abs} \simeq \left(\frac{(j-s)!(j+s)!}{(2j)!(2j+1)!!}\right)^2
\prod_{n=1}^j \left[1 + \left(\frac{\omega
- m \Omega}{n \kappa}\right)^2\right]
2 \left(\frac{\omega
- m \Omega}{\kappa}\right) \left(\frac{A_H \kappa}{2 \pi}
\omega\right)^{2j + 1} \,,
	\label{eq:pabs}
\end{equation}
\end{widetext}
where $A_H$ is the BH horizon area,
$\kappa \equiv (4 \pi (r_+ - r_g))/A_H$, and a negative
absorption probability corresponds to superradiant amplification.
A massless wave of frequency $\omega \simeq \mu$
has an absorption probability scaling as $\alpha^{2j+1}$,
to leading order in $\alpha$.
As we will see below, part (in some cases all)
of the energy flux into the BH can be obtained
by matching the hydrogenic bound state at small $r$
onto a massless wave of this frequency.
The relevant scaling of the hydrogenic bound state's squared
amplitude at small $r$ is $\alpha^{2 \ell}$,
giving an overall growth rate of
$\Gamma_{\mathrm{SR}}\propto\alpha^{2j + 2\ell + 4} \mu$ (where the 4
comes from choosing $\mu$ as our normalization; see below).  

For a scalar field, where $j=\ell$, this reproduces the known
$\alpha^{4\ell + 4}\mu$ scaling.  As reviewed in
Section~\ref{sec:srcompare}, previous works observed the
$\alpha^{2j + 2\ell + 4}\mu$ scaling in numerical solutions of the
Proca equation in a BH background, and derived it analytically for the
$j=\ell$ mode in the small-$a_*$ limit.  Here, we demonstrate how the
scaling arises for all of the bound states, and analytically determine
the leading-$\alpha$ growth/decay rates for some of the low-$j$ modes,
including the fastest-growing $\ell = 0, j=m=1$ mode.  In the main
text, when we use higher-$\ell$ superradiance rates for
phenomenological purposes, we adopt an estimate that should be
$\OO(1)$ correct at small $\alpha$ --- the full leading-order
expression could be derived in a similar fashion to our calculations
here.


\subsection{Hydrogenic wavefunctions}

For $r \gg r_g$, the bound
state wavefunctions are hydrogen-like, 
\begin{align} 
	\Psi_i \simeq \Psi_i^{H} &\equiv R^{n\ell}(r)Y^{\ell,jm}_i  \,,\\
  \Psi_0 \simeq \Psi_0^{H} &\equiv f_0(r)Y^{jm}\,,
\label{eq:psih}
\end{align}
 where $f_0 (r)$ is fixed
 by the Lorentz condition. The vector spherical harmonics are given by 
\begin{align}
&Y_i^{\ell,jm}(\theta,\phi) = \\
&\quad\sum_{m_{\ell}= -\ell}^{\ell} \sum_{m_{s}= -1}^{1} \langle (1,m_{s}),(\ell,m_{\ell}) | j,m \rangle \xi_i^{m_{s}} Y^{\ell m_{\ell}} (\theta,\phi)\,,\nonumber
\end{align}
where
$\xi^{0} = \hat{z},\,\xi^{\pm 1} = \mp \frac{1}{\sqrt{2}} (\hat{x} \pm
i \hat{y})$
are unit vectors \cite{thorne}. The labels run over
$\ell = 0,1,2,\dots$, with $j = \ell-1,\ell,\ell+1$ ($j = 0,1$ for
$\ell=0$), and $m = -j,\dots,j$; $\ell$ can be identified as the
orbital angular momentum, since $Y_i^{\ell,jm} (\theta,\phi)$ are
eigenfunctions of the orbital angular momentum operator,
$-r^2 \nabla^2 Y_i^{\ell,jm} = \ell(\ell+1)
Y_i^{\ell,jm}$. 

For $a \gg r\gg r_g$,
\begin{align}
	\Psi_i^{H}&\simeq R_0 \left(
	\frac{r}{a}\right)^{\ell} Y^{\ell,jm}_i(1 + \OO(r/a) + \OO(r_g/r)) \,,
\end{align}
where $R_0$ is the first coefficient in the Taylor expansion of $R^{n\ell}(r)$
around the origin.


\subsection{`Massless' wavefunctions}

As noted above, for $r \ll \mu^{-1}$ the mass term is a sub-leading
correction to the Proca wave equation.
This is particularly easy to see in a Schwarzschild
background, where the Proca equation separates into angular and radial
parts. The angular equations do not depend on the mass, and the mass
only enters into the radial equations via the
operator~\cite{Rosa:2011my}
\begin{equation}
	\hat{\mathcal{D}}_2 \equiv -\frac{\partial^2}{\partial t^2}
	+ \frac{\partial^2}{\partial r_*^2} - \left(1 - \frac{2 r_g}{r}\right)
	\left(\frac{j(j+1)}{r^2} + \mu^2\right) \,,
\end{equation}
where $r_*$ is the standard tortoise coordinate.  As a result, for
$r \ll \mu^{-1}$, the $\mu^2$ term is a sub-leading correction to the
EoM. This is also true in the full Kerr background.

For $r \gg r_g$, the solutions of the massless wave equation are close
to the flat-space solutions.  A (flat-space) transverse plane wave has
$\partial_i A_i = 0$, so by the Lorentz condition, $A_0 = 0$. A basis
for the transverse modes with frequency $\omega$ that are regular
at the origin is given by 
\begin{equation}
	\Psi^B_i \equiv j_j(\omega r) Y^{j,jm}\,,
\label{eq:transb}
\end{equation}
\begin{equation}
	\Psi^E_i \equiv j_{j-1}(\omega r) Y^{j-1,jm} - \sqrt{\frac{j}{j+1}} j_{j+1}(\omega r)
	Y^{j+1,jm}\,,
	\label{eq:transe}
\end{equation}
where the $j_n$ are spherical Bessel functions of the first kind.
Near the origin, for $r\ll \omega^{-1}$,
\begin{equation}
	j_{n}(r\omega) \simeq \frac{1}{(2 n + 1)!!} (r\omega)^{n} + \OO(r\omega)^{n+2} \,,
\end{equation}
where $!!$ denotes the double factorial. 

The longitudinal solutions of the flat-space wave equation are, in the
massless limit, pure-gauge; that is,
$-\partial_i A_0 - \partial_t A_i = 0$ and
$\nabla \times \vec{A} = 0$. In terms of the pure-orbital vector
spherical harmonics, the regular-at-the-origin longitudinal solutions
are
\begin{equation}
	\Psi_i^R \equiv \frac{\sqrt{j} j_{j-1}(\omega r) Y^{j-1,jm}_i
	+ \sqrt{j+1} j_{j+1}(\omega r) Y^{j+1,jm}_i}{\sqrt{2j + 1}}\,,
\label{eq:longr}
\end{equation}
\begin{equation}
	\Psi_0^R = j_j(k r) Y^{jm}.
\label{eq:longr0}
\end{equation}
Together with the transverse modes $\Psi^E$ and $\Psi^B$, these form a
complete basis for regular-at-the-origin solutions to the flat-space wave
equation for vanishing mass.

The reason for taking the solutions that are regular at the origin
is that these correspond to an ingoing wave which `passes through the origin'
undisturbed to become an outgoing wave. At large radii, $r \gg \omega^{-1}$,
we can decompose $j_n (\omega r)$ into ingoing and outgoing parts,
\begin{equation}
	j_{n}(r\omega) \simeq \frac{e^{i\omega r - (n+1)i \pi/2 }+e^{-i\omega r + (n+1)i\pi/2}}{2\omega r} \,.
\end{equation}


\subsection{$j = \ell$ bound states}

The $j=\ell$ hydrogenic bound states have wavefunction
\begin{align}
	\Psi_i &= R^{nj}(r) Y^{j,jm}_i, \quad	\Psi_0 = 0\,.
\end{align}
In the interval $r_g \ll r \ll \mu^{-1}$ we can match this form onto the $\Psi^B$ transverse solution, with
\begin{equation}
	\Psi_i^{H} \simeq C_B j_j(\mu r) Y^{j,jm}_i,\,\, C_B = \frac{(2j+1)!! \, R_0}{(\mu a)^j}\,.
\end{equation}
Thus, near the BH, the wavefunction looks like a superposition of
ingoing and outgoing massless waves. In that physical situation,
the energy flux through the horizon is given by the energy flux
in the ingoing wave, multiplied by the BH absorption probability
from equation~\ref{eq:pabs}. This gives us the energy flux into the BH
from the `gauge-invariant'
components of the stress-energy tensor, i.e.\ the first and third terms
in 
\begin{equation}
	T_{\mu\nu} = F_{\mu\alpha}F_\nu^{\,\alpha} + \mu^2 A_\mu A_\nu
	+ g_{\mu\nu}\left(-\frac{1}{4} F_{\rho\lambda}F^{\rho\lambda} - \frac{1}{2}\mu^2 A_\rho A^\rho \right) \,.
\end{equation}
Since $A_0 \simeq 0$ in the hydrogenic regime, the
$\mu^2 A_\mu A_\nu$ contribution to the energy flux
into the BH can be shown to be sub-leading in $\alpha$
compared to the $F_{\mu\alpha} F_\nu^{\,\, \alpha}$
`Poynting' contribution (the $g_{\mu\nu}$
parts of $T_{\mu\nu}$ do not give an ingoing energy flux).

The magnitude of the Poynting flux can be found via equation~\ref{eq:pabs}.
To set the normalization of the `massless waves', we can evaluate the
energy flux in the ingoing part of $\Psi^B$
by calculating the Poynting vector at large radii. There,
the ingoing part has wavefunction $\Psi_i  \simeq C \frac{e^{i \omega r}}{2 \omega r} Y^{j,jm}_i$, for which the time-independent part of the Poynting vector
is
\begin{equation}
	S = E \times B = |C|^2 \left( \frac{1}{4\omega} \frac{1}{r^2}
	\left| Y^{j,jm} \right|^2
	\hat{r} \, + \perp \right) + \dots \,,
\end{equation}
where $\perp$ indicates components perpendicular to the radial
direction $\hat r$.
This gives an energy flux through a sphere of large
radius of $\sim |C|^2/(4 \omega)$.
Thus, the energy flux across the BH horizon for the bound state
is, to leading order in $\alpha$, $|C_B|^2 \mathbb{P}_{s=1,j}/(4 \mu)$.
Performing the analogous calculations for a scalar bound
state, we obtain a flux of $|C_B|^2 \mathbb{P}_{s=0,j}/(4 \mu)$,
so
\begin{equation}
	\Gamma_{j=l} = \frac{\PP_{s=1,j}}{\PP_{s=0,j}} \Gamma_{{\rm scalar},j}
	= \frac{(j+1)^2}{j^2} \Gamma_{{\rm scalar}, j} \,.
\end{equation}
This agrees with the expression derived by~\cite{Pani:2012bp}, and is
accurate to leading order in $\alpha$ for any $a_*$.
Our matching calculation gives a scalar bound state decay
rate that is two times smaller than that found
in~\cite{Detweiler:1980uk}, though it agrees with the expression given
in~\cite{Pani:2012bp}, and also matches the numerical results
of~\cite{Dolan:2007mj}. 


\subsection{$\ell = j-1$ bound states}

Vector bound states with $j \neq \ell$ have $A_0 \neq 0$ at the level
of the hydrogenic wavefunctions, so the matching procedure is 
slightly more complicated than for the $j=\ell$ states.
For the $\ell=j-1$ hydrogenic bound state, we have
\begin{align}
	\partial_i \Psi_i^H &= \sqrt{\frac{j}{2j+1}} \left(R'(r) - (j-1) \frac{R(r)}{r}\right) Y^{jm} \,.
\end{align}
Since $\Psi_i^H \sim (r/a)^{j-1} Y^{j-1,jm}_i$ near the origin,
the leading term in the brackets cancels, and we obtain
\begin{equation}
	\Psi_0 \sim \frac{1}{\mu a} \left(\frac{r}{a}\right)^{j-1} Y^{jm}\,.
\end{equation}
Taking the example of the $\ell = 0, j = 1, n=0$, mode, we have
\begin{equation}
	\Psi_i^H = \frac{2}{a^{3/2}} e^{-r/a} Y_i^{0,11}
	\quad , \quad
	\Psi_0 \simeq \frac{2 i}{\sqrt{3} \omega a^{5/2}} e^{-r/a} Y^{11} \,.
	\label{eq:l0}
\end{equation}
In the matching region, where only the first term of \eqref{eq:transe}
is significant at leading order in $\alpha$, the bound state
can be approximated by the purely-transverse even-parity mode $\Psi^E$
\eqref{eq:transe}.
As for the $j=\ell$ modes, this matching tells us the 
energy flux into the BH from the $E$ and $B$ fields.
However,
for the $\ell=0,j=1$ mode, the contribution from $\mu^2 A_\mu A_\nu$
is no longer sub-leading.
In a Schwarzschild background, we can see this from the fact that
the radial dependence of $\Psi_0$ is approximately constant
near to the BH, taking the form from equation~\ref{eq:l0}
even at small radii.\footnote{Using the Proca equations
of motion, which are separable in Schwarzschild coordinates~\cite{Rosa:2011my},
we have $\hat{\mathcal{D}}_2 (r A_0) - \frac{2 r_g f}{r} \left(E_r + (r A_0)/r^2\right) = 0$, where $f \equiv 1 - 2 r_g/r$,
and $E_r$ is the radial electric field. Since $A_0$ is approximately
constant at small radii in the hydrogenic regime, $A_0 \simeq -r_g E_r$
there.
In a Schwarzschild background, 
$E_r$ obeys its own wave equation, and 
it can be checked that $A_0 + r_g E_r$ remains small even
when $r$ becomes comparable to $2 r_g$. 
Thus, $A_0$ approximately obeys the $\ell=0$ scalar radial
equation, so is approximately constant all the way to the horizon.
} The energy flux through the BH horizon is
\begin{align}
	\frac{dE}{dt} &= \int dA_H T_{\mu\nu} k^\mu \xi^\nu \\
	&= \int_H dA \left(F_{\mu\alpha} F_\nu^{\,\,\alpha}
	+ \mu^2 A_\mu A_\nu \right) k^\mu \xi^\nu \,,
\end{align}
where $k^\mu$ is the timelike Killing vector field, and $\xi^\mu$ ($= k^\mu$
in Schwarzschild) is the normal
to the horizon. Using ingoing Eddington-Finkelstein coordinates, we obtain
\begin{equation}
	\left\langle \frac{dE}{dt} \right\rangle 
	= \int_H dA \, \mu^2 \langle A_v^2 \rangle
	\simeq \frac{16}{3} \alpha^7 \mu^2 \,,
\end{equation}
(where the angle brackets indicate time averaging)
to leading order in $\alpha$, compared to $\frac{32}{3} \alpha^7 \mu^2$ for the Poynting term from
the $\Psi^E$ matching. 
Numerical simulations of Proca bound states in a Schwarzschild background
show a ratio of Poynting to $\mu^2 A_\mu A_\nu$ fluxes
that is very close to 2, as expected.
The combined decay rate for the bound
state is $\Gamma \simeq 16 \alpha^7 \mu$ \,,
(where we take `decay rate' to mean the energy density decay rate,
i.e.\ twice the field amplitude decay rate).

The $\Psi^E$ matching computation does not depend on whether the background is 
Schwarzschild or Kerr, since the $\sim (r_g/r)^2$ differences
between the two metrics give corrections that are sub-leading in $\alpha$.
So, to obtain the Poynting energy flux into a Kerr BH, we can multiply the Schwarzschild flux by the pre-factor
$\frac{\PP_{j=1}|_{a_*}}{\PP_{j=1}|_{a* = 0}}$.
For the $\mu^2 A_\mu A_\nu$ contribution, the energy flux into the BH
scales like
$\frac{\PP_{j=0,m=1}|_{a_*}}{\PP_{j=0,m=1}|_{a* = 0}}$
(where we have abused notation slightly by
evaluating~\eqref{eq:pabs} with $j=0$ but $m=1$).
Combining these scalings, we can obtain the leading-$\alpha$
growth rate for the bound state around a Kerr BH of any spin,
\begin{align}
	\Gamma_{l=0,j=m=1} &\simeq -\frac{32}{3} \alpha^7 \mu
\frac{\PP_{j=1}|_{a_*}}{\PP_{j=1}|_{a* = 0}}
	- \frac{16}{3} \alpha^7 \mu 
\frac{\PP_{j=0,m=1}|_{a_*}}{\PP_{j=0,m=1}|_{a* = 0}}
\\
	&\simeq 4 a_* \alpha^6 \mu
\end{align}
where the last equality is in the $\alpha \ll a_*$ limit.
While the $\omega$ terms in the $\prod_{n=1}^j$ part of $\PP$
(equation~\ref{eq:pabs})
are technically higher order in $\alpha$, we keep them in the expressions
we use, since they give the correct behavior when $\omega - m \Omega$ is
a small parameter (i.e.\ near the superradiance boundary for that level).

For $\ell = j + 1 > 0$ bound states, the situation is similar:
the Poynting flux can be obtained by matching onto the transverse
$\Psi^E$ mode, while there is an additional `scalar' flux
that can be determined from the behavior of the $A_0$ part 
of the wavefunction. It can be demonstrated that both
of these result in a $\Gamma \sim \alpha^{2j + 5} \mu$
scaling (for the Schwarzschild decay rate), while their coefficients
could be calculated by arguments similar to those above.


\subsection{$j=0$, $\ell=1$ bound states}

For $j=0$, the longitudinal mode is
\begin{align}
	\Psi_0^R &= j_0(\omega r) Y^{00} \,, \\
	\Psi_i^R &= j_1(\omega r) Y^{1,00} \simeq \frac{\omega r}{3} Y^{1,00}_i
	+ \OO((\omega r)^3) \,,
\end{align}
(there being no $Y^{-1,00}$ harmonic)
which matches the form of the hydrogenic bound
state, 
\begin{equation}
	\Psi_i^H \simeq R_0 (r/a) Y^{1,00}_i \Rightarrow \Psi_0^H \sim \frac{-3iR_0}{\mu a} Y^{00}\,.
\end{equation}
This is as expected, since massless
transverse states must have $j \ge 1$, so cannot
match the form of a $j=0$ state. Matching onto the massless
scalar absorption rate,
we obtain
\begin{equation}
	\Gamma \simeq 	\frac{16 \alpha ^7 \mu  (n+1) (n+3)}{(n+2)^5}
\end{equation}
as the leading-$\alpha$ Schwarzschild absorption rate.
We can derive the Kerr rates via multiplying by the 
$\frac{\PP_{j=0,m=0}|_{a_*}}{\PP_{j=0,m=0}|_{a* = 0}}$ factor as above.


\subsection{$\ell = j+1$ bound states}

Similarly to the $\ell = j-1$ bound states, the
$\ell = j+1$ bound state growth/decay rates receive same-order
contributions from the Poynting flux obtained by matching onto the
$\Psi^E$ mode, and from the `scalar' flux. The $\Psi_i^H \sim (r/a)^{j+1}$
behavior near the origin results in a Schwarzschild decay rate
$\Gamma \sim \alpha^{4j + 7} \mu$, which can then by translated
in a Kerr growth/decay rate as above.


\subsection{Comparison to literature}
\label{sec:srcompare}

\begin{figure}
\includegraphics[width =.8 \columnwidth]{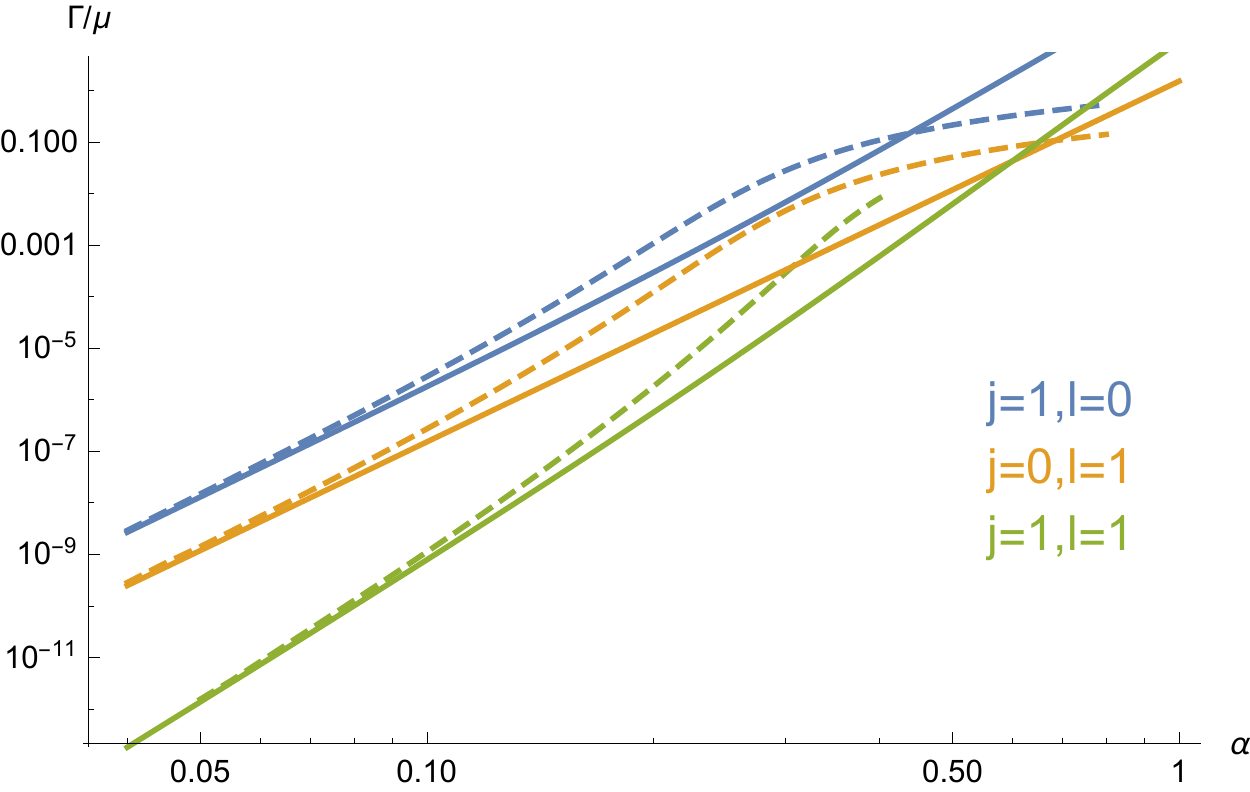}
\includegraphics[width =.8 \columnwidth]{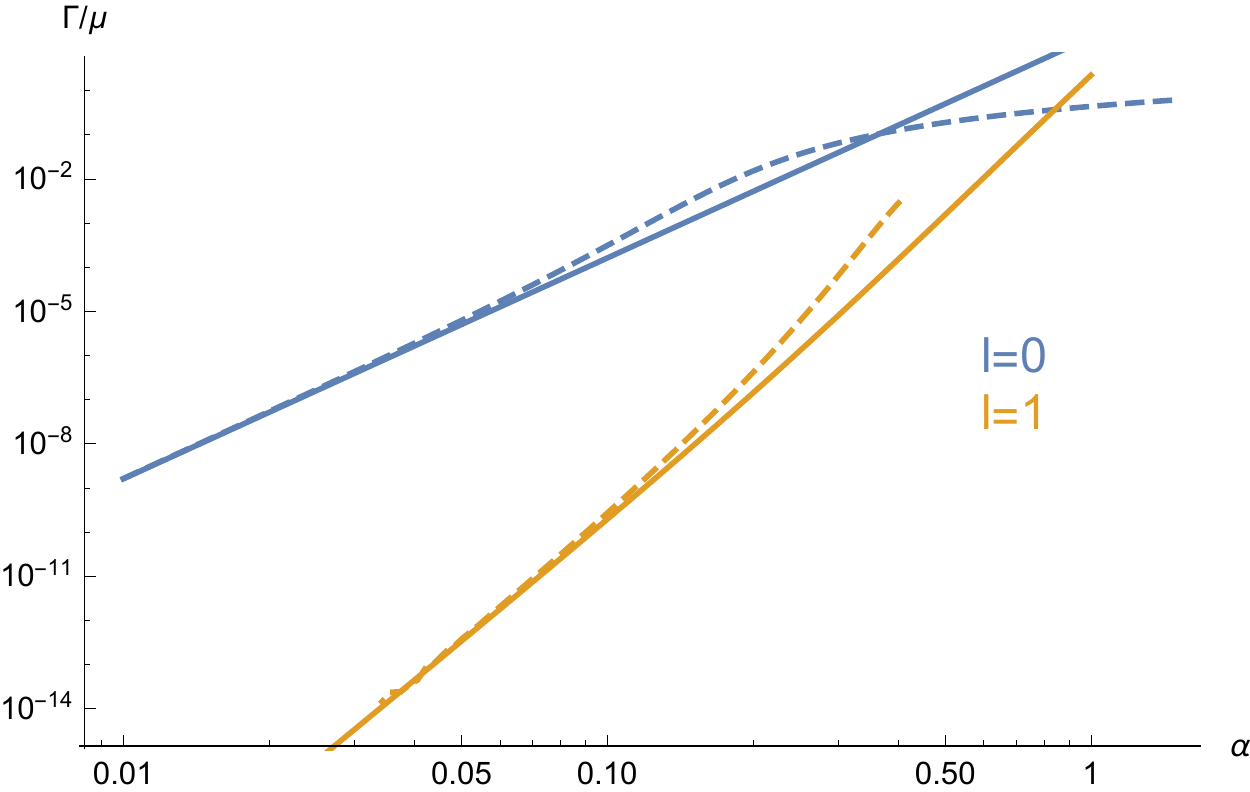}
	\caption{Decay rates of vector (top panel) and scalar (bottom panel) bound states around a Schwarzschild BH.
	Dotted lines are numerical calculations from~\cite{Rosa:2011my}
	and~\cite{Dolan:2007mj}, while
	solid lines are our analytic rates.
	As expected, the analytic and numerical results agree at
	small $\alpha$.
}\label{fig:dolan}
\end{figure}

A number of papers have considered the growth/decay rates of vector
bound states around BHs. The first work to notice the $\Gamma
\sim \alpha^{2j + 2\ell + 5} \mu$ scaling, \cite{Rosa:2011my},
performed numerical computations of bound state decay rates around a
Schwarzschild BH, where the Proca equation is separable (as noted in
Section~\ref{sec:boundstates}, their labels for $j$ and $\ell$ are
switched relative to ours). They observed that the numerical results
were well fit by the form we derived above, and gave an analytic
argument for the $j=\ell$ case. In Figure~\ref{fig:dolan}, we compare
our analytic leading-$\alpha$ decay rates for the Schwarzschild
background, to~\cite{Rosa:2011my}'s numerical results. The numerical
results converge to the analytic expressions at small $\alpha$, and
deviate at finite $\alpha$ in a manner similar to the scalar case, which is
plotted in
the lower panel of Figure~\ref{fig:dolan}.

A different analytic matching approach was pursued in
\cite{Endlich:2016jgc}. While we match at the level of wavefunctions,
\cite{Endlich:2016jgc} posit an effective Hamiltonian that describes
the interaction of long-wavelength field modes with a small BH, and
then use this to compute bound state growth/decay rates in the
small-$a_*$ approximation. Given the assumed form of their effective
Hamiltonian, there appears to be no choice for the operator
coefficients that leads to the correct scaling of growth rates / amplification factors
as a function of $\alpha$ for all of the bound states / scattering
states. For example, the choice of
coefficients presented in~\cite{Endlich:2016jgc} gives the correct
scaling $\alpha^{2\ell+2j+5} \mu$ for $j\geq 1$ but gives a decay rate of
$\sim \alpha^9 \mu$ in Schwarzschild for the $j=0, \ell=1$ mode, whereas
we find $\sim \alpha^7 \mu$, which matches
the numerical computations of~\cite{Rosa:2011my} and \cite{Pani:2012bp}. 
Moreover, though this choice
gives the correct $\alpha$ dependence for other bound states,
it gives different numerical coefficients to ours. 
Comparing the 
leading-$a_*$, leading-$\alpha$ superradiance rates presented
in~\cite{Endlich:2016jgc} with those derived in this work, we have
\begin{align}
	&\Gamma \simeq 4 a_* \alpha^6 \mu \quad 
	&\Gamma_{\rm EP} \simeq  \frac{20}{3} a_* \alpha^6 \mu \quad &(j = 1, \ell = 0) \\
	&\Gamma \simeq \frac{1}{6} a_* \alpha^8 \mu \quad 
	&\Gamma_{\rm EP} \simeq \frac{1}{3} a_* \alpha^8 \mu \quad &(j = 1, \ell = 1)  \,,
\end{align}
where $\Gamma_{\rm EP}$ denotes the rate from~\cite{Endlich:2016jgc}.
In
particular, \cite{Endlich:2016jgc} obtains a rate for the $j=\ell=1$ level
a factor of 2 larger than this work, where the latter agrees
with the analytic calculations of \cite{Rosa:2011my} and
\cite{Pani:2012bp}. 

\begin{figure}
	\includegraphics[width = \columnwidth]{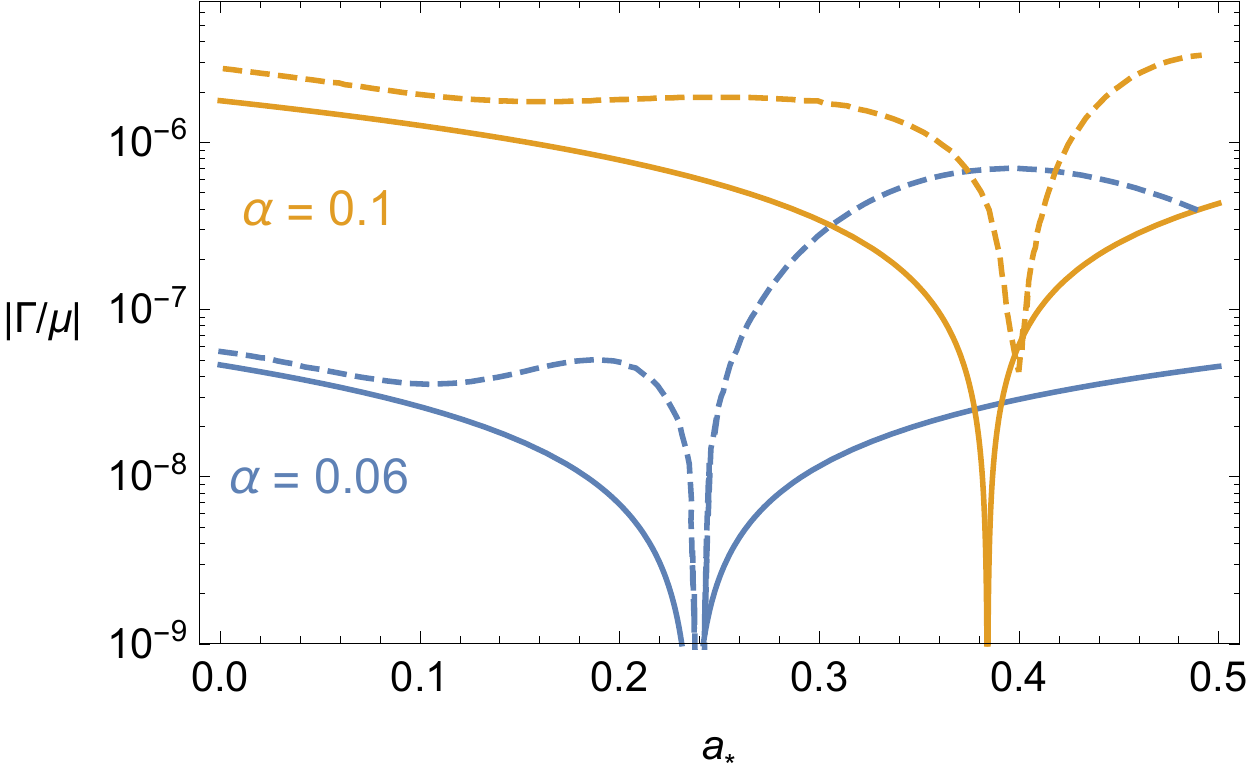}
	\caption{Comparison between the second-order in $a_*$ growth/decay rates
	for the $\ell=0$, $j=m=1$ mode computed in~\cite{Pani:2012bp} (dashed lines), 
	and our leading-order in $\alpha$ approximation (solid lines), for two values of $\alpha$.
	As expected, the decay rates agree to within a factor $\sim 2$
	when both $a_*$ and $\alpha$ are small, but are more different when
	$a_*$ is larger.
	}
\label{fig:pani}
\end{figure}

Another approach is to expand in small $a_*$, obtaining a coupled set
of one-dimensional equations are then solved numerically
\cite{Pani:2012bp,Pani:2012vp} . Whereas our analytic approach is
under control at small $\alpha$ but arbitrary $a_*$, this semi-analytic
method \cite{Pani:2012bp,Pani:2012vp} is theoretically under control at arbitrary
$\alpha$ but small $a_*$.  The numerics are challenging at small
$\alpha$, resulting in the $\alpha^{2\ell + 2j+4}$ scaling being
obscured due to numerical errors for the leading $j=1,\ell=0$ level
\cite{Pani:2012bp}.  Comparing the two
approaches, we find good agreement in the small-$\alpha$, small-$a_*$
regime, as expected.  Figure~\ref{fig:pani} compares the results for
the $\ell=0$, $j=m=1$ level.  At $a_* = 0$, the small difference
between the result from \cite{Pani:2012bp}, which is a numerical
calculation to all orders in $\alpha$, and our leading-$\alpha$
result, comes from $\alpha$ needing to be smaller for our expansion to
accurately converge, as per Figure~\ref{fig:dolan}.  As
Figure~\ref{fig:pani} illustrates, the difference is significantly
greater at larger $a_*$. It would be interesting to compare to a full
numerical calculation, to see how small $\alpha$ needs to be before
our leading-$\alpha$ form gives the correct $a_*$ dependence.

To perform a controlled calculation at large $a_*$ and large $\alpha$,
it appears to be necessary to do a full 2D (in $r$ and $\theta$)
computation in the Kerr background. 
\cite{Witek:2012tr} performs a numerical time-domain simulation of
this kind, and finds that the amplitude of an initial wave packet
grows over time around a fast-spinning BH, though they do not
appear to cleanly separate the different bound states.
\cite{East:2017mrj} performs a similar numerical simulation,
and extracts the 
fastest-growing vector bound states
in a Kerr background. As described in Section~\ref{sec:srrates}
and illustrated in Figure~\ref{fig:vrates},
their results correspond well to our analytic
approximation.

\section{Perturbations due to companion}
\label{ap:mixing}

The presence of a star or BH as a binary companion to a
superradiating BH can hinder superradiance of a growing level because
such an asymmetric perturbation can cause the growing level to mix with
decaying ones. In this appendix, we show that this perturbation does not
affect the bounds set with BH spin measurements, and estimate the size
of the effect on spins in BBH systems.

The orbits of companion stars in X-ray binaries have a period on the
order of days or more, at distances of $R\sim 10^{7}$ km away from the
BH. Since the orbital periods are much smaller than the energies of
the bound states, the perturbing potential changes adiabatically, and we
can consider a static perturbing potential $\delta V(\vec{r})$. This is also
the case for BBH systems at the early stages of their formation, where
they are separated by distance
\begin{align}
	R\sim 10^6 r_g \left(\frac{T}{10^{10}\mathrm{yr}}\right)^{1/4} \left(\frac{2 M_{\odot}}{M}\right)^{1/4} \,,
\end{align} 
assuming two equal mass BHs that merge in time $T$ through GW emission.

The condition that a particular superradiant level has a net positive growth is given by
\begin{align}
\left|
\frac{\Gamma_{\text{decay}}^{n'j'\ell'm'}}{\Gamma_{\text{sr}}^{nj\ell
m}} \right|^{\frac{1}{2}} \left|\frac{\langle
A_{\text{decay}}^{n'j'\ell'm'} | \delta V(\vec{r}) |
A_{\text{sr}}^{nj\ell m}\rangle}{\Delta E} \right| <1 \,,
\end{align}
where $\Gamma_{\text{decay}}$ and $\Gamma_{\text{sr}}$ are the decay and superradiance rates, $A_{\text{decay}}$ and $A_{\text{sr}}$ are the bound states, and $\Delta E$ is the energy difference between the states~\cite{Arvanitaki:2014wva}.

At small $\alpha$ (large cloud size) where this perturbation is most
important, the ratio of the rates is dominated by the $\alpha$ power,
and scales as
\begin{equation}
\frac{\Gamma_{\text{decay}}^{n'j'\ell'm'}}{\Gamma_{\text{sr}}^{njlm}} \sim \alpha^{2(j'-j) + 2(\ell'-\ell)}\,.
\end{equation}

To estimate the perturbation by the companion on the bound state, we decompose its
gravitational potential into multipoles:
\begin{align}
&\delta V(r,\theta,\phi) \simeq \\
&\, \frac{G M'}{R} \left(1+ \frac{r}{R} \sum_{-1
  \leq m \leq 1}\frac{4\pi}{3}Y_{1m}(\theta,\phi) Y_{1m}^*(\theta',\phi')  + ... \right) \,, \nonumber
\end{align}
where $M'$ is the companion mass, $R$ the distance between BH and
companion, and primed coordinates correspond to the companion's
location.  The leading contribution comes from the dipole term. The
dipole mixing term allows mixing of the superradiating ($N=\ell +1$,
$j$, $\ell=j-1$, $m=j$) state with any other bound state with
$\Delta \ell = 1$.

The fastest superradiating level has $j=m=1,\ell =0, N=1$ and
can undergo dipole mixing with the $N=2$, $j=m=0$, $\ell=1$
level. Since $\ell+j$ is the same for both states,
$\Gamma_{\text{decay}}/\Gamma_{\text{sr}} \sim 1$. Because
$\Delta N \neq 0$, the energy difference is set by the difference in
the bound state energies $\Delta E \sim \frac{3\mu
  \alpha^{2}}{8}$.
For perturbations to be unimportant for this level, the condition is
\begin{align}
\alpha \gtrsim \left( \frac{r_{g}}{R}\right)^{1/2} \left(\frac{M'}{M_{BH}} \right)^{1/4} \,.
\end{align}
For typical X-ray binary and BBH systems, this condition saturates at
$\alpha \sim 10^{-3}$. At such low values of $\alpha$, the
superradiance rates are too slow to form a maximally filled cloud
within a binary lifetime, so mixing does not impose an additional
constraint.

The story for the $j>1$ levels is slightly different. For the
strongest perturbation, consider mixing with the ``lower'' ($N'=N$,
$j'=j-1$, $\ell'=\ell-1$, $m'=m-1$) bound state. Because the energy
quantum number $N$ stays the same but $\ell$ changes, the energy
difference is given by the ``fine'' splitting $\sim \mu \alpha^{4}$ in
the hydrogenic limit; including higher-order corrections to the bound
states and metric may lift degeneracies and increase the
splitting, which would reduce the effect of the mixing. The ratio of decay to SR
rate is $\sim \alpha^{4}$.

For $j>1$ levels, the resulting mixing does not affect superradiant growth if 
\begin{align}
\alpha \gtrsim\left( \frac{N  \, r_{g}}{R} \right)^{1/4} \left(\frac{M'}{M_{BH}}  \right) ^{1/8} \,.
\end{align}

For the X-ray binary systems used to constrain vector masses, the values
of $\alpha$ that saturate this condition are small enough such that the
``lower'' bound state still satisfies the SR condition. (Note that these
``lower'' bound states we considered are not the same as the fastest
superradiating bound states with a lower $j$; they have different
energy quantum number $N$.) Thus, for these BHs, perturbations due to a
companion star are not the limiting factor. Instead, the limiting factor
is that the superradiance time has to be faster than the BH lifetime.

For BBH systems, the typical time for superradiance to occur is longer,
$10^7$ -- $10^{10}$ years, and the BH companion may start at a closer
distance. This means that for the $j>1$ levels the mixing is important in
limiting the spin-down of BHs in black hole binaries, and we take this into
account when estimating the power of statistical searches at aLIGO
(Section~\ref{sec:spinstats}).

For $j=1$, the values of $\alpha$ affected by companions are also
typically too small to produce observable GW signals, so blind searches
for GWs may be sensitive to BHs in binary systems as well as isolated
BHs; the populations are expected to be comparable \cite{shakura1973black}.

\section{Gravitational radiation rates}
\label{ap:gwrates}

From~\cite{Misner:1974qy}, if the metric $g_{\mu\nu}^{(B)}$ corresponds
to a background vacuum ($R_{\mu\nu} = 0$) spacetime, 
then small metric perturbations
\begin{equation}
	g_{\mu\nu} = g^{(B)}_{\mu\nu} + h_{\mu\nu}
\end{equation}
are sourced by a small stress-energy tensor $T_{\mu\nu}$
through the linearized wave equation
\begin{equation}
	D^\alpha D_\alpha \bar{h}_{\mu\nu} + 2 R^{(B)}_{\alpha\mu\beta\nu} \bar{h}^{\alpha\beta} 
	= - 16 \pi G T_{\mu\nu} \,,
\end{equation}
where $\bar{h}_{\mu\nu} = h_{\mu\nu} - \frac{1}{2} h g^B_{\mu\nu}$ is the
trace-reversed metric perturbation
(working in `Lorentz gauge' $\bar{h}_{\mu\nu}^{;\nu} = 0$).
Indices here are raised and lowered by $g^{(B)}_{\mu\nu}$,
and $\nabla$s denote covariant derivatives.

In flat space, this reduces to the usual wave equation
\begin{equation}
	\partial^\alpha \partial_\alpha \bar{h}_{\mu\nu}  
	= - 16 \pi G T_{\mu\nu} \,.
\end{equation}
This can be solved using the flat-space Green's function.
For a periodic $T_{\mu\nu}$, we obtain an emitted power~\cite{Weinberg:1972kfs}
\begin{equation}
	\frac{dP}{d\Omega} = \frac{G \omega^2}{\pi} \tilde{T}_{TT}^{ij}(\omega,k)
	\left(\tilde{T}_{TT}^{ij}(\omega,k)
	\right)^* \,,
\end{equation}
where $\tilde{T}_{\mu\nu}$ is the spatial Fourier transform of
$T_{\mu\nu}$, and $\tilde{T}_{TT}^{ij}$ is the transverse (relative to
$\vec{k}$) traceless part of $\tilde{T}^{ij}$.

If the background space is weakly curved,
\begin{equation}
	g^{(B)}_{\mu\nu} = \eta_{\mu\nu} + \epsilon_{\mu\nu} \,,
\end{equation}
where $\epsilon_{\mu\nu}$ is small, then we can expand $h_{\mu\nu}$ to
each order in $\epsilon$, writing
	$h_{\mu\nu} = h^{(0)}_{\mu\nu}
+ h^{(1)}_{\mu\nu} + \dots$.
The homogeneous wave equation at first order in $\epsilon$ is then (schematically)
\begin{equation}
 \partial^2 \bar{h}^{(1)} = - (\nabla \partial + \partial \nabla - 2
\partial \partial ) \bar{h}^{(0)} - 2 R^{(B)} \bar{h}^{(0)} \,,
	\label{eq:h1}
\end{equation}
where we have suppressed indices.

For gravitational radiation from transitions between levels, the above
story works well. It can be checked that,
in the small-$\alpha$ limit, the contribution to gravitational
radiation from the strongly-curved space near the BH 
($r \sim {\rm few \, } \times r_g \ll \omega^{-1}$) is small.
In the weakly-curved background far away from the BH,
the $h^{(1)}$ correction is sub-leading in $\alpha$ to $h^{(0)}$,
where $h^{(0)}$ is the result obtained from the non-relativistic bound state $T_{\mu\nu}$
sourcing gravitational waves through their flat-space Green's function.

However, the small-curvature expansion breaks down for gravitational
waves from bound state annihilations. The problem is that we have only a
single small parameter, $\alpha$, which controls both the small curvature
and the properties of $T_{\mu\nu}$. From the continuity properties of
the (non-relativistic) $T_{\mu\nu}$ near the origin, its Fourier transform must
fall off at least as fast as some particular power of $k$, at large $k \gg a^{-1}$.
However, due to a cancellation that occurs for hydrogenic bound states, it actually falls off faster than this,
resulting in $h^{(0)}$ being suppressed by extra powers
of $1/(a k) \sim \alpha$.
If we try to solve Eq.~\eqref{eq:h1} for $h^{(1)}$,
there is a suppression from the small non-flat parts of the metric,
but the cancellation from the radial profile of the source term
is lifted. The result is that
$h^{(1)}$ is of the same order or lower order in $\alpha$ as $h^{(0)}$,
and the expansion 
	$h_{\mu\nu} = h^{(0)}_{\mu\nu}
+ h^{(1)}_{\mu\nu} + \dots$ stops making sense.

In the case of the $2p$ scalar bound state~\cite{Brito:2014wla}, the cancellation
suppresses the flat space amplitude by two powers of $\alpha$ relative
to the `naive' expectation, while (in the formal expansion) the $h^{(1)}$ part
is only suppressed by $\alpha$ from the small curvature.
The result is that the `Schwarzschild background corrections'
actually determine the leading-order-in-$\alpha$ annihilation rate.
For higher-$\ell$ scalar bound states, the cancellation only
suppresses the flat space amplitude by one power of $\alpha$,
resulting in the Schwarzschild correction being of the same order.

For all of the vector bound states, the latter situation applies;
the Schwarzschild correction gives a same-order contribution to
the annihilation rate. To actually compute the 
leading-$\alpha$ form of the rate, it is generally simpler
to use the wave equation for perturbations to the Riemann tensor, rather than 
to the metric as above. 
Using the Newman-Penrose scalar $\Psi_4$,
which corresponds to the appropriate components of the Weyl tensor
describing gravitational radiation, it is possible to write
down a one-dimensional radial equation, sourced
by appropriate combinations of $T_{\mu\nu}$ and its
derivatives~\cite{Brito:2014wla,Poisson:1993vp}. 
We have calculated an approximation to this solution for the $\ell=0, j=1,
m=1$ level (the fastest-growing vector bound state), finding that
it leads to a rate $\sim 10$ times the naive flat-space estimate
in the small-$\alpha$ limit.
This gives an emitted GW power of
\begin{equation}
	P \simeq 60 \alpha^{12} \frac{G N^2}{r_g^4}
	\simeq N_M \mu \left(\frac{N}{N_M}\right)^2 60 \alpha^{12} \mu \,,
\end{equation}
where $N_M \equiv G M^2$ is of order the maximum occupation
number of the level (as per Eq.~\eqref{eq:nmax}).
Thus, for small $\alpha$, the cloud annihilates on a timescale
much larger than its oscillation time. However, as discussed in
Section~\ref{sec:history}, this timescale can be shorter than other
astrophysically-relevant ones, including the growth timescale
for the next superradiant level.
For a general bound state,
the gravitational radiation power scales like
$\alpha^{4\ell + 12} G N^2 r_g^{-4}$. 
Table~\ref{tab:ann_power} shows the `flat-space' results for the first
few levels; as described above, one would expect
the full leading-$\alpha$ results to have
the same $\alpha$ dependence, but different (generally larger)
numerical coefficients.

For a spherically-symmetric background metric, the
angular distribution of the emitted gravitational
radiation is set by the angular form of the $T_{ij}$ source.
This will be the case for the leading-$\alpha$ GW emission from bound
states, since as per above, the leading-$\alpha$ annihilation power can
be obtained purely from the Schwarzschild part of the metric.
Thus, the emitted tensor modes from a $Y^{\ell,jm}$ state
will be of the $T^{E2,(2j)(2m)}$ form (in the notation of~\cite{thorne}),
giving the same polarization and angular emission profiles
as a standard quadrupole GW source, such as a spinning asymmetric neutron
star.
For a $j=m=1$ state, this corresponds to an angular emission power
profile
of
\begin{equation}
	\frac{dP}{d\Omega} \propto (1 + 6 \cos^2\theta_k + \cos^4 \theta_k) \,.
\end{equation}
In particular, the emitted power in the most intense direction is $5/2$
times the angle-averaged power, and in the least intense direction is
$5/16$ times the average power. Since this angular variation is not too strong,
we have simply used the averaged value for the calculations in the 
main text.

\begin{table}
\centering
\begin{tabular}{c | c c c}
\hline
 $j$ & $\ell=j-1$ & $\ell=j$ & $\ell=j+1$ \\ [0.5ex] 
\hline
 1 & $\frac{32 \alpha^{12}}{5}$ & $\frac{\alpha^{16}}{2560}$ & $(5\times 10^{-8}) \alpha^{20}$ \\ 
 2 & $\frac{\alpha^{16}}{126}$ & $(1\times10^{-6}) \alpha^{20}$ & $(1\times10^{-11}) \alpha^{24}$ \\
 3 & $(6\times10^{-6}) \alpha^{20}$ & $(7\times10^{-10})\alpha^{24}$ & $(4\times10^{-15})\alpha^{28}$ \\
 4 & $(2\times10^{-9})\alpha^{24}$ & $(2\times10^{-13})\alpha^{28}$ & $(2\times10^{-13})\alpha^{32}$ \\
 5 & $(4\times10^{-13})\alpha^{28}$ & $(2\times10^{-17})\alpha^{32}$ & $(6\times10^{-23})\alpha^{36}$ \\ [1ex] 
\hline
\end{tabular}
 \caption{`Flat-space' results for the leading-$\alpha$ gravitational wave emission power/$(N_m^{2}\frac{G}{r_{g}^{4}})$ for vector bound states. For given $j$ and $\ell$, and we take $m=j$, and $n=0$.}
  \label{tab:ann_power}
\end{table}

\section{Calculation for number of detectable sources}
\label{ap:events}

For each vector mass $\mu$, we estimate the number of detectable sources by integrating over various astrophysical distributions of stellar mass BHs:
\begin{align}
\begin{split}
	\text{Number of sources } = &  \int_{M_{\text{min}}}^{M_{\text{max}}} \text{d}M \, P(M) \times \int_{0}^{1} \text{d} a_{*} \, P(a_{*}) \\
&\times \int_{0}^{\text{reach}(M,a_{*})} \text{d}r \, P(r) \\
&\times \tau_{\text{sig}}(M,a_{*},r) \times \text{BHFR} \,,
\end{split}
\end{align}
where $P(M)$, $P(a_{*})$, and $P(r)$ are the normalized probability distributions of finding a BH with mass $M$ and spin $a_{*}$ at a distance $r$ away, reach($M,a_{*}$) is the maximum distance of such a BH whose signal is above the detector noise floor, $\tau_{\text{sig}}(M,a_{*},r)$ is the signal duration, and BHFR is the BH formation rate.

We follow the astrophysical distributions used in Appendix C of
\cite{Arvanitaki:2014wva} with a few changes to the BH mass
distributions in light of recent LIGO observations confirming the
existence of $\sim  30 M_{\odot}$ BHs. To account for
astrophysical uncertainties, the parameters in the distributions are
varied to produce optimistic, realistic, and pessimistic estimates
(see Table \ref{tab:distributions}).

For the mass distribution, we use
\begin{align}
P(M) = M_{0}e^{\frac{M_{\text{min}}}{M_{0}}} e^{-\frac{M}{M_{0}}} \,,
\end{align}
with $M_{\text{min}} = 4.1 M_{\odot}$, and width $M_{0}$ that is
varied in our estimates. We impose a maximum cutoff mass with two
different values: $M_{\text{max}} = (80,160)M_{\odot}$. Using a
distribution that falls off more slowly, e.g. power law, increases the
number of detectable sources at low vector mass. We use the spin
distributions stated in Table \ref{tab:distributions}. For the
distance distribution, we use a $\delta$-function at 8 kpc for BHs in
the Milky Way, and follow the blue luminosity distribution (Figure 6
in \cite{0004-637X-675-2-1459}) for those beyond our galaxy.

\begin{table}[h]
  \begin{center}
    \begin{tabular}{ | c| c  c  c |}
      \hline
      Quantity & Optimistic & Realistic & Pessimistic \\ 
      \hline
    $M_{0}$ ($M_{\odot}$) & 7.9 & 4.7 & 4.7 \\
    $P(a_{*})$ & $90\%$ above 0.8  & $30\%$ above 0.8 & uniform \\
   BHFR (per century) & 0.9 & 0.38 & 0.08 \\
      \hline
    \end{tabular}
  \end{center}
  \caption{
Parameters of astrophysical distributions used to estimate number of events.}
  \label{tab:distributions}
\end{table}

\section{Frequency drift}
\label{ap:freqdrift}

As mentioned in Section~\ref{sec:anligo}, the gravitational self-energy
of the cloud will affect the frequency of GWs emitted through annihilations.
As the cloud annihilates away, this contribution to the binding energy
decreases, resulting in the emitted frequency increasing with time.
The rate at which the frequency
changes can affect the detectability of the annihilation signal.

LIGO continuous wave searches currently cover a range of positive to
negative frequency derivatives, $3.39\times 10^{-10}$ Hz/s through
$-2.67\times 10^{-9}$ Hz/s with $T_{\text{coh}} = 60$ hours
\cite{TheLIGOScientific:2016uns}. For $T_{\text{coh}} = 250$ hours,
and the same amount of computational time, the maximum positive
frequency drift searched over is
$\dot f_{\text{max}} \sim 2\times 10^{-11}$ Hz/s \cite{Wette:2009uea}.

To estimate the expected frequency drift $\dot f$, we start by calculating the
gravitational self-energy (per particle) of a bound state,
\begin{align}
U_{\text{cloud}} = -G\int{\frac{\rho \, m_{N}(r)}{r} \text{d}^{3}\mathbf{x}} \,,
\end{align}
where
\begin{align}
m_{N}(r) = \int_{0}^{r} N_m \rho(r) \text{ d}^{3}\mathbf{x} \quad \text{and} \quad
\rho(r) = \mu |\psi|^{2} \,.
\end{align}
For the $j=1$ bound state, $\psi$ is the $n=0$, $\ell=0$ hydrogen wave
function, and the frequency drift is
\begin{align}
\dot f & \simeq \frac{1}{2\pi} \times 2 \dot U_{\text{j=1 cloud}} \\
& \simeq \frac{5\, \mu \,\alpha^3}{16 \pi}(\Gamma_{\rm ann} N_m) \left(\frac{N_m}{G M^{2}} \right)\,,
\end{align}
where we have used $\dot N_m = - \Gamma_{\rm ann} N_m^{2}$. For annihilation
signals that come from beyond the Milky Way, which are stronger and
have shorter annihilation times than those from the Milky Way, the
frequency drifts are too large ($\gtrsim 10^{-8}$ Hz/s), and are thus
beyond the range searched by aLIGO.  Without this constraint on the
frequency drift, there would be a comparable number of expected events
from outside the Milky Way peaking at $\sim 10^{-12}$~eV.


\bibliography{SR-mergers}

\begin{thebibliography}{85}%
\makeatletter
\providecommand \@ifxundefined [1]{%
 \@ifx{#1\undefined}
}%
\providecommand \@ifnum [1]{%
 \ifnum #1\expandafter \@firstoftwo
 \else \expandafter \@secondoftwo
 \fi
}%
\providecommand \@ifx [1]{%
 \ifx #1\expandafter \@firstoftwo
 \else \expandafter \@secondoftwo
 \fi
}%
\providecommand \natexlab [1]{#1}%
\providecommand \enquote  [1]{``#1''}%
\providecommand \bibnamefont  [1]{#1}%
\providecommand \bibfnamefont [1]{#1}%
\providecommand \citenamefont [1]{#1}%
\providecommand \href@noop [0]{\@secondoftwo}%
\providecommand \href [0]{\begingroup \@sanitize@url \@href}%
\providecommand \@href[1]{\@@startlink{#1}\@@href}%
\providecommand \@@href[1]{\endgroup#1\@@endlink}%
\providecommand \@sanitize@url [0]{\catcode `\\12\catcode `\$12\catcode
  `\&12\catcode `\#12\catcode `\^12\catcode `\_12\catcode `\%12\relax}%
\providecommand \@@startlink[1]{}%
\providecommand \@@endlink[0]{}%
\providecommand \url  [0]{\begingroup\@sanitize@url \@url }%
\providecommand \@url [1]{\endgroup\@href {#1}{\urlprefix }}%
\providecommand \urlprefix  [0]{URL }%
\providecommand \Eprint [0]{\href }%
\providecommand \doibase [0]{http://dx.doi.org/}%
\providecommand \selectlanguage [0]{\@gobble}%
\providecommand \bibinfo  [0]{\@secondoftwo}%
\providecommand \bibfield  [0]{\@secondoftwo}%
\providecommand \translation [1]{[#1]}%
\providecommand \BibitemOpen [0]{}%
\providecommand \bibitemStop [0]{}%
\providecommand \bibitemNoStop [0]{.\EOS\space}%
\providecommand \EOS [0]{\spacefactor3000\relax}%
\providecommand \BibitemShut  [1]{\csname bibitem#1\endcsname}%
\let\auto@bib@innerbib\@empty
\bibitem [{\citenamefont {Essig}\ \emph {et~al.}(2013)\citenamefont {Essig}
  \emph {et~al.}}]{Essig:2013lka}%
  \BibitemOpen
  \bibfield  {author} {\bibinfo {author} {\bibfnamefont {R.}~\bibnamefont
  {Essig}} \emph {et~al.},\ }in\ \href
  {http://inspirehep.net/record/1263039/files/arXiv:1311.0029.pdf} {\emph
  {\bibinfo {booktitle} {{Proceedings, 2013 Community Summer Study on the
  Future of U.S. Particle Physics: Snowmass on the Mississippi (CSS2013):
  Minneapolis, MN, USA, July 29-August 6, 2013}}}}\ (\bibinfo {year} {2013})\
  \Eprint {http://arxiv.org/abs/1311.0029} {arXiv:1311.0029 [hep-ph]}
  \BibitemShut {NoStop}%
\bibitem [{\citenamefont {Arvanitaki}\ \emph
  {et~al.}(2010{\natexlab{a}})\citenamefont {Arvanitaki}, \citenamefont
  {Dimopoulos}, \citenamefont {Dubovsky}, \citenamefont {Kaloper},\ and\
  \citenamefont {March-Russell}}]{Arvanitaki:2009fg}%
  \BibitemOpen
  \bibfield  {author} {\bibinfo {author} {\bibfnamefont {A.}~\bibnamefont
  {Arvanitaki}}, \bibinfo {author} {\bibfnamefont {S.}~\bibnamefont
  {Dimopoulos}}, \bibinfo {author} {\bibfnamefont {S.}~\bibnamefont
  {Dubovsky}}, \bibinfo {author} {\bibfnamefont {N.}~\bibnamefont {Kaloper}}, \
  and\ \bibinfo {author} {\bibfnamefont {J.}~\bibnamefont {March-Russell}},\
  }\href {\doibase 10.1103/PhysRevD.81.123530} {\bibfield  {journal} {\bibinfo
  {journal} {Phys. Rev.}\ }\textbf {\bibinfo {volume} {D81}},\ \bibinfo {pages}
  {123530} (\bibinfo {year} {2010}{\natexlab{a}})},\ \Eprint
  {http://arxiv.org/abs/0905.4720} {arXiv:0905.4720 [hep-th]} \BibitemShut
  {NoStop}%
\bibitem [{\citenamefont {Arvanitaki}\ \emph
  {et~al.}(2010{\natexlab{b}})\citenamefont {Arvanitaki}, \citenamefont
  {Craig}, \citenamefont {Dimopoulos}, \citenamefont {Dubovsky},\ and\
  \citenamefont {March-Russell}}]{Arvanitaki:2009hb}%
  \BibitemOpen
  \bibfield  {author} {\bibinfo {author} {\bibfnamefont {A.}~\bibnamefont
  {Arvanitaki}}, \bibinfo {author} {\bibfnamefont {N.}~\bibnamefont {Craig}},
  \bibinfo {author} {\bibfnamefont {S.}~\bibnamefont {Dimopoulos}}, \bibinfo
  {author} {\bibfnamefont {S.}~\bibnamefont {Dubovsky}}, \ and\ \bibinfo
  {author} {\bibfnamefont {J.}~\bibnamefont {March-Russell}},\ }\href {\doibase
  10.1103/PhysRevD.81.075018} {\bibfield  {journal} {\bibinfo  {journal} {Phys.
  Rev.}\ }\textbf {\bibinfo {volume} {D81}},\ \bibinfo {pages} {075018}
  (\bibinfo {year} {2010}{\natexlab{b}})},\ \Eprint
  {http://arxiv.org/abs/0909.5440} {arXiv:0909.5440 [hep-ph]} \BibitemShut
  {NoStop}%
\bibitem [{\citenamefont {Zeldovich}(1971)}]{Zeldovich}%
  \BibitemOpen
  \bibfield  {author} {\bibinfo {author} {\bibfnamefont {Y.~B.}\ \bibnamefont
  {Zeldovich}},\ }\href@noop {} {\bibfield  {journal} {\bibinfo  {journal}
  {JETP Lett.}\ }\textbf {\bibinfo {volume} {14}},\ \bibinfo {pages} {180}
  (\bibinfo {year} {1971})}\BibitemShut {NoStop}%
\bibitem [{\citenamefont {Misner}(1972)}]{Misner:1972kx}%
  \BibitemOpen
  \bibfield  {author} {\bibinfo {author} {\bibfnamefont {C.~W.}\ \bibnamefont
  {Misner}},\ }\href {\doibase 10.1103/PhysRevLett.28.994} {\bibfield
  {journal} {\bibinfo  {journal} {Phys. Rev. Lett.}\ }\textbf {\bibinfo
  {volume} {28}},\ \bibinfo {pages} {994} (\bibinfo {year} {1972})}\BibitemShut
  {NoStop}%
\bibitem [{\citenamefont {Starobinskii}(1973)}]{Starobinskii}%
  \BibitemOpen
  \bibfield  {author} {\bibinfo {author} {\bibfnamefont {A.}~\bibnamefont
  {Starobinskii}},\ }\href@noop {} {\bibfield  {journal} {\bibinfo  {journal}
  {Soviet Phys. JETP}\ }\textbf {\bibinfo {volume} {37}},\ \bibinfo {pages}
  {28} (\bibinfo {year} {1973})}\BibitemShut {NoStop}%
\bibitem [{\citenamefont {Ternov}\ \emph {et~al.}(1978)\citenamefont {Ternov},
  \citenamefont {Khalilov}, \citenamefont {Chizhov},\ and\ \citenamefont
  {Gaina}}]{Ternov:1978gq}%
  \BibitemOpen
  \bibfield  {author} {\bibinfo {author} {\bibfnamefont {I.~M.}\ \bibnamefont
  {Ternov}}, \bibinfo {author} {\bibfnamefont {V.~R.}\ \bibnamefont
  {Khalilov}}, \bibinfo {author} {\bibfnamefont {G.~A.}\ \bibnamefont
  {Chizhov}}, \ and\ \bibinfo {author} {\bibfnamefont {A.~B.}\ \bibnamefont
  {Gaina}},\ }\href {\doibase 10.1007/BF00894575} {\bibfield  {journal}
  {\bibinfo  {journal} {Sov. Phys. J.}\ }\textbf {\bibinfo {volume} {21}},\
  \bibinfo {pages} {1200} (\bibinfo {year} {1978})},\ \bibinfo {note} {[Izv.
  Vuz. Fiz.21N9,109(1978)]}\BibitemShut {NoStop}%
\bibitem [{\citenamefont {Zouros}\ and\ \citenamefont
  {Eardley}(1979)}]{Zouros:1979iw}%
  \BibitemOpen
  \bibfield  {author} {\bibinfo {author} {\bibfnamefont {T.~J.~M.}\
  \bibnamefont {Zouros}}\ and\ \bibinfo {author} {\bibfnamefont {D.~M.}\
  \bibnamefont {Eardley}},\ }\href {\doibase 10.1016/0003-4916(79)90237-9}
  {\bibfield  {journal} {\bibinfo  {journal} {Annals Phys.}\ }\textbf {\bibinfo
  {volume} {118}},\ \bibinfo {pages} {139} (\bibinfo {year}
  {1979})}\BibitemShut {NoStop}%
\bibitem [{\citenamefont {Detweiler}(1980)}]{Detweiler:1980uk}%
  \BibitemOpen
  \bibfield  {author} {\bibinfo {author} {\bibfnamefont {S.~L.}\ \bibnamefont
  {Detweiler}},\ }\href {\doibase 10.1103/PhysRevD.22.2323} {\bibfield
  {journal} {\bibinfo  {journal} {Phys. Rev.}\ }\textbf {\bibinfo {volume}
  {D22}},\ \bibinfo {pages} {2323} (\bibinfo {year} {1980})}\BibitemShut
  {NoStop}%
\bibitem [{\citenamefont {Dolan}(2007)}]{Dolan:2007mj}%
  \BibitemOpen
  \bibfield  {author} {\bibinfo {author} {\bibfnamefont {S.~R.}\ \bibnamefont
  {Dolan}},\ }\href {\doibase 10.1103/PhysRevD.76.084001} {\bibfield  {journal}
  {\bibinfo  {journal} {Phys. Rev.}\ }\textbf {\bibinfo {volume} {D76}},\
  \bibinfo {pages} {084001} (\bibinfo {year} {2007})},\ \Eprint
  {http://arxiv.org/abs/0705.2880} {arXiv:0705.2880 [gr-qc]} \BibitemShut
  {NoStop}%
\bibitem [{\citenamefont {Arvanitaki}\ and\ \citenamefont
  {Dubovsky}(2011)}]{Arvanitaki:2010sy}%
  \BibitemOpen
  \bibfield  {author} {\bibinfo {author} {\bibfnamefont {A.}~\bibnamefont
  {Arvanitaki}}\ and\ \bibinfo {author} {\bibfnamefont {S.}~\bibnamefont
  {Dubovsky}},\ }\href {\doibase 10.1103/PhysRevD.83.044026} {\bibfield
  {journal} {\bibinfo  {journal} {Phys. Rev.}\ }\textbf {\bibinfo {volume}
  {D83}},\ \bibinfo {pages} {044026} (\bibinfo {year} {2011})},\ \Eprint
  {http://arxiv.org/abs/1004.3558} {arXiv:1004.3558 [hep-th]} \BibitemShut
  {NoStop}%
\bibitem [{\citenamefont {Yoshino}\ and\ \citenamefont
  {Kodama}(2014)}]{Yoshino:2013ofa}%
  \BibitemOpen
  \bibfield  {author} {\bibinfo {author} {\bibfnamefont {H.}~\bibnamefont
  {Yoshino}}\ and\ \bibinfo {author} {\bibfnamefont {H.}~\bibnamefont
  {Kodama}},\ }\href {\doibase 10.1093/ptep/ptu029} {\bibfield  {journal}
  {\bibinfo  {journal} {PTEP}\ }\textbf {\bibinfo {volume} {2014}},\ \bibinfo
  {pages} {043E02} (\bibinfo {year} {2014})},\ \Eprint
  {http://arxiv.org/abs/1312.2326} {arXiv:1312.2326 [gr-qc]} \BibitemShut
  {NoStop}%
\bibitem [{\citenamefont {Arvanitaki}\ \emph {et~al.}(2015)\citenamefont
  {Arvanitaki}, \citenamefont {Baryakhtar},\ and\ \citenamefont
  {Huang}}]{Arvanitaki:2014wva}%
  \BibitemOpen
  \bibfield  {author} {\bibinfo {author} {\bibfnamefont {A.}~\bibnamefont
  {Arvanitaki}}, \bibinfo {author} {\bibfnamefont {M.}~\bibnamefont
  {Baryakhtar}}, \ and\ \bibinfo {author} {\bibfnamefont {X.}~\bibnamefont
  {Huang}},\ }\href {\doibase 10.1103/PhysRevD.91.084011} {\bibfield  {journal}
  {\bibinfo  {journal} {Phys. Rev.}\ }\textbf {\bibinfo {volume} {D91}},\
  \bibinfo {pages} {084011} (\bibinfo {year} {2015})},\ \Eprint
  {http://arxiv.org/abs/1411.2263} {arXiv:1411.2263 [hep-ph]} \BibitemShut
  {NoStop}%
\bibitem [{\citenamefont {Brito}\ \emph
  {et~al.}(2015{\natexlab{a}})\citenamefont {Brito}, \citenamefont {Cardoso},\
  and\ \citenamefont {Pani}}]{Brito:2014wla}%
  \BibitemOpen
  \bibfield  {author} {\bibinfo {author} {\bibfnamefont {R.}~\bibnamefont
  {Brito}}, \bibinfo {author} {\bibfnamefont {V.}~\bibnamefont {Cardoso}}, \
  and\ \bibinfo {author} {\bibfnamefont {P.}~\bibnamefont {Pani}},\ }\href
  {\doibase 10.1088/0264-9381/32/13/134001} {\bibfield  {journal} {\bibinfo
  {journal} {Class. Quant. Grav.}\ }\textbf {\bibinfo {volume} {32}},\ \bibinfo
  {pages} {134001} (\bibinfo {year} {2015}{\natexlab{a}})},\ \Eprint
  {http://arxiv.org/abs/1411.0686} {arXiv:1411.0686 [gr-qc]} \BibitemShut
  {NoStop}%
\bibitem [{\citenamefont {Brito}\ \emph
  {et~al.}(2015{\natexlab{b}})\citenamefont {Brito}, \citenamefont {Cardoso},\
  and\ \citenamefont {Pani}}]{Brito:2015oca}%
  \BibitemOpen
  \bibfield  {author} {\bibinfo {author} {\bibfnamefont {R.}~\bibnamefont
  {Brito}}, \bibinfo {author} {\bibfnamefont {V.}~\bibnamefont {Cardoso}}, \
  and\ \bibinfo {author} {\bibfnamefont {P.}~\bibnamefont {Pani}},\ }\href
  {\doibase 10.1007/978-3-319-19000-6} {\bibfield  {journal} {\bibinfo
  {journal} {Lect. Notes Phys.}\ }\textbf {\bibinfo {volume} {906}},\ \bibinfo
  {pages} {pp.1} (\bibinfo {year} {2015}{\natexlab{b}})},\ \Eprint
  {http://arxiv.org/abs/1501.06570} {arXiv:1501.06570 [gr-qc]} \BibitemShut
  {NoStop}%
\bibitem [{\citenamefont {Arvanitaki}\ \emph {et~al.}(2017)\citenamefont
  {Arvanitaki}, \citenamefont {Baryakhtar}, \citenamefont {Dimopoulos},
  \citenamefont {Dubovsky},\ and\ \citenamefont
  {Lasenby}}]{Arvanitaki:2016qwi}%
  \BibitemOpen
  \bibfield  {author} {\bibinfo {author} {\bibfnamefont {A.}~\bibnamefont
  {Arvanitaki}}, \bibinfo {author} {\bibfnamefont {M.}~\bibnamefont
  {Baryakhtar}}, \bibinfo {author} {\bibfnamefont {S.}~\bibnamefont
  {Dimopoulos}}, \bibinfo {author} {\bibfnamefont {S.}~\bibnamefont
  {Dubovsky}}, \ and\ \bibinfo {author} {\bibfnamefont {R.}~\bibnamefont
  {Lasenby}},\ }\href {\doibase 10.1103/PhysRevD.95.043001} {\bibfield
  {journal} {\bibinfo  {journal} {Phys. Rev.}\ }\textbf {\bibinfo {volume}
  {D95}},\ \bibinfo {pages} {043001} (\bibinfo {year} {2017})},\ \Eprint
  {http://arxiv.org/abs/1604.03958} {arXiv:1604.03958 [hep-ph]} \BibitemShut
  {NoStop}%
\bibitem [{\citenamefont {Goodsell}\ \emph {et~al.}(2009)\citenamefont
  {Goodsell}, \citenamefont {Jaeckel}, \citenamefont {Redondo},\ and\
  \citenamefont {Ringwald}}]{Goodsell:2009xc}%
  \BibitemOpen
  \bibfield  {author} {\bibinfo {author} {\bibfnamefont {M.}~\bibnamefont
  {Goodsell}}, \bibinfo {author} {\bibfnamefont {J.}~\bibnamefont {Jaeckel}},
  \bibinfo {author} {\bibfnamefont {J.}~\bibnamefont {Redondo}}, \ and\
  \bibinfo {author} {\bibfnamefont {A.}~\bibnamefont {Ringwald}},\ }\href
  {\doibase 10.1088/1126-6708/2009/11/027} {\bibfield  {journal} {\bibinfo
  {journal} {JHEP}\ }\textbf {\bibinfo {volume} {11}},\ \bibinfo {pages} {027}
  (\bibinfo {year} {2009})},\ \Eprint {http://arxiv.org/abs/0909.0515}
  {arXiv:0909.0515 [hep-ph]} \BibitemShut {NoStop}%
\bibitem [{\citenamefont {Camara}\ \emph {et~al.}(2011)\citenamefont {Camara},
  \citenamefont {Ibanez},\ and\ \citenamefont {Marchesano}}]{Camara:2011jg}%
  \BibitemOpen
  \bibfield  {author} {\bibinfo {author} {\bibfnamefont {P.~G.}\ \bibnamefont
  {Camara}}, \bibinfo {author} {\bibfnamefont {L.~E.}\ \bibnamefont {Ibanez}},
  \ and\ \bibinfo {author} {\bibfnamefont {F.}~\bibnamefont {Marchesano}},\
  }\href {\doibase 10.1007/JHEP09(2011)110} {\bibfield  {journal} {\bibinfo
  {journal} {JHEP}\ }\textbf {\bibinfo {volume} {09}},\ \bibinfo {pages} {110}
  (\bibinfo {year} {2011})},\ \Eprint {http://arxiv.org/abs/1106.0060}
  {arXiv:1106.0060 [hep-th]} \BibitemShut {NoStop}%
\bibitem [{\citenamefont {Rosa}\ and\ \citenamefont
  {Dolan}(2012)}]{Rosa:2011my}%
  \BibitemOpen
  \bibfield  {author} {\bibinfo {author} {\bibfnamefont {J.~G.}\ \bibnamefont
  {Rosa}}\ and\ \bibinfo {author} {\bibfnamefont {S.~R.}\ \bibnamefont
  {Dolan}},\ }\href {\doibase 10.1103/PhysRevD.85.044043} {\bibfield  {journal}
  {\bibinfo  {journal} {Phys. Rev.}\ }\textbf {\bibinfo {volume} {D85}},\
  \bibinfo {pages} {044043} (\bibinfo {year} {2012})},\ \Eprint
  {http://arxiv.org/abs/1110.4494} {arXiv:1110.4494 [hep-th]} \BibitemShut
  {NoStop}%
\bibitem [{\citenamefont {Pani}\ \emph
  {et~al.}(2012{\natexlab{a}})\citenamefont {Pani}, \citenamefont {Cardoso},
  \citenamefont {Gualtieri}, \citenamefont {Berti},\ and\ \citenamefont
  {Ishibashi}}]{Pani:2012bp}%
  \BibitemOpen
  \bibfield  {author} {\bibinfo {author} {\bibfnamefont {P.}~\bibnamefont
  {Pani}}, \bibinfo {author} {\bibfnamefont {V.}~\bibnamefont {Cardoso}},
  \bibinfo {author} {\bibfnamefont {L.}~\bibnamefont {Gualtieri}}, \bibinfo
  {author} {\bibfnamefont {E.}~\bibnamefont {Berti}}, \ and\ \bibinfo {author}
  {\bibfnamefont {A.}~\bibnamefont {Ishibashi}},\ }\href {\doibase
  10.1103/PhysRevD.86.104017} {\bibfield  {journal} {\bibinfo  {journal} {Phys.
  Rev.}\ }\textbf {\bibinfo {volume} {D86}},\ \bibinfo {pages} {104017}
  (\bibinfo {year} {2012}{\natexlab{a}})},\ \Eprint
  {http://arxiv.org/abs/1209.0773} {arXiv:1209.0773 [gr-qc]} \BibitemShut
  {NoStop}%
\bibitem [{\citenamefont {Pani}\ \emph
  {et~al.}(2012{\natexlab{b}})\citenamefont {Pani}, \citenamefont {Cardoso},
  \citenamefont {Gualtieri}, \citenamefont {Berti},\ and\ \citenamefont
  {Ishibashi}}]{Pani:2012vp}%
  \BibitemOpen
  \bibfield  {author} {\bibinfo {author} {\bibfnamefont {P.}~\bibnamefont
  {Pani}}, \bibinfo {author} {\bibfnamefont {V.}~\bibnamefont {Cardoso}},
  \bibinfo {author} {\bibfnamefont {L.}~\bibnamefont {Gualtieri}}, \bibinfo
  {author} {\bibfnamefont {E.}~\bibnamefont {Berti}}, \ and\ \bibinfo {author}
  {\bibfnamefont {A.}~\bibnamefont {Ishibashi}},\ }\href {\doibase
  10.1103/PhysRevLett.109.131102} {\bibfield  {journal} {\bibinfo  {journal}
  {Phys. Rev. Lett.}\ }\textbf {\bibinfo {volume} {109}},\ \bibinfo {pages}
  {131102} (\bibinfo {year} {2012}{\natexlab{b}})},\ \Eprint
  {http://arxiv.org/abs/1209.0465} {arXiv:1209.0465 [gr-qc]} \BibitemShut
  {NoStop}%
\bibitem [{\citenamefont {Endlich}\ and\ \citenamefont
  {Penco}(2016)}]{Endlich:2016jgc}%
  \BibitemOpen
  \bibfield  {author} {\bibinfo {author} {\bibfnamefont {S.}~\bibnamefont
  {Endlich}}\ and\ \bibinfo {author} {\bibfnamefont {R.}~\bibnamefont
  {Penco}},\ }\href@noop {} {\  (\bibinfo {year} {2016})},\ \Eprint
  {http://arxiv.org/abs/1609.06723} {arXiv:1609.06723 [hep-th]} \BibitemShut
  {NoStop}%
\bibitem [{\citenamefont {East}(2017)}]{East:2017mrj}%
  \BibitemOpen
  \bibfield  {author} {\bibinfo {author} {\bibfnamefont {W.~E.}\ \bibnamefont
  {East}},\ }\href@noop {} {\  (\bibinfo {year} {2017})},\ \Eprint
  {http://arxiv.org/abs/1705.01544} {arXiv:1705.01544 [gr-qc]} \BibitemShut
  {NoStop}%
\bibitem [{\citenamefont {Witek}\ \emph {et~al.}(2013)\citenamefont {Witek},
  \citenamefont {Cardoso}, \citenamefont {Ishibashi},\ and\ \citenamefont
  {Sperhake}}]{Witek:2012tr}%
  \BibitemOpen
  \bibfield  {author} {\bibinfo {author} {\bibfnamefont {H.}~\bibnamefont
  {Witek}}, \bibinfo {author} {\bibfnamefont {V.}~\bibnamefont {Cardoso}},
  \bibinfo {author} {\bibfnamefont {A.}~\bibnamefont {Ishibashi}}, \ and\
  \bibinfo {author} {\bibfnamefont {U.}~\bibnamefont {Sperhake}},\ }\href
  {\doibase 10.1103/PhysRevD.87.043513} {\bibfield  {journal} {\bibinfo
  {journal} {Phys. Rev.}\ }\textbf {\bibinfo {volume} {D87}},\ \bibinfo {pages}
  {043513} (\bibinfo {year} {2013})},\ \Eprint {http://arxiv.org/abs/1212.0551}
  {arXiv:1212.0551 [gr-qc]} \BibitemShut {NoStop}%
\bibitem [{\citenamefont {Teukolsky}\ and\ \citenamefont
  {Press}(1974)}]{Teukolsky:1974yv}%
  \BibitemOpen
  \bibfield  {author} {\bibinfo {author} {\bibfnamefont {S.~A.}\ \bibnamefont
  {Teukolsky}}\ and\ \bibinfo {author} {\bibfnamefont {W.~H.}\ \bibnamefont
  {Press}},\ }\href {\doibase 10.1086/153180} {\bibfield  {journal} {\bibinfo
  {journal} {Astrophys. J.}\ }\textbf {\bibinfo {volume} {193}},\ \bibinfo
  {pages} {443} (\bibinfo {year} {1974})}\BibitemShut {NoStop}%
\bibitem [{\citenamefont {Bekenstein}(1973)}]{PhysRevD.7.949}%
  \BibitemOpen
  \bibfield  {author} {\bibinfo {author} {\bibfnamefont {J.~D.}\ \bibnamefont
  {Bekenstein}},\ }\href {\doibase 10.1103/PhysRevD.7.949} {\bibfield
  {journal} {\bibinfo  {journal} {Phys. Rev. D}\ }\textbf {\bibinfo {volume}
  {7}},\ \bibinfo {pages} {949} (\bibinfo {year} {1973})}\BibitemShut {NoStop}%
\bibitem [{\citenamefont {Press}\ and\ \citenamefont
  {Teukolsky}(1972)}]{Press:1972zz}%
  \BibitemOpen
  \bibfield  {author} {\bibinfo {author} {\bibfnamefont {W.~H.}\ \bibnamefont
  {Press}}\ and\ \bibinfo {author} {\bibfnamefont {S.~A.}\ \bibnamefont
  {Teukolsky}},\ }\href {\doibase 10.1038/238211a0} {\bibfield  {journal}
  {\bibinfo  {journal} {Nature}\ }\textbf {\bibinfo {volume} {238}},\ \bibinfo
  {pages} {211} (\bibinfo {year} {1972})}\BibitemShut {NoStop}%
\bibitem [{\citenamefont {Page}(1976)}]{Page:1976df}%
  \BibitemOpen
  \bibfield  {author} {\bibinfo {author} {\bibfnamefont {D.~N.}\ \bibnamefont
  {Page}},\ }\href {\doibase 10.1103/PhysRevD.13.198} {\bibfield  {journal}
  {\bibinfo  {journal} {Phys. Rev.}\ }\textbf {\bibinfo {volume} {D13}},\
  \bibinfo {pages} {198} (\bibinfo {year} {1976})}\BibitemShut {NoStop}%
\bibitem [{\citenamefont {Thorne}(1980)}]{thorne}%
  \BibitemOpen
  \bibfield  {author} {\bibinfo {author} {\bibfnamefont {K.~S.}\ \bibnamefont
  {Thorne}},\ }\href {\doibase 10.1103/RevModPhys.52.299} {\bibfield  {journal}
  {\bibinfo  {journal} {Rev. Mod. Phys.}\ }\textbf {\bibinfo {volume} {52}},\
  \bibinfo {pages} {299} (\bibinfo {year} {1980})}\BibitemShut {NoStop}%
\bibitem [{\citenamefont {East}\ and\ \citenamefont
  {Pretorius}(2017)}]{East:2017ovw}%
  \BibitemOpen
  \bibfield  {author} {\bibinfo {author} {\bibfnamefont {W.~E.}\ \bibnamefont
  {East}}\ and\ \bibinfo {author} {\bibfnamefont {F.}~\bibnamefont
  {Pretorius}},\ }\href@noop {} {\  (\bibinfo {year} {2017})},\ \Eprint
  {http://arxiv.org/abs/1704.04791} {arXiv:1704.04791 [gr-qc]} \BibitemShut
  {NoStop}%
\bibitem [{\citenamefont {Bosch}\ \emph {et~al.}(2016)\citenamefont {Bosch},
  \citenamefont {Green},\ and\ \citenamefont {Lehner}}]{Bosch:2016vcp}%
  \BibitemOpen
  \bibfield  {author} {\bibinfo {author} {\bibfnamefont {P.}~\bibnamefont
  {Bosch}}, \bibinfo {author} {\bibfnamefont {S.~R.}\ \bibnamefont {Green}}, \
  and\ \bibinfo {author} {\bibfnamefont {L.}~\bibnamefont {Lehner}},\ }\href
  {\doibase 10.1103/PhysRevLett.116.141102} {\bibfield  {journal} {\bibinfo
  {journal} {Phys. Rev. Lett.}\ }\textbf {\bibinfo {volume} {116}},\ \bibinfo
  {pages} {141102} (\bibinfo {year} {2016})},\ \Eprint
  {http://arxiv.org/abs/1601.01384} {arXiv:1601.01384 [gr-qc]} \BibitemShut
  {NoStop}%
\bibitem [{\citenamefont {Sanchis-Gual}\ \emph {et~al.}(2016)\citenamefont
  {Sanchis-Gual}, \citenamefont {Degollado}, \citenamefont {Herdeiro},
  \citenamefont {Font},\ and\ \citenamefont {Montero}}]{Sanchis-Gual:2016tcm}%
  \BibitemOpen
  \bibfield  {author} {\bibinfo {author} {\bibfnamefont {N.}~\bibnamefont
  {Sanchis-Gual}}, \bibinfo {author} {\bibfnamefont {J.~C.}\ \bibnamefont
  {Degollado}}, \bibinfo {author} {\bibfnamefont {C.}~\bibnamefont {Herdeiro}},
  \bibinfo {author} {\bibfnamefont {J.~A.}\ \bibnamefont {Font}}, \ and\
  \bibinfo {author} {\bibfnamefont {P.~J.}\ \bibnamefont {Montero}},\ }\href
  {\doibase 10.1103/PhysRevD.94.044061} {\bibfield  {journal} {\bibinfo
  {journal} {Phys. Rev.}\ }\textbf {\bibinfo {volume} {D94}},\ \bibinfo {pages}
  {044061} (\bibinfo {year} {2016})},\ \Eprint
  {http://arxiv.org/abs/1607.06304} {arXiv:1607.06304 [gr-qc]} \BibitemShut
  {NoStop}%
\bibitem [{\citenamefont {Miller}\ and\ \citenamefont
  {Miller}(2014)}]{Miller:2014aaa}%
  \BibitemOpen
  \bibfield  {author} {\bibinfo {author} {\bibfnamefont {M.~C.}\ \bibnamefont
  {Miller}}\ and\ \bibinfo {author} {\bibfnamefont {J.~M.}\ \bibnamefont
  {Miller}},\ }\href {\doibase 10.1016/j.physrep.2014.09.003} {\bibfield
  {journal} {\bibinfo  {journal} {Phys. Rept.}\ }\textbf {\bibinfo {volume}
  {548}},\ \bibinfo {pages} {1} (\bibinfo {year} {2014})},\ \Eprint
  {http://arxiv.org/abs/1408.4145} {arXiv:1408.4145 [astro-ph.HE]} \BibitemShut
  {NoStop}%
\bibitem [{\citenamefont {Reynolds}(2014)}]{Reynolds:2013qqa}%
  \BibitemOpen
  \bibfield  {author} {\bibinfo {author} {\bibfnamefont {C.~S.}\ \bibnamefont
  {Reynolds}},\ }\href {\doibase 10.1007/s11214-013-0006-6} {\bibfield
  {journal} {\bibinfo  {journal} {Space Sci. Rev.}\ }\textbf {\bibinfo {volume}
  {183}},\ \bibinfo {pages} {277} (\bibinfo {year} {2014})},\ \Eprint
  {http://arxiv.org/abs/1302.3260} {arXiv:1302.3260 [astro-ph.HE]} \BibitemShut
  {NoStop}%
\bibitem [{\citenamefont {McClintock}\ \emph {et~al.}(2014)\citenamefont
  {McClintock}, \citenamefont {Narayan},\ and\ \citenamefont
  {Steiner}}]{McClintock:2013vwa}%
  \BibitemOpen
  \bibfield  {author} {\bibinfo {author} {\bibfnamefont {J.~E.}\ \bibnamefont
  {McClintock}}, \bibinfo {author} {\bibfnamefont {R.}~\bibnamefont {Narayan}},
  \ and\ \bibinfo {author} {\bibfnamefont {J.~F.}\ \bibnamefont {Steiner}},\
  }\href {\doibase 10.1007/s11214-013-0003-9} {\bibfield  {journal} {\bibinfo
  {journal} {Space Sci. Rev.}\ }\textbf {\bibinfo {volume} {183}},\ \bibinfo
  {pages} {295} (\bibinfo {year} {2014})},\ \Eprint
  {http://arxiv.org/abs/1303.1583} {arXiv:1303.1583 [astro-ph.HE]} \BibitemShut
  {NoStop}%
\bibitem [{\citenamefont {Shankar}\ \emph {et~al.}(2009)\citenamefont
  {Shankar}, \citenamefont {Weinberg},\ and\ \citenamefont
  {Miralda-Escude}}]{Shankar:2007zg}%
  \BibitemOpen
  \bibfield  {author} {\bibinfo {author} {\bibfnamefont {F.}~\bibnamefont
  {Shankar}}, \bibinfo {author} {\bibfnamefont {D.~H.}\ \bibnamefont
  {Weinberg}}, \ and\ \bibinfo {author} {\bibfnamefont {J.}~\bibnamefont
  {Miralda-Escude}},\ }\href {\doibase 10.1088/0004-637X/690/1/20} {\bibfield
  {journal} {\bibinfo  {journal} {Astrophys. J.}\ }\textbf {\bibinfo {volume}
  {690}},\ \bibinfo {pages} {20} (\bibinfo {year} {2009})},\ \Eprint
  {http://arxiv.org/abs/0710.4488} {arXiv:0710.4488 [astro-ph]} \BibitemShut
  {NoStop}%
\bibitem [{\citenamefont {Shapiro}\ and\ \citenamefont
  {Teukolsky}(1983)}]{Shapiro:1983du}%
  \BibitemOpen
  \bibfield  {author} {\bibinfo {author} {\bibfnamefont {S.~L.}\ \bibnamefont
  {Shapiro}}\ and\ \bibinfo {author} {\bibfnamefont {S.~A.}\ \bibnamefont
  {Teukolsky}},\ }\href@noop {} {\emph {\bibinfo {title} {{Black holes, white
  dwarfs, and neutron stars: The physics of compact objects}}}}\ (\bibinfo
  {year} {1983})\BibitemShut {NoStop}%
\bibitem [{\citenamefont {Reid}\ \emph
  {et~al.}(2014{\natexlab{a}})\citenamefont {Reid}, \citenamefont {McClintock},
  \citenamefont {Steiner}, \citenamefont {Steeghs}, \citenamefont {Remillard},
  \citenamefont {Dhawan},\ and\ \citenamefont {Narayan}}]{0004-637X-796-1-2}%
  \BibitemOpen
  \bibfield  {author} {\bibinfo {author} {\bibfnamefont {M.~J.}\ \bibnamefont
  {Reid}}, \bibinfo {author} {\bibfnamefont {J.~E.}\ \bibnamefont
  {McClintock}}, \bibinfo {author} {\bibfnamefont {J.~F.}\ \bibnamefont
  {Steiner}}, \bibinfo {author} {\bibfnamefont {D.}~\bibnamefont {Steeghs}},
  \bibinfo {author} {\bibfnamefont {R.~A.}\ \bibnamefont {Remillard}}, \bibinfo
  {author} {\bibfnamefont {V.}~\bibnamefont {Dhawan}}, \ and\ \bibinfo {author}
  {\bibfnamefont {R.}~\bibnamefont {Narayan}},\ }\href
  {http://stacks.iop.org/0004-637X/796/i=1/a=2} {\bibfield  {journal} {\bibinfo
   {journal} {The Astrophysical Journal}\ }\textbf {\bibinfo {volume} {796}},\
  \bibinfo {pages} {2} (\bibinfo {year} {2014}{\natexlab{a}})}\BibitemShut
  {NoStop}%
\bibitem [{\citenamefont {Reid}\ \emph
  {et~al.}(2014{\natexlab{b}})\citenamefont {Reid}, \citenamefont {McClintock},
  \citenamefont {Steiner}, \citenamefont {Steeghs}, \citenamefont {Remillard},
  \citenamefont {Dhawan},\ and\ \citenamefont {Narayan}}]{Reid:2014ywa}%
  \BibitemOpen
  \bibfield  {author} {\bibinfo {author} {\bibfnamefont {M.~J.}\ \bibnamefont
  {Reid}}, \bibinfo {author} {\bibfnamefont {J.~E.}\ \bibnamefont
  {McClintock}}, \bibinfo {author} {\bibfnamefont {J.~F.}\ \bibnamefont
  {Steiner}}, \bibinfo {author} {\bibfnamefont {D.}~\bibnamefont {Steeghs}},
  \bibinfo {author} {\bibfnamefont {R.~A.}\ \bibnamefont {Remillard}}, \bibinfo
  {author} {\bibfnamefont {V.}~\bibnamefont {Dhawan}}, \ and\ \bibinfo {author}
  {\bibfnamefont {R.}~\bibnamefont {Narayan}},\ }\href {\doibase
  10.1088/0004-637X/796/1/2} {\bibfield  {journal} {\bibinfo  {journal}
  {Astrophys. J.}\ }\textbf {\bibinfo {volume} {796}},\ \bibinfo {pages} {2}
  (\bibinfo {year} {2014}{\natexlab{b}})},\ \Eprint
  {http://arxiv.org/abs/1409.2453} {arXiv:1409.2453 [astro-ph.GA]} \BibitemShut
  {NoStop}%
\bibitem [{\citenamefont {Steiner}\ and\ \citenamefont
  {McClintock}(tion)}]{Steiner}%
  \BibitemOpen
  \bibfield  {author} {\bibinfo {author} {\bibfnamefont {J.}~\bibnamefont
  {Steiner}}\ and\ \bibinfo {author} {\bibfnamefont {J.}~\bibnamefont
  {McClintock}},\ }\href@noop {} {} (\bibinfo {year} {private
  communication})\BibitemShut {NoStop}%
\bibitem [{\citenamefont {Dhawan}\ \emph {et~al.}(2007)\citenamefont {Dhawan},
  \citenamefont {Mirabel}, \citenamefont {Ribó},\ and\ \citenamefont
  {Rodrigues}}]{dhawan2007kinematics}%
  \BibitemOpen
  \bibfield  {author} {\bibinfo {author} {\bibfnamefont {V.}~\bibnamefont
  {Dhawan}}, \bibinfo {author} {\bibfnamefont {I.~F.}\ \bibnamefont {Mirabel}},
  \bibinfo {author} {\bibfnamefont {M.}~\bibnamefont {Ribó}}, \ and\ \bibinfo
  {author} {\bibfnamefont {I.}~\bibnamefont {Rodrigues}},\ }\href
  {http://stacks.iop.org/0004-637X/668/i=1/a=430} {\bibfield  {journal}
  {\bibinfo  {journal} {The Astrophysical Journal}\ }\textbf {\bibinfo {volume}
  {668}},\ \bibinfo {pages} {430} (\bibinfo {year} {2007})}\BibitemShut
  {NoStop}%
\bibitem [{\citenamefont {Steeghs}\ \emph {et~al.}(2013)\citenamefont
  {Steeghs}, \citenamefont {McClintock}, \citenamefont {Parsons}, \citenamefont
  {Reid}, \citenamefont {Littlefair},\ and\ \citenamefont
  {Dhillon}}]{Steeghs:2013ksa}%
  \BibitemOpen
  \bibfield  {author} {\bibinfo {author} {\bibfnamefont {D.}~\bibnamefont
  {Steeghs}}, \bibinfo {author} {\bibfnamefont {J.~E.}\ \bibnamefont
  {McClintock}}, \bibinfo {author} {\bibfnamefont {S.~G.}\ \bibnamefont
  {Parsons}}, \bibinfo {author} {\bibfnamefont {M.~J.}\ \bibnamefont {Reid}},
  \bibinfo {author} {\bibfnamefont {S.}~\bibnamefont {Littlefair}}, \ and\
  \bibinfo {author} {\bibfnamefont {V.~S.}\ \bibnamefont {Dhillon}},\ }\href
  {\doibase 10.1088/0004-637X/768/2/185} {\bibfield  {journal} {\bibinfo
  {journal} {Astrophys. J.}\ }\textbf {\bibinfo {volume} {768}},\ \bibinfo
  {pages} {185} (\bibinfo {year} {2013})},\ \Eprint
  {http://arxiv.org/abs/1304.1808} {arXiv:1304.1808 [astro-ph.HE]} \BibitemShut
  {NoStop}%
\bibitem [{\citenamefont {Gou}\ \emph {et~al.}(2014)\citenamefont {Gou},
  \citenamefont {McClintock}, \citenamefont {Remillard}, \citenamefont
  {Steiner}, \citenamefont {Reid}, \citenamefont {Orosz}, \citenamefont
  {Narayan}, \citenamefont {Hanke},\ and\ \citenamefont
  {García}}]{2014ApJ...790...29G}%
  \BibitemOpen
  \bibfield  {author} {\bibinfo {author} {\bibfnamefont {L.}~\bibnamefont
  {Gou}}, \bibinfo {author} {\bibfnamefont {J.~E.}\ \bibnamefont {McClintock}},
  \bibinfo {author} {\bibfnamefont {R.~A.}\ \bibnamefont {Remillard}}, \bibinfo
  {author} {\bibfnamefont {J.~F.}\ \bibnamefont {Steiner}}, \bibinfo {author}
  {\bibfnamefont {M.~J.}\ \bibnamefont {Reid}}, \bibinfo {author}
  {\bibfnamefont {J.~A.}\ \bibnamefont {Orosz}}, \bibinfo {author}
  {\bibfnamefont {R.}~\bibnamefont {Narayan}}, \bibinfo {author} {\bibfnamefont
  {M.}~\bibnamefont {Hanke}}, \ and\ \bibinfo {author} {\bibfnamefont
  {J.}~\bibnamefont {García}},\ }\href {\doibase 10.1088/0004-637X/790/1/29}
  {\bibfield  {journal} {\bibinfo  {journal} {Astrophys. J.}\ }\textbf
  {\bibinfo {volume} {790}},\ \bibinfo {pages} {29} (\bibinfo {year} {2014})},\
  \Eprint {http://arxiv.org/abs/1308.4760} {arXiv:1308.4760 [astro-ph.HE]}
  \BibitemShut {NoStop}%
\bibitem [{\citenamefont {Wong}\ \emph {et~al.}(2012)\citenamefont {Wong},
  \citenamefont {Valsecchi}, \citenamefont {Fragos},\ and\ \citenamefont
  {Kalogera}}]{Wong:2011eg}%
  \BibitemOpen
  \bibfield  {author} {\bibinfo {author} {\bibfnamefont {T.-W.}\ \bibnamefont
  {Wong}}, \bibinfo {author} {\bibfnamefont {F.}~\bibnamefont {Valsecchi}},
  \bibinfo {author} {\bibfnamefont {T.}~\bibnamefont {Fragos}}, \ and\ \bibinfo
  {author} {\bibfnamefont {V.}~\bibnamefont {Kalogera}},\ }\href {\doibase
  10.1088/0004-637X/747/2/111} {\bibfield  {journal} {\bibinfo  {journal}
  {Astrophys. J.}\ }\textbf {\bibinfo {volume} {747}},\ \bibinfo {pages} {111}
  (\bibinfo {year} {2012})},\ \Eprint {http://arxiv.org/abs/1107.5585}
  {arXiv:1107.5585 [astro-ph.HE]} \BibitemShut {NoStop}%
\bibitem [{\citenamefont {Gou}\ \emph {et~al.}(2011)\citenamefont {Gou},
  \citenamefont {McClintock}, \citenamefont {Reid}, \citenamefont {Orosz},
  \citenamefont {Steiner}, \citenamefont {Narayan}, \citenamefont {Xiang},
  \citenamefont {Remillard}, \citenamefont {Arnaud},\ and\ \citenamefont
  {Davis}}]{Gou:2011nq}%
  \BibitemOpen
  \bibfield  {author} {\bibinfo {author} {\bibfnamefont {L.}~\bibnamefont
  {Gou}}, \bibinfo {author} {\bibfnamefont {J.~E.}\ \bibnamefont {McClintock}},
  \bibinfo {author} {\bibfnamefont {M.~J.}\ \bibnamefont {Reid}}, \bibinfo
  {author} {\bibfnamefont {J.~A.}\ \bibnamefont {Orosz}}, \bibinfo {author}
  {\bibfnamefont {J.~F.}\ \bibnamefont {Steiner}}, \bibinfo {author}
  {\bibfnamefont {R.}~\bibnamefont {Narayan}}, \bibinfo {author} {\bibfnamefont
  {J.}~\bibnamefont {Xiang}}, \bibinfo {author} {\bibfnamefont {R.~A.}\
  \bibnamefont {Remillard}}, \bibinfo {author} {\bibfnamefont {K.~A.}\
  \bibnamefont {Arnaud}}, \ and\ \bibinfo {author} {\bibfnamefont {S.~W.}\
  \bibnamefont {Davis}},\ }\href {\doibase 10.1088/0004-637X/742/2/85}
  {\bibfield  {journal} {\bibinfo  {journal} {Astrophys. J.}\ }\textbf
  {\bibinfo {volume} {742}},\ \bibinfo {pages} {85} (\bibinfo {year} {2011})},\
  \Eprint {http://arxiv.org/abs/1106.3690} {arXiv:1106.3690 [astro-ph.HE]}
  \BibitemShut {NoStop}%
\bibitem [{\citenamefont {Willems}\ \emph {et~al.}(2005)\citenamefont
  {Willems}, \citenamefont {Henninger}, \citenamefont {Levin}, \citenamefont
  {Ivanova}, \citenamefont {Kalogera}, \citenamefont {Timmes},\ and\
  \citenamefont {Fryer}}]{Willems:2004kk}%
  \BibitemOpen
  \bibfield  {author} {\bibinfo {author} {\bibfnamefont {B.}~\bibnamefont
  {Willems}}, \bibinfo {author} {\bibfnamefont {M.}~\bibnamefont {Henninger}},
  \bibinfo {author} {\bibfnamefont {T.}~\bibnamefont {Levin}}, \bibinfo
  {author} {\bibfnamefont {N.}~\bibnamefont {Ivanova}}, \bibinfo {author}
  {\bibfnamefont {V.}~\bibnamefont {Kalogera}}, \bibinfo {author}
  {\bibfnamefont {F.~X.}\ \bibnamefont {Timmes}}, \ and\ \bibinfo {author}
  {\bibfnamefont {C.~L.}\ \bibnamefont {Fryer}},\ }\href {\doibase
  10.1086/429557} {\bibfield  {journal} {\bibinfo  {journal} {Astrophys. J.}\
  }\textbf {\bibinfo {volume} {625}},\ \bibinfo {pages} {324} (\bibinfo {year}
  {2005})},\ \Eprint {http://arxiv.org/abs/astro-ph/0411423}
  {arXiv:astro-ph/0411423 [astro-ph]} \BibitemShut {NoStop}%
\bibitem [{\citenamefont {Gou}\ \emph {et~al.}(2009)\citenamefont {Gou},
  \citenamefont {McClintock}, \citenamefont {Liu}, \citenamefont {Narayan},
  \citenamefont {Steiner}, \citenamefont {Remillard}, \citenamefont {Orosz},\
  and\ \citenamefont {Davis}}]{Gou:2009ks}%
  \BibitemOpen
  \bibfield  {author} {\bibinfo {author} {\bibfnamefont {L.}~\bibnamefont
  {Gou}}, \bibinfo {author} {\bibfnamefont {J.~E.}\ \bibnamefont {McClintock}},
  \bibinfo {author} {\bibfnamefont {J.}~\bibnamefont {Liu}}, \bibinfo {author}
  {\bibfnamefont {R.}~\bibnamefont {Narayan}}, \bibinfo {author} {\bibfnamefont
  {J.~F.}\ \bibnamefont {Steiner}}, \bibinfo {author} {\bibfnamefont {R.~A.}\
  \bibnamefont {Remillard}}, \bibinfo {author} {\bibfnamefont {J.~A.}\
  \bibnamefont {Orosz}}, \ and\ \bibinfo {author} {\bibfnamefont {S.~W.}\
  \bibnamefont {Davis}},\ }\href {\doibase 10.1088/0004-637X/701/2/1076}
  {\bibfield  {journal} {\bibinfo  {journal} {Astrophys. J.}\ }\textbf
  {\bibinfo {volume} {701}},\ \bibinfo {pages} {1076} (\bibinfo {year}
  {2009})},\ \Eprint {http://arxiv.org/abs/0901.0920} {arXiv:0901.0920
  [astro-ph.HE]} \BibitemShut {NoStop}%
\bibitem [{\citenamefont {Orosz}\ \emph {et~al.}(2009)\citenamefont {Orosz}
  \emph {et~al.}}]{Orosz:2008kk}%
  \BibitemOpen
  \bibfield  {author} {\bibinfo {author} {\bibfnamefont {J.~A.}\ \bibnamefont
  {Orosz}} \emph {et~al.},\ }\href {\doibase 10.1088/0004-637X/697/1/573}
  {\bibfield  {journal} {\bibinfo  {journal} {Astrophys. J.}\ }\textbf
  {\bibinfo {volume} {697}},\ \bibinfo {pages} {573} (\bibinfo {year}
  {2009})},\ \Eprint {http://arxiv.org/abs/0810.3447} {arXiv:0810.3447
  [astro-ph]} \BibitemShut {NoStop}%
\bibitem [{\citenamefont {{Ruhlen}}\ \emph {et~al.}(2011)\citenamefont
  {{Ruhlen}}, \citenamefont {{Smith}},\ and\ \citenamefont
  {{Swank}}}]{2011ApJ...742...75R}%
  \BibitemOpen
  \bibfield  {author} {\bibinfo {author} {\bibfnamefont {L.}~\bibnamefont
  {{Ruhlen}}}, \bibinfo {author} {\bibfnamefont {D.~M.}\ \bibnamefont
  {{Smith}}}, \ and\ \bibinfo {author} {\bibfnamefont {J.~H.}\ \bibnamefont
  {{Swank}}},\ }\href {\doibase 10.1088/0004-637X/742/2/75} {\bibfield
  {journal} {\bibinfo  {journal} {The Astrophysical Journal}\ }\textbf
  {\bibinfo {volume} {742}},\ \bibinfo {eid} {75} (\bibinfo {year} {2011})},\
  \Eprint {http://arxiv.org/abs/1107.5100} {arXiv:1107.5100 [astro-ph.HE]}
  \BibitemShut {NoStop}%
\bibitem [{\citenamefont {Orosz}\ \emph {et~al.}(2007)\citenamefont {Orosz}
  \emph {et~al.}}]{Orosz:2007ng}%
  \BibitemOpen
  \bibfield  {author} {\bibinfo {author} {\bibfnamefont {J.~A.}\ \bibnamefont
  {Orosz}} \emph {et~al.},\ }\href {\doibase 10.1038/nature06218} {\bibfield
  {journal} {\bibinfo  {journal} {Nature}\ }\textbf {\bibinfo {volume} {449}},\
  \bibinfo {pages} {872} (\bibinfo {year} {2007})},\ \Eprint
  {http://arxiv.org/abs/0710.3165} {arXiv:0710.3165 [astro-ph]} \BibitemShut
  {NoStop}%
\bibitem [{\citenamefont {Pietsch}\ \emph {et~al.}(2006)\citenamefont
  {Pietsch}, \citenamefont {Haberl}, \citenamefont {Sasaki}, \citenamefont
  {Gaetz}, \citenamefont {Plucinsky}, \citenamefont {Ghavamian}, \citenamefont
  {Long},\ and\ \citenamefont {Pannuti}}]{0004-637X-646-1-420}%
  \BibitemOpen
  \bibfield  {author} {\bibinfo {author} {\bibfnamefont {W.}~\bibnamefont
  {Pietsch}}, \bibinfo {author} {\bibfnamefont {F.}~\bibnamefont {Haberl}},
  \bibinfo {author} {\bibfnamefont {M.}~\bibnamefont {Sasaki}}, \bibinfo
  {author} {\bibfnamefont {T.~J.}\ \bibnamefont {Gaetz}}, \bibinfo {author}
  {\bibfnamefont {P.~P.}\ \bibnamefont {Plucinsky}}, \bibinfo {author}
  {\bibfnamefont {P.}~\bibnamefont {Ghavamian}}, \bibinfo {author}
  {\bibfnamefont {K.~S.}\ \bibnamefont {Long}}, \ and\ \bibinfo {author}
  {\bibfnamefont {T.~G.}\ \bibnamefont {Pannuti}},\ }\href
  {http://stacks.iop.org/0004-637X/646/i=1/a=420} {\bibfield  {journal}
  {\bibinfo  {journal} {The Astrophysical Journal}\ }\textbf {\bibinfo {volume}
  {646}},\ \bibinfo {pages} {420} (\bibinfo {year} {2006})}\BibitemShut
  {NoStop}%
\bibitem [{\citenamefont {Reynolds}(2013)}]{Reynolds:2013rva}%
  \BibitemOpen
  \bibfield  {author} {\bibinfo {author} {\bibfnamefont {C.~S.}\ \bibnamefont
  {Reynolds}},\ }\href {\doibase 10.1088/0264-9381/30/24/244004} {\bibfield
  {journal} {\bibinfo  {journal} {Class. Quant. Grav.}\ }\textbf {\bibinfo
  {volume} {30}},\ \bibinfo {pages} {244004} (\bibinfo {year} {2013})},\
  \Eprint {http://arxiv.org/abs/1307.3246} {arXiv:1307.3246 [astro-ph.HE]}
  \BibitemShut {NoStop}%
\bibitem [{\citenamefont {Abbott}\ \emph
  {et~al.}(2016{\natexlab{a}})\citenamefont {Abbott} \emph
  {et~al.}}]{TheLIGOScientific:2016pea}%
  \BibitemOpen
  \bibfield  {author} {\bibinfo {author} {\bibfnamefont {B.~P.}\ \bibnamefont
  {Abbott}} \emph {et~al.} (\bibinfo {collaboration} {Virgo, LIGO
  Scientific}),\ }\href {\doibase 10.1103/PhysRevX.6.041015} {\bibfield
  {journal} {\bibinfo  {journal} {Phys. Rev.}\ }\textbf {\bibinfo {volume}
  {X6}},\ \bibinfo {pages} {041015} (\bibinfo {year} {2016}{\natexlab{a}})},\
  \Eprint {http://arxiv.org/abs/1606.04856} {arXiv:1606.04856 [gr-qc]}
  \BibitemShut {NoStop}%
\bibitem [{\citenamefont {Abbott}\ \emph
  {et~al.}(2016{\natexlab{b}})\citenamefont {Abbott} \emph
  {et~al.}}]{Abbott:2016nhf}%
  \BibitemOpen
  \bibfield  {author} {\bibinfo {author} {\bibfnamefont {B.~P.}\ \bibnamefont
  {Abbott}} \emph {et~al.} (\bibinfo {collaboration} {Virgo, LIGO
  Scientific}),\ }\href@noop {} {\  (\bibinfo {year} {2016}{\natexlab{b}})},\
  \Eprint {http://arxiv.org/abs/1602.03842} {arXiv:1602.03842 [astro-ph.HE]}
  \BibitemShut {NoStop}%
\bibitem [{\citenamefont {Abbott}\ \emph
  {et~al.}(2016{\natexlab{c}})\citenamefont {Abbott} \emph
  {et~al.}}]{TheLIGOScientific:2016htt}%
  \BibitemOpen
  \bibfield  {author} {\bibinfo {author} {\bibfnamefont {B.~P.}\ \bibnamefont
  {Abbott}} \emph {et~al.} (\bibinfo {collaboration} {Virgo, LIGO
  Scientific}),\ }\href {\doibase 10.3847/2041-8205/818/2/L22} {\bibfield
  {journal} {\bibinfo  {journal} {Astrophys. J.}\ }\textbf {\bibinfo {volume}
  {818}},\ \bibinfo {pages} {L22} (\bibinfo {year} {2016}{\natexlab{c}})},\
  \Eprint {http://arxiv.org/abs/1602.03846} {arXiv:1602.03846 [astro-ph.HE]}
  \BibitemShut {NoStop}%
\bibitem [{\citenamefont {Vitale}\ \emph {et~al.}(2014)\citenamefont {Vitale},
  \citenamefont {Lynch}, \citenamefont {Veitch}, \citenamefont {Raymond},\ and\
  \citenamefont {Sturani}}]{Vitale:2014mka}%
  \BibitemOpen
  \bibfield  {author} {\bibinfo {author} {\bibfnamefont {S.}~\bibnamefont
  {Vitale}}, \bibinfo {author} {\bibfnamefont {R.}~\bibnamefont {Lynch}},
  \bibinfo {author} {\bibfnamefont {J.}~\bibnamefont {Veitch}}, \bibinfo
  {author} {\bibfnamefont {V.}~\bibnamefont {Raymond}}, \ and\ \bibinfo
  {author} {\bibfnamefont {R.}~\bibnamefont {Sturani}},\ }\href {\doibase
  10.1103/PhysRevLett.112.251101} {\bibfield  {journal} {\bibinfo  {journal}
  {Phys. Rev. Lett.}\ }\textbf {\bibinfo {volume} {112}},\ \bibinfo {pages}
  {251101} (\bibinfo {year} {2014})},\ \Eprint {http://arxiv.org/abs/1403.0129}
  {arXiv:1403.0129 [gr-qc]} \BibitemShut {NoStop}%
\bibitem [{\citenamefont {Vitale}(tion)}]{Vitale}%
  \BibitemOpen
  \bibfield  {author} {\bibinfo {author} {\bibfnamefont {S.}~\bibnamefont
  {Vitale}},\ }\href@noop {} {} (\bibinfo {year} {private
  communication})\BibitemShut {NoStop}%
\bibitem [{\citenamefont {Vitale}\ \emph {et~al.}(2016)\citenamefont {Vitale},
  \citenamefont {Lynch}, \citenamefont {Raymond}, \citenamefont {Sturani},
  \citenamefont {Veitch},\ and\ \citenamefont {Graff}}]{Vitale:2016avz}%
  \BibitemOpen
  \bibfield  {author} {\bibinfo {author} {\bibfnamefont {S.}~\bibnamefont
  {Vitale}}, \bibinfo {author} {\bibfnamefont {R.}~\bibnamefont {Lynch}},
  \bibinfo {author} {\bibfnamefont {V.}~\bibnamefont {Raymond}}, \bibinfo
  {author} {\bibfnamefont {R.}~\bibnamefont {Sturani}}, \bibinfo {author}
  {\bibfnamefont {J.}~\bibnamefont {Veitch}}, \ and\ \bibinfo {author}
  {\bibfnamefont {P.}~\bibnamefont {Graff}},\ }\href@noop {} {\  (\bibinfo
  {year} {2016})},\ \Eprint {http://arxiv.org/abs/1611.01122} {arXiv:1611.01122
  [gr-qc]} \BibitemShut {NoStop}%
\bibitem [{\citenamefont {Vitale}\ and\ \citenamefont
  {Evans}(2016)}]{Vitale:2016icu}%
  \BibitemOpen
  \bibfield  {author} {\bibinfo {author} {\bibfnamefont {S.}~\bibnamefont
  {Vitale}}\ and\ \bibinfo {author} {\bibfnamefont {M.}~\bibnamefont {Evans}},\
  }\href@noop {} {\  (\bibinfo {year} {2016})},\ \Eprint
  {http://arxiv.org/abs/1610.06917} {arXiv:1610.06917 [gr-qc]} \BibitemShut
  {NoStop}%
\bibitem [{\citenamefont {Yoshino}\ and\ \citenamefont
  {Kodama}(2015)}]{Yoshino:2015nsa}%
  \BibitemOpen
  \bibfield  {author} {\bibinfo {author} {\bibfnamefont {H.}~\bibnamefont
  {Yoshino}}\ and\ \bibinfo {author} {\bibfnamefont {H.}~\bibnamefont
  {Kodama}},\ }\href {\doibase 10.1088/0264-9381/32/21/214001} {\bibfield
  {journal} {\bibinfo  {journal} {Class. Quant. Grav.}\ }\textbf {\bibinfo
  {volume} {32}},\ \bibinfo {pages} {214001} (\bibinfo {year} {2015})},\
  \Eprint {http://arxiv.org/abs/1505.00714} {arXiv:1505.00714 [gr-qc]}
  \BibitemShut {NoStop}%
\bibitem [{\citenamefont {Abbott}\ \emph
  {et~al.}(2016{\natexlab{d}})\citenamefont {Abbott} \emph
  {et~al.}}]{Abbott:2016blz}%
  \BibitemOpen
  \bibfield  {author} {\bibinfo {author} {\bibfnamefont {B.~P.}\ \bibnamefont
  {Abbott}} \emph {et~al.} (\bibinfo {collaboration} {Virgo, LIGO
  Scientific}),\ }\href {\doibase 10.1103/PhysRevLett.116.061102} {\bibfield
  {journal} {\bibinfo  {journal} {Phys. Rev. Lett.}\ }\textbf {\bibinfo
  {volume} {116}},\ \bibinfo {pages} {061102} (\bibinfo {year}
  {2016}{\natexlab{d}})},\ \Eprint {http://arxiv.org/abs/1602.03837}
  {arXiv:1602.03837 [gr-qc]} \BibitemShut {NoStop}%
\bibitem [{\citenamefont {Wette}(2009)}]{Wette:2009uea}%
  \BibitemOpen
  \bibfield  {author} {\bibinfo {author} {\bibfnamefont {K.~W.}\ \bibnamefont
  {Wette}},\ }\emph {\bibinfo {title} {{Gravitational waves from accreting
  neutron stars and Cassiopeia A}}},\ \href
  {http://inspirehep.net/record/1262892/files/wette_thesis.pdf} {Ph.D.
  thesis},\ \bibinfo  {school} {Australian Natl. U., Canberra} (\bibinfo {year}
  {2009})\BibitemShut {NoStop}%
\bibitem [{\citenamefont {Abbott}\ \emph
  {et~al.}(2016{\natexlab{e}})\citenamefont {Abbott} \emph
  {et~al.}}]{TheLIGOScientific:2016uns}%
  \BibitemOpen
  \bibfield  {author} {\bibinfo {author} {\bibfnamefont {B.~P.}\ \bibnamefont
  {Abbott}} \emph {et~al.} (\bibinfo {collaboration} {Virgo, LIGO
  Scientific}),\ }\href {\doibase 10.1103/PhysRevD.94.102002} {\bibfield
  {journal} {\bibinfo  {journal} {Phys. Rev.}\ }\textbf {\bibinfo {volume}
  {D94}},\ \bibinfo {pages} {102002} (\bibinfo {year} {2016}{\natexlab{e}})},\
  \Eprint {http://arxiv.org/abs/1606.09619} {arXiv:1606.09619 [gr-qc]}
  \BibitemShut {NoStop}%
\bibitem [{\citenamefont {Amaro-Seoane}\ \emph {et~al.}(2012)\citenamefont
  {Amaro-Seoane} \emph {et~al.}}]{AmaroSeoane:2012je}%
  \BibitemOpen
  \bibfield  {author} {\bibinfo {author} {\bibfnamefont {P.}~\bibnamefont
  {Amaro-Seoane}} \emph {et~al.},\ }\href {\doibase
  10.1088/0264-9381/29/12/124016} {\bibfield  {journal} {\bibinfo  {journal}
  {Class. Quant. Grav.}\ }\textbf {\bibinfo {volume} {29}},\ \bibinfo {pages}
  {124016} (\bibinfo {year} {2012})},\ \Eprint {http://arxiv.org/abs/1202.0839}
  {arXiv:1202.0839 [gr-qc]} \BibitemShut {NoStop}%
\bibitem [{\citenamefont {Belczynski}\ \emph {et~al.}(2015)\citenamefont
  {Belczynski}, \citenamefont {Repetto}, \citenamefont {Holz}, \citenamefont
  {O'Shaughnessy}, \citenamefont {Bulik}, \citenamefont {Berti}, \citenamefont
  {Fryer},\ and\ \citenamefont {Dominik}}]{Belczynski:2015tba}%
  \BibitemOpen
  \bibfield  {author} {\bibinfo {author} {\bibfnamefont {K.}~\bibnamefont
  {Belczynski}}, \bibinfo {author} {\bibfnamefont {S.}~\bibnamefont {Repetto}},
  \bibinfo {author} {\bibfnamefont {D.}~\bibnamefont {Holz}}, \bibinfo {author}
  {\bibfnamefont {R.}~\bibnamefont {O'Shaughnessy}}, \bibinfo {author}
  {\bibfnamefont {T.}~\bibnamefont {Bulik}}, \bibinfo {author} {\bibfnamefont
  {E.}~\bibnamefont {Berti}}, \bibinfo {author} {\bibfnamefont
  {C.}~\bibnamefont {Fryer}}, \ and\ \bibinfo {author} {\bibfnamefont
  {M.}~\bibnamefont {Dominik}},\ }\href@noop {} {\  (\bibinfo {year} {2015})},\
  \Eprint {http://arxiv.org/abs/1510.04615} {arXiv:1510.04615 [astro-ph.HE]}
  \BibitemShut {NoStop}%
\bibitem [{\citenamefont {Abbott}\ \emph {et~al.}(2017)\citenamefont {Abbott}
  \emph {et~al.}}]{Abbott:2017iws}%
  \BibitemOpen
  \bibfield  {author} {\bibinfo {author} {\bibfnamefont {B.~P.}\ \bibnamefont
  {Abbott}} \emph {et~al.} (\bibinfo {collaboration} {Virgo, LIGO
  Scientific}),\ }\href@noop {} {\  (\bibinfo {year} {2017})},\ \Eprint
  {http://arxiv.org/abs/1704.04628} {arXiv:1704.04628 [gr-qc]} \BibitemShut
  {NoStop}%
\bibitem [{\citenamefont {Abadie}\ \emph {et~al.}(2010)\citenamefont {Abadie}
  \emph {et~al.}}]{Abadie:2010cf}%
  \BibitemOpen
  \bibfield  {author} {\bibinfo {author} {\bibfnamefont {J.}~\bibnamefont
  {Abadie}} \emph {et~al.} (\bibinfo {collaboration} {VIRGO, LIGO
  Scientific}),\ }\href {\doibase 10.1088/0264-9381/27/17/173001} {\bibfield
  {journal} {\bibinfo  {journal} {Class. Quant. Grav.}\ }\textbf {\bibinfo
  {volume} {27}},\ \bibinfo {pages} {173001} (\bibinfo {year} {2010})},\
  \Eprint {http://arxiv.org/abs/1003.2480} {arXiv:1003.2480 [astro-ph.HE]}
  \BibitemShut {NoStop}%
\bibitem [{\citenamefont {Abbott}\ \emph
  {et~al.}(2016{\natexlab{f}})\citenamefont {Abbott} \emph
  {et~al.}}]{Abbott:2016iqz}%
  \BibitemOpen
  \bibfield  {author} {\bibinfo {author} {\bibfnamefont {B.~P.}\ \bibnamefont
  {Abbott}} \emph {et~al.} (\bibinfo {collaboration} {InterPlanetary Network,
  DES, INTEGRAL, La Silla-QUEST Survey, MWA, Fermi-LAT, J-GEM, DEC, Zadko,
  GRAWITA, Pi of the Sky, MASTER, Swift, iPTF, VISTA, ASKAP, SkyMapper, PESSTO,
  TOROS, Pan-STARRS, Virgo, Algerian National Observatory, Liverpool Telescope,
  BOOTES, LIGO Scientific, LOFAR, TAROT, C2PU, MAXI, Fermi-GBM}),\ }\href
  {\doibase 10.3847/0067-0049/225/1/8} {\bibfield  {journal} {\bibinfo
  {journal} {Astrophys. J. Suppl.}\ }\textbf {\bibinfo {volume} {225}},\
  \bibinfo {pages} {8} (\bibinfo {year} {2016}{\natexlab{f}})},\ \Eprint
  {http://arxiv.org/abs/1604.07864} {arXiv:1604.07864 [astro-ph.HE]}
  \BibitemShut {NoStop}%
\bibitem [{\citenamefont {Abbott}\ \emph
  {et~al.}(2016{\natexlab{g}})\citenamefont {Abbott} \emph
  {et~al.}}]{Abbott:2016izl}%
  \BibitemOpen
  \bibfield  {author} {\bibinfo {author} {\bibfnamefont {B.}~\bibnamefont
  {Abbott}} \emph {et~al.} (\bibinfo {collaboration} {Virgo, LIGO
  Scientific}),\ }\href {\doibase 10.1103/PhysRevX.6.041014} {\bibfield
  {journal} {\bibinfo  {journal} {Phys. Rev.}\ }\textbf {\bibinfo {volume}
  {X6}},\ \bibinfo {pages} {041014} (\bibinfo {year} {2016}{\natexlab{g}})},\
  \Eprint {http://arxiv.org/abs/1606.01210} {arXiv:1606.01210 [gr-qc]}
  \BibitemShut {NoStop}%
\bibitem [{\citenamefont {Buonanno}\ \emph {et~al.}(2008)\citenamefont
  {Buonanno}, \citenamefont {Kidder},\ and\ \citenamefont
  {Lehner}}]{Buonanno:2007sv}%
  \BibitemOpen
  \bibfield  {author} {\bibinfo {author} {\bibfnamefont {A.}~\bibnamefont
  {Buonanno}}, \bibinfo {author} {\bibfnamefont {L.~E.}\ \bibnamefont
  {Kidder}}, \ and\ \bibinfo {author} {\bibfnamefont {L.}~\bibnamefont
  {Lehner}},\ }\href {\doibase 10.1103/PhysRevD.77.026004} {\bibfield
  {journal} {\bibinfo  {journal} {Phys. Rev.}\ }\textbf {\bibinfo {volume}
  {D77}},\ \bibinfo {pages} {026004} (\bibinfo {year} {2008})},\ \Eprint
  {http://arxiv.org/abs/0709.3839} {arXiv:0709.3839 [astro-ph]} \BibitemShut
  {NoStop}%
\bibitem [{\citenamefont {Dimopoulos}\ \emph {et~al.}(2008)\citenamefont
  {Dimopoulos}, \citenamefont {Graham}, \citenamefont {Hogan}, \citenamefont
  {Kasevich},\ and\ \citenamefont {Rajendran}}]{Dimopoulos:2008sv}%
  \BibitemOpen
  \bibfield  {author} {\bibinfo {author} {\bibfnamefont {S.}~\bibnamefont
  {Dimopoulos}}, \bibinfo {author} {\bibfnamefont {P.~W.}\ \bibnamefont
  {Graham}}, \bibinfo {author} {\bibfnamefont {J.~M.}\ \bibnamefont {Hogan}},
  \bibinfo {author} {\bibfnamefont {M.~A.}\ \bibnamefont {Kasevich}}, \ and\
  \bibinfo {author} {\bibfnamefont {S.}~\bibnamefont {Rajendran}},\ }\href
  {\doibase 10.1103/PhysRevD.78.122002} {\bibfield  {journal} {\bibinfo
  {journal} {Phys. Rev.}\ }\textbf {\bibinfo {volume} {D78}},\ \bibinfo {pages}
  {122002} (\bibinfo {year} {2008})},\ \Eprint {http://arxiv.org/abs/0806.2125}
  {arXiv:0806.2125 [gr-qc]} \BibitemShut {NoStop}%
\bibitem [{\citenamefont {Graham}\ \emph {et~al.}(2013)\citenamefont {Graham},
  \citenamefont {Hogan}, \citenamefont {Kasevich},\ and\ \citenamefont
  {Rajendran}}]{Graham:2012sy}%
  \BibitemOpen
  \bibfield  {author} {\bibinfo {author} {\bibfnamefont {P.~W.}\ \bibnamefont
  {Graham}}, \bibinfo {author} {\bibfnamefont {J.~M.}\ \bibnamefont {Hogan}},
  \bibinfo {author} {\bibfnamefont {M.~A.}\ \bibnamefont {Kasevich}}, \ and\
  \bibinfo {author} {\bibfnamefont {S.}~\bibnamefont {Rajendran}},\ }\href
  {\doibase 10.1103/PhysRevLett.110.171102} {\bibfield  {journal} {\bibinfo
  {journal} {Phys. Rev. Lett.}\ }\textbf {\bibinfo {volume} {110}},\ \bibinfo
  {pages} {171102} (\bibinfo {year} {2013})},\ \Eprint
  {http://arxiv.org/abs/1206.0818} {arXiv:1206.0818 [quant-ph]} \BibitemShut
  {NoStop}%
\bibitem [{\citenamefont {Audley}\ \emph {et~al.}(2017)\citenamefont {Audley}
  \emph {et~al.}}]{Audley:2017drz}%
  \BibitemOpen
  \bibfield  {author} {\bibinfo {author} {\bibfnamefont {H.}~\bibnamefont
  {Audley}} \emph {et~al.},\ }\href@noop {} {\  (\bibinfo {year} {2017})},\
  \Eprint {http://arxiv.org/abs/1702.00786} {arXiv:1702.00786 [astro-ph.IM]}
  \BibitemShut {NoStop}%
\bibitem [{\citenamefont {Pani}\ and\ \citenamefont
  {Loeb}(2013)}]{Pani:2013hpa}%
  \BibitemOpen
  \bibfield  {author} {\bibinfo {author} {\bibfnamefont {P.}~\bibnamefont
  {Pani}}\ and\ \bibinfo {author} {\bibfnamefont {A.}~\bibnamefont {Loeb}},\
  }\href {\doibase 10.1103/PhysRevD.88.041301} {\bibfield  {journal} {\bibinfo
  {journal} {Phys. Rev.}\ }\textbf {\bibinfo {volume} {D88}},\ \bibinfo {pages}
  {041301} (\bibinfo {year} {2013})},\ \Eprint {http://arxiv.org/abs/1307.5176}
  {arXiv:1307.5176 [astro-ph.CO]} \BibitemShut {NoStop}%
\bibitem [{\citenamefont {Conlon}\ and\ \citenamefont
  {Herdeiro}(2017)}]{Conlon:2017hhi}%
  \BibitemOpen
  \bibfield  {author} {\bibinfo {author} {\bibfnamefont {J.~P.}\ \bibnamefont
  {Conlon}}\ and\ \bibinfo {author} {\bibfnamefont {C.~A.~R.}\ \bibnamefont
  {Herdeiro}},\ }\href@noop {} {\  (\bibinfo {year} {2017})},\ \Eprint
  {http://arxiv.org/abs/1701.02034} {arXiv:1701.02034 [astro-ph.HE]}
  \BibitemShut {NoStop}%
\bibitem [{\citenamefont {Raffelt}(1996)}]{Raffelt:1996wa}%
  \BibitemOpen
  \bibfield  {author} {\bibinfo {author} {\bibfnamefont {G.~G.}\ \bibnamefont
  {Raffelt}},\ }\href
  {http://wwwth.mpp.mpg.de/members/raffelt/mypapers/199613.pdf} {\emph
  {\bibinfo {title} {{Stars as laboratories for fundamental physics}}}}\
  (\bibinfo {year} {1996})\BibitemShut {NoStop}%
\bibitem [{\citenamefont {Blandford}\ and\ \citenamefont
  {Znajek}(1977)}]{Blandford:1977ds}%
  \BibitemOpen
  \bibfield  {author} {\bibinfo {author} {\bibfnamefont {R.~D.}\ \bibnamefont
  {Blandford}}\ and\ \bibinfo {author} {\bibfnamefont {R.~L.}\ \bibnamefont
  {Znajek}},\ }\href@noop {} {\bibfield  {journal} {\bibinfo  {journal} {Mon.
  Not. Roy. Astron. Soc.}\ }\textbf {\bibinfo {volume} {179}},\ \bibinfo
  {pages} {433} (\bibinfo {year} {1977})}\BibitemShut {NoStop}%
\bibitem [{\citenamefont {Adelberger}\ \emph {et~al.}(2003)\citenamefont
  {Adelberger}, \citenamefont {Heckel},\ and\ \citenamefont
  {Nelson}}]{Adelberger:2003zx}%
  \BibitemOpen
  \bibfield  {author} {\bibinfo {author} {\bibfnamefont {E.~G.}\ \bibnamefont
  {Adelberger}}, \bibinfo {author} {\bibfnamefont {B.~R.}\ \bibnamefont
  {Heckel}}, \ and\ \bibinfo {author} {\bibfnamefont {A.~E.}\ \bibnamefont
  {Nelson}},\ }\href {\doibase 10.1146/annurev.nucl.53.041002.110503}
  {\bibfield  {journal} {\bibinfo  {journal} {Ann. Rev. Nucl. Part. Sci.}\
  }\textbf {\bibinfo {volume} {53}},\ \bibinfo {pages} {77} (\bibinfo {year}
  {2003})},\ \Eprint {http://arxiv.org/abs/hep-ph/0307284}
  {arXiv:hep-ph/0307284 [hep-ph]} \BibitemShut {NoStop}%
\bibitem [{\citenamefont {Wagner}\ \emph {et~al.}(2012)\citenamefont {Wagner},
  \citenamefont {Schlamminger}, \citenamefont {Gundlach},\ and\ \citenamefont
  {Adelberger}}]{Wagner:2012ui}%
  \BibitemOpen
  \bibfield  {author} {\bibinfo {author} {\bibfnamefont {T.~A.}\ \bibnamefont
  {Wagner}}, \bibinfo {author} {\bibfnamefont {S.}~\bibnamefont
  {Schlamminger}}, \bibinfo {author} {\bibfnamefont {J.~H.}\ \bibnamefont
  {Gundlach}}, \ and\ \bibinfo {author} {\bibfnamefont {E.~G.}\ \bibnamefont
  {Adelberger}},\ }\href {\doibase 10.1088/0264-9381/29/18/184002} {\bibfield
  {journal} {\bibinfo  {journal} {Class. Quant. Grav.}\ }\textbf {\bibinfo
  {volume} {29}},\ \bibinfo {pages} {184002} (\bibinfo {year} {2012})},\
  \Eprint {http://arxiv.org/abs/1207.2442} {arXiv:1207.2442 [gr-qc]}
  \BibitemShut {NoStop}%
\bibitem [{\citenamefont {Teukolsky}(1973)}]{Teukolsky:1973ha}%
  \BibitemOpen
  \bibfield  {author} {\bibinfo {author} {\bibfnamefont {S.~A.}\ \bibnamefont
  {Teukolsky}},\ }\href {\doibase 10.1086/152444} {\bibfield  {journal}
  {\bibinfo  {journal} {Astrophys. J.}\ }\textbf {\bibinfo {volume} {185}},\
  \bibinfo {pages} {635} (\bibinfo {year} {1973})}\BibitemShut {NoStop}%
\bibitem [{\citenamefont {Shakura}\ and\ \citenamefont
  {Sunyaev}(1973)}]{shakura1973black}%
  \BibitemOpen
  \bibfield  {author} {\bibinfo {author} {\bibfnamefont {N.}~\bibnamefont
  {Shakura}}\ and\ \bibinfo {author} {\bibfnamefont {R.}~\bibnamefont
  {Sunyaev}},\ }\href@noop {} {\bibfield  {journal} {\bibinfo  {journal}
  {Astronomy and Astrophysics}\ }\textbf {\bibinfo {volume} {24}},\ \bibinfo
  {pages} {337} (\bibinfo {year} {1973})}\BibitemShut {NoStop}%
\bibitem [{\citenamefont {Misner}\ \emph {et~al.}(1973)\citenamefont {Misner},
  \citenamefont {Thorne},\ and\ \citenamefont {Wheeler}}]{Misner:1974qy}%
  \BibitemOpen
  \bibfield  {author} {\bibinfo {author} {\bibfnamefont {C.~W.}\ \bibnamefont
  {Misner}}, \bibinfo {author} {\bibfnamefont {K.~S.}\ \bibnamefont {Thorne}},
  \ and\ \bibinfo {author} {\bibfnamefont {J.~A.}\ \bibnamefont {Wheeler}},\
  }\href@noop {} {\emph {\bibinfo {title} {{Gravitation}}}}\ (\bibinfo
  {publisher} {W. H. Freeman},\ \bibinfo {address} {San Francisco},\ \bibinfo
  {year} {1973})\BibitemShut {NoStop}%
\bibitem [{\citenamefont {Weinberg}(1972)}]{Weinberg:1972kfs}%
  \BibitemOpen
  \bibfield  {author} {\bibinfo {author} {\bibfnamefont {S.}~\bibnamefont
  {Weinberg}},\ }\href
  {http://www-spires.fnal.gov/spires/find/books/www?cl=QC6.W431} {\emph
  {\bibinfo {title} {{Gravitation and Cosmology}}}}\ (\bibinfo  {publisher}
  {John Wiley and Sons},\ \bibinfo {address} {New York},\ \bibinfo {year}
  {1972})\BibitemShut {NoStop}%
\bibitem [{\citenamefont {Poisson}(1993)}]{Poisson:1993vp}%
  \BibitemOpen
  \bibfield  {author} {\bibinfo {author} {\bibfnamefont {E.}~\bibnamefont
  {Poisson}},\ }\href {\doibase 10.1103/PhysRevD.47.1497} {\bibfield  {journal}
  {\bibinfo  {journal} {Phys. Rev.}\ }\textbf {\bibinfo {volume} {D47}},\
  \bibinfo {pages} {1497} (\bibinfo {year} {1993})}\BibitemShut {NoStop}%
\bibitem [{\citenamefont {Kopparapu}\ \emph {et~al.}(2008)\citenamefont
  {Kopparapu}, \citenamefont {Hanna}, \citenamefont {Kalogera}, \citenamefont
  {O’Shaughnessy}, \citenamefont {González}, \citenamefont {Brady},\ and\
  \citenamefont {Fairhurst}}]{0004-637X-675-2-1459}%
  \BibitemOpen
  \bibfield  {author} {\bibinfo {author} {\bibfnamefont {R.~K.}\ \bibnamefont
  {Kopparapu}}, \bibinfo {author} {\bibfnamefont {C.}~\bibnamefont {Hanna}},
  \bibinfo {author} {\bibfnamefont {V.}~\bibnamefont {Kalogera}}, \bibinfo
  {author} {\bibfnamefont {R.}~\bibnamefont {O’Shaughnessy}}, \bibinfo
  {author} {\bibfnamefont {G.}~\bibnamefont {González}}, \bibinfo {author}
  {\bibfnamefont {P.~R.}\ \bibnamefont {Brady}}, \ and\ \bibinfo {author}
  {\bibfnamefont {S.}~\bibnamefont {Fairhurst}},\ }\href
  {http://stacks.iop.org/0004-637X/675/i=2/a=1459} {\bibfield  {journal}
  {\bibinfo  {journal} {The Astrophysical Journal}\ }\textbf {\bibinfo {volume}
  {675}},\ \bibinfo {pages} {1459} (\bibinfo {year} {2008})}\BibitemShut
  {NoStop}%
\end{thebibliography}%

\end{document}